\pdfoutput=1
\documentclass[11pt,twoside,a4paper,cmspaper,final,collab]{cms-tdr}
\def\svnVersion{ddf7f8d}\def\svnDate{2023/07/10}\def\cmsCernNoTag{CERN-EP-2022-175}\def\cmsCernDate{\today}\def\cmsMessage{Published in the Journal of High Energy Physics as \href{http://dx.doi.org/10.1007/JHEP07(2023)020}{\doi{10.1007/JHEP07(2023)020}.}}
\begin{document}\cmsNoteHeader{B2G-20-011}

\providecommand{\PQT}{\ensuremath{\cmsSymbolFace{T}}\xspace}
\providecommand{\PQB}{\ensuremath{\cmsSymbolFace{B}}\xspace}
\providecommand{\PAQT}{\ensuremath{\overline{\cmsSymbolFace{T}}}\xspace}
\providecommand{\PAQB}{\ensuremath{\overline{\cmsSymbolFace{B}}}\xspace}
\newcommand{\ST}{\ensuremath{S_\mathrm{T}}\xspace}
\newcommand{\TTbar}{\ensuremath{\PQT\PAQT}\xspace}
\newcommand{\BBbar}{\ensuremath{\PQB\PAQB}\xspace}
\newcommand{\HTlep}{\ensuremath{\HT^{\mathrm{lep}}}\xspace}
\newcommand{\wjet}{{\PW}+jets\xspace}
\providecommand{\cmsTable}[1]{\resizebox{\textwidth}{!}{#1}}
\providecommand{\NA}{\ensuremath{\text{---}}}
\newcommand{\eee}{\ensuremath{\Pe\Pe\Pe}}
\newcommand{\eem}{\ensuremath{\Pe\Pe\PGm}}
\newcommand{\emm}{\ensuremath{\Pe\PGm\PGm}}
\newcommand{\mmm}{\ensuremath{\PGm\PGm\PGm}}

\cmsNoteHeader{B2G-20-011}
\title{Search for pair production of vector-like quarks in leptonic final states in proton-proton collisions at \texorpdfstring{$\sqrt{s} = 13\TeV$}{sqrt(s) = 13 TeV}}

\author*[cern]{The CMS Collaboration}
\date{\today}

\abstract{
  A search is presented for vector-like \PQT and \PQB quark-antiquark pairs produced in proton-proton collisions at a center-of-mass energy of 13\TeV.
  Data were collected by the CMS experiment at the CERN LHC in 2016--2018, with an integrated luminosity of 138\fbinv.
  Events are separated into single-lepton, same-sign charge dilepton, and multilepton channels.
  In the analysis of the single-lepton channel a multilayer neural network and jet identification techniques are employed to select signal events, while the same-sign dilepton and multilepton channels rely on the high-energy signature of the signal to distinguish it from standard model backgrounds.
  The data are consistent with standard model background predictions, and the production of vector-like quark pairs is excluded at 95\% confidence level for \PQT quark masses up to 1.54\TeV and \PQB quark masses up to 1.56\TeV, depending on the branching fractions assumed, with maximal sensitivity to decay modes that include multiple top quarks.
  The limits obtained in this search are the strongest limits to date for \TTbar production, excluding masses below 1.48\TeV for all decays to third generation quarks, and are the strongest limits to date for \BBbar production with \PQB quark decays to {\PQt}\PW.
}

\hypersetup{
  pdfauthor={CMS Collaboration},
  pdftitle={Search for pair production of vector-like quarks in leptonic final states at \texorpdfstring{$\sqrt{s} = 13\TeV$}{sqrt(s) = 13 TeV}},
  pdfsubject={CMS},
  pdfkeywords={CMS, B2G, VLQ}
}

\maketitle

\section{Introduction}

A decade ago, the ATLAS and CMS Collaborations announced the discovery of a Higgs boson (\PH) with a mass near 125\GeV at the CERN LHC \cite{ATLAS:2012yve_Higgs,CMS:2012qbp_Higgs,CMS:2013btf_Higgs}.
The discovery experimentally confirmed the last fundamental piece of the standard model (SM) of particle physics, and subsequent precision measurements have remained consistent with the SM description of the properties and interactions of elementary particles.
However, the apparent fine-tuning of SM parameters and other unexplained phenomena indicate the incompleteness of the SM~\cite{PDGNeutrino}, and thus motivate searches for beyond-the-SM (BSM) physics.
To restore ``naturalness'' to the SM, many BSM theories (Little Higgs~\cite{PhysRevD.69.075002,Matsedonskyi2013}, Composite Higgs~\cite{CH1,CH2}, etc.) introduce new heavy fermions, such as ``vector-like quarks'' (VLQs).
The presence of VLQs is an extension of the SM that is not currently excluded by experiments.

The VLQs are hypothetical fermions whose left- and right-handed components transform identically under the SM electroweak gauge group $\text{SU}(2)_L \otimes \text{U}(1)_Y$, in contrast to the behavior of the SM chiral quarks.
This chiral symmetry allows a mass term to be included in the Lagrangian, which means that the VLQ masses are not dependent on Higgs Yukawa couplings.
The existence of such VLQs could cancel out the leading quantum loop corrections to the observed \PH mass from top quarks, thus stabilizing it~\cite{VLQ1,VLQ2}.

\begin{figure}[htbp]
  \begin{center}
    \includegraphics[width=0.38\textwidth]{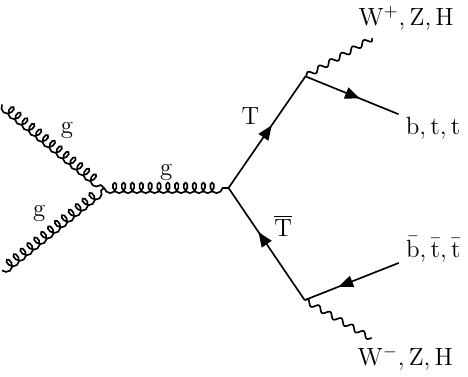}
    \hspace{1.5cm}
    \includegraphics[width=0.38\textwidth]{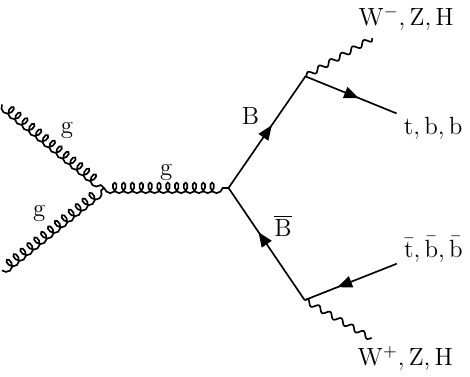}
    \caption{Representative Feynman diagrams of the pair production of \TTbar (left) and \BBbar (right), with decays to third generation quarks and SM bosons.}
    \label{fig:diagrams}
  \end{center}
\end{figure}

In most of the relevant models, VLQs are assumed to mix primarily with the third-generation SM quarks, as is consistent with the expected large top quark coupling~\cite{VLQmixing}.
At the LHC, VLQs can be produced in pairs via the strong interaction or singly via an electroweak interaction. At high masses, VLQ pair production typically has a lower cross section than single production, but for the narrow-width VLQs assumed by this analysis the pair production cross section is independent of the electroweak couplings, simplifying interpretation. 
A vector-like top quark ($\PQT$) with an electric charge of $2e/3$, analogous to the SM top quark, can decay in three modes that produce characteristic high-momentum signatures: $\PQT\to \PQt\PH$, $\PQt\PZ$, and $\PQb\PW$~\cite{VLQ1}.
Similarly, a vector-like bottom quark (\PQB) with an electric charge of $-e/3$ can decay through $\PQB \to \PQb\PH$, $\PQb\PZ$, and $\PQt\PW$.
Examples of tree-level Feynman diagrams are shown in Fig.~\ref{fig:diagrams}.

In minimal models, these VLQs may only exist as electroweak singlets $\PQT$ and $\PQB$, in a doublet ($\PQT$, $\PQB$), or in doublets and triplets with further VLQs that would have exotic charges.
Each scenario results in different $\PQT$ and $\PQB$ branching fractions.
For singlets, the branching fractions are 50\% for $\PQT \to \PQb\PW$ and $\PQB \to \PQt\PW$, and 25\% for $\PQT \to \PQt\PH$, $\PQT\to\PQt\PZ$, $\PQB \to\PQb\PH$, and $\PQB\to\PQb\PZ$.
In various doublet scenarios, the $\PQT$ decays only to $\PQt\PH$ and $\PQt\PZ$ with equal branching fractions of 50\%, and similarly the $\PQB$ decays only to $\PQb\PH$ and $\PQb\PZ$ with equal branching fractions~\cite{delAguila:1989rq, VLQ2}.
These singlet and doublet branching fraction scenarios are used as benchmarks.

In this paper, a search for pair production of $\PQT$ and $\PQB$ quarks at the LHC is presented using three final states containing charged electrons or muons: a single-lepton channel, a same-sign charge (SS) dilepton channel, and a ``multilepton'' channel with at least three leptons.
The data were collected from proton-proton ($\Pp\Pp$) collisions at $\sqrt{s}=13\TeV$ by the CMS experiment at the LHC from 2016--2018, with a total integrated luminosity of 138\fbinv.
It is assumed that only one flavor of VLQ is present, and results are independently derived for the production of \PQT and \PQB quarks.
Previous searches for $\PQT$ and $\PQB$ quark pairs by the ATLAS and CMS Collaborations at $\sqrt{s} = 7\TeV$~\cite{EXO-11-005,EXO-11-099,Aad:2012bdq}, 8\TeV~\cite{CMScombo2014,Run1anal,PhysRevD.92.112007,Aad2015}, and 13\TeV~\cite{B2G-17-003,B2G-17-011,B2G-17-012,B2G-18-005,B2G-19-005,Atlas2016combo} have excluded \PQT quark masses below 1.31\TeV (singlet) to 1.42\TeV (100\% {\PQt}\PH) and \PQB quark masses below 1.22\TeV (singlet) to 1.58\TeV (100\% {\PQb}\PH) at 95\% confidence level (\CL).
For decays to \PW bosons, pair production of either VLQ flavor is currently excluded for masses below $\sim$1.35\TeV~\cite{B2G-17-003,Atlas2016combo}.

In the following, Section~\ref{sec:CMS} describes the CMS detector and reconstruction algorithms for common physics objects.
Section~\ref{sec:samples} describes the simulations used in the search, and Section~\ref{sec:reco} presents physics object and event selection requirements shared by all channels.
The overall analysis strategy is described in Section~\ref{sec:strategy}, with the event selection, background estimation, and event categorization presented in detail for each channel in Sections~\ref{sec:1lep}--\ref{sec:3lep}.
Treatment of systematic uncertainties for the combined search and the search results are presented in Sections~\ref{sec:syst} and~\ref{sec:results}, respectively.
The search is summarized in Section~\ref{sec:summary}.
Tabulated results are provided in the HEPData record for this search~\cite{hepdata}.

\section{The CMS detector and event reconstruction}
\label{sec:CMS}

The central feature of the CMS apparatus is a superconducting solenoid of 6\unit{m} internal diameter, providing a magnetic field of 3.8\unit{T}.
Within the solenoid volume are a silicon pixel and strip tracker, a lead tungstate crystal electromagnetic calorimeter (ECAL), and a brass and scintillator hadron calorimeter (HCAL), each composed of a barrel and two endcap sections.
Forward calorimeters extend the pseudorapidity ($\eta$) coverage provided by the barrel and endcap detectors.
Muons are detected in gas-ionization chambers embedded in the steel flux-return yoke outside the solenoid.
A more detailed description of the CMS detector, together with a definition of the coordinate system used and the relevant kinematic variables, can be found in Ref.~\cite{Chatrchyan:2008zzk}.

The particle-flow (PF) algorithm~\cite{CMS-PRF-14-001} aims to reconstruct and identify each individual particle in an event, with an optimized combination of information from the various elements of the CMS detector.
The energy of photons is obtained from the ECAL measurement.
The energy of electrons is determined from a combination of the electron momentum as determined by the tracker, the energy of the corresponding ECAL cluster, and the energy sum of all bremsstrahlung photons spatially compatible with originating from the electron track.
Muon reconstruction is based on extrapolating between hits in the inner tracker and the outermost muon chamber.
The energy of muons is obtained from the curvature of the reconstructed track.
The energy of charged hadrons is determined from a combination of their momentum measured in the tracker and the matching ECAL and HCAL energy deposits, corrected for the response function of the calorimeters to hadronic showers.
Finally, the energy of neutral hadrons is obtained from the corresponding corrected ECAL and HCAL energies.

For each event, hadronic jets are clustered from these reconstructed particles using the infrared and collinear safe anti-\kt algorithm~\cite{Cacciari:2008gp} implemented by the  \FASTJET package~\cite{Cacciari:2011ma}, with distance parameters of 0.4 (``small-radius'') and 0.8 (``large-radius'').
Jet momentum is determined as the vectorial sum of all particle momenta in the jet, and is found from simulation to be, on average, within 5--10\% of the true momentum over the entire \pt spectrum and detector acceptance.
Additional $\Pp\Pp$ interactions within the same or nearby bunch crossings (pileup) can contribute additional tracks and calorimetric energy deposits, increasing the apparent jet momentum.
To mitigate this effect in small-radius jets, tracks identified as originating from pileup vertices are discarded and an offset correction is applied to correct for remaining contributions.
In large-radius jets, the pileup-per-particle identification algorithm~\cite{Bertolini:2014bba,Sirunyan:2020foa} is used to mitigate the effect of pileup at the reconstructed-particle level, making use of local shape information, event pileup properties, and tracking information.
Jet energy corrections are derived from simulation studies so that the average measured energy of jets becomes identical to that of particle-level jets.
In situ measurements of the momentum balance in dijet, photon+jet, {\PZ}+jet, and multijet events are used to determine any residual differences between the jet energy scale in data and in simulation, and appropriate corrections are made~\cite{Khachatryan:2016kdb}.
The jet energy resolution amounts typically to 15--20\% at 30\GeV, 10\% at 100\GeV, and 5\% at 1\TeV~\cite{Khachatryan:2016kdb}, and the energy of jets in simulated samples is corrected such that the energy resolution agrees with data.

The missing transverse momentum vector \ptvecmiss is computed as the negative vector \pt sum of all the PF candidates in an event.
Its magnitude is denoted as \ptmiss.
Anomalous high-\ptmiss events can be due to a variety of reconstruction failures, detector malfunctions or noncollision backgrounds.
Such events are rejected by event filters that are designed to identify more than 85--90\% of the spurious high-\ptmiss events with a mistagging rate less than 0.1\%~\cite{Sirunyan:2019kia}.
The \ptvecmiss is modified to account for corrections to the energy scale of the reconstructed jets in the event.

Events of interest are selected using a two-tiered trigger system.
The first level (L1), composed of custom hardware processors, uses information from the calorimeters and muon detectors to select events at a rate of around 100\unit{kHz} within a fixed latency of about 4\mus~\cite{Sirunyan:2020zal}.
The second level, known as the high-level trigger, consists of a farm of processors running a version of the full event reconstruction software optimized for fast processing, and reduces the event rate to around 1\unit{kHz} before data storage~\cite{Khachatryan:2016bia}.

\section{Simulated samples}
\label{sec:samples}

Signal and background processes are simulated using the Monte Carlo (MC) method with different matrix element generators.
Simulations with conditions appropriate for the 2016 data are generated using the NNPDF~3.0 parton distribution function (PDF) set at leading order (LO) or next-to-LO (NLO)~\cite{NNPDF30}.
Background simulations with 2017--2018 conditions are generated using the NNPDF~3.1 PDF set at next-to-NLO (NNLO)~\cite{NNPDF31}, and signal simulations for 2017--2018 conditions use the PDF4LHC15 PDF set at NLO~\cite{PDF4LHC}.

The \POWHEG~v2 generator~\cite{POWHEG1,POWHEG2,POWHEG3} is used to simulate \ttbar~\cite{POWHEGttbar}, $\ttbar\PH$, and most single top quark production~\cite{POWHEGsingletopST,POWHEGsingletopTW}, as well as $\PW\PZ$ (2016 and 2018) and $\PZ\PZ$ production~\cite{POWHEGWZ1,POWHEGWZ2} for the SS dilepton and multilepton channels.
For the 2016 generation of single top quark production in the $t$ channel, \POWHEG is interfaced with \textsc{MadSpin}~\cite{MADSPIN}.
The \MGvATNLO generator~\cite{MadGraph} versions 2.2.2 (2016) and 2.4.2 (2017--2018) are used at LO with the MLM matching scheme~\cite{MLM} to simulate Drell--Yan and W boson production with leptonic decays and up to four additional partons, as well as production of multijet events from quantum chromodynamics (QCD) interactions.
They are also used at NLO to simulate $\ttbar\PW$, $\ttbar\PZ$, $\ttbar\ttbar$, $\PW^+\PW^+$, and triboson production, as well as single top quark $s$-channel production with 2016 and 2018 conditions, and $\PW\PZ$ production with 2017 conditions.
The FxFx~\cite{FXFX} matching scheme is applied to the $\ttbar\PW$ and $\PW\PZ$ samples, and the $\ttbar\PW$ samples are also interfaced with \textsc{MadSpin}.
An alternate sample of $\ttbar$ production for neural network training is also simulated at LO with \MGvATNLO.

Pair production of vector-like \PQT and \PQB quarks is simulated at LO using \MGvATNLO 2.2.2 (2016) and 2.4.2 (2017--2018).
Samples with masses in the range 0.9--1.8\TeV are generated with a narrow width, chosen to be 10\GeV.
The theoretical cross sections of vector-like \PQT and \PQB quark pair production via the strong interaction, calculated at NNLO with the \textsc{Top++}2.0 program~\cite{TPRIMEXSEC}, range from $90\pm4\unit{fb}$ at 0.9\TeV to $0.39\pm^{0.04}_{0.03}\unit{fb}$ at 1.8\TeV.

All MC generators are interfaced with \PYTHIA8~\cite{PYTHIA8} to simulate the parton showering and underlying event kinematics, with versions 8.212, 8.226, and 8.230 used for 2016, 2017, and 2018 conditions, respectively.
In the 2016 simulation, the CUETP8M1 underlying event tune~\cite{CUETP8M1} is applied to all processes except \PQt quark production processes, which use the dedicated tune CUETP8M2T4~\cite{CUETP8M2T4}.
In the 2017--2018 simulation, the tune CP5 is applied to all processes~\cite{CP5}.
\PYTHIA8 is also used as a generator to simulate diboson production at LO in the single-lepton channel.
The \GEANTfour~\cite{GEANT4} program is used to simulate the response of the CMS detector.
The effects of pileup are included in the simulation, and for each year simulated events are weighted to ensure that the mean number of interactions per bunch crossing agrees with that observed in the data.

For the single-lepton channel, background simulations are grouped into ``QCD'', ``TOP'' and ``EW'' categories.
The QCD category consists of multijet production; the TOP category consists of \ttbar, single \PQt, $\ttbar\PH$, and $\ttbar\PV$ processes, where $\PV$ denotes $\PW$ and $\PZ$; and the EW category consists of {\PW}+jets, Drell--Yan, and diboson processes.
For the SS dilepton and multi-lepton channel, the ``$\ttbar\PV$' group consists of $\ttbar\PW$ and $\ttbar\PZ$ processes, and the ``$\PV\PV$(\PV)'' group consists of all vector diboson and triboson processes.

\section{Physics object and event selection}
\label{sec:reco}

The primary vertex (PV) is taken to be the $\Pp\Pp$ collision vertex corresponding to the hardest scattering in the event, evaluated using tracking information alone, as described in Section 9.4.1 of Ref.~\cite{CMS-TDR-15-02}.
Events selected for this search are required to have at least one reconstructed vertex with longitudinal position $\abs{z}<24 \unit{cm}$ and radial position $r<2 \unit{cm}$ relative to the mean collision point and the nominal beam axis, respectively.

All selected events must have at least one electron or one muon candidate.
For the single-lepton channel, events are required to have satisfied a trigger requiring the presence of an electron or muon.
The primary trigger selected events with an electron or muon of $\pt > 15\GeV$ that is very loosely isolated from surrounding energy deposits.
These events also have transverse hadronic energy greater than 350\GeV (2016) or 450\GeV (2017--2018).
Additional triggers selected events with a muon of $\pt > 50\GeV$, or an isolated electron with $\pt > 32\, (38)\GeV$ in 2016 (2017--2018) data.
In 2017--2018 data, the SS dilepton and multilepton channels required events to have passed dielectron, dimuon, or electron-muon triggers.
For the dielectron trigger, the leading electron was required to have $\pt > 23\GeV$ and the subleading electron $\pt > 12\GeV$.
The dimuon triggers selected events with two muons of $\pt > 17$ and 8\GeV, respectively.
For triggers selecting mixed-flavor events, the leading lepton was required to have $\pt > 23\GeV$ and the subleading electron (muon) $\pt > 12\,(8)\GeV$.

During 2016 and 2017, a gradual shift in the timing of the inputs of the ECAL L1 trigger in the region $\abs{\eta} > 2.0$ caused a specific trigger inefficiency.
The effect is a function \pt, $\eta$, and time, and for events containing an electron (a jet) with \pt larger than $\approx$50 ($\approx$100\GeV), in the region $2.5 < \abs{\eta} < 3.0$ the efficiency loss is 10--20\%. To model this effect in simulation, correction factors are computed from data.

Reconstructed electrons are required to satisfy several quality criteria on observables such as the shower shape and the ratio of energy deposition in the ECAL to that in the HCAL~\cite{Khachatryan:2015hwa}.
A multivariate discriminator is used to determine the quality of reconstructed electrons with two different degrees of stringency: ``tight'' and ``loose'', with  average electron identification efficiencies of 90\% and 98\%, respectively. Misidentification rates are in the ranges 2--3\% for tight identification and 5--15\% for loose identification, depending on the electron's $\eta$~\cite{EGamma}.
The reconstructed electrons are required to lie within the range $\abs{\eta} < 2.5$, excluding the barrel-endcap interface region of $1.44 < \abs{\eta} < 1.57$.
During 2018 data taking, a detector failure in the HCAL caused jets to be misidentified as electrons, so electrons within the affected region (a range of $1.02$ in $\eta$ and $0.65$ in azimuthal angle $\phi$) are rejected in both the 2018 data and the corresponding simulation.

In the SS dilepton channel, different measurements of each electron's charge must be consistent. Three measurements are considered: two are based on electron track reconstruction, and a third is based on the $\phi$ angle difference between the pixel detector seed hits of the electron track and the linked ECAL cluster~\cite{B2G-17-014}. For each electron, the two charge measurements based on track reconstruction must agree, and for electrons with $\pt < 100\GeV$ all three measurements must agree.

Muons are selected at two quality levels, ``tight'' and ``loose'',  based on the number of muon chamber hits, the inner track impact parameter with respect to the PV, and the track fit $\chi^2$. The tight level has an efficiency of 95--99\% for identifying reconstructed muons and the loose level is over 99\% efficient, with misidentification rates for hadrons below 0.5\% in all cases~\cite{Sirunyan:2018}.
Selected muons are required to be located within the detector acceptance region of $\abs{\eta} < 2.4$.

To reduce background from leptons produced within jets, usually from semileptonic decays of hadrons, an isolation requirement is applied to leptons using an observable $I_\mathrm{mini}$.
This quantity is defined as
the scalar \pt sum of charged hadron, neutral hadron, and photon PF candidates within a cone in $\eta$-$\phi$ space around the lepton is calculated, corrected for pileup using an effective area method~\cite{Cacciari:2008gn}, and then divided by the lepton \pt.
The cone size depends on the lepton \pt and has radius $\mathcal{R}$, defined as
\begin{linenomath*}\begin{equation}\label{eqn:riso}
    \mathcal{R}\,=\frac{10\GeV}{\min(\max(\pt,50\GeV), 200\GeV)}.
  \end{equation}\end{linenomath*}
Using a \pt-dependent cone radius improves the efficiency of selecting isolated leptons from highly Lorentz-boosted particles, such as VLQ decay products.
Tight (loose) leptons require $I_\mathrm{mini} < 0.1$ (0.4), with efficiencies $>$95\% (98\%) for both lepton flavors.
To account for observed small differences in reconstruction, identification, and isolation between data and simulation, the simulation is corrected by factors estimated from data using the ``tag-and-probe'' method~\cite{Sirunyan:2018}.

Small-radius (large-radius) jets with $\pt > 30\, (200)\GeV$ and $\abs{\eta} < 2.4$ are selected if they pass selection criteria designed to remove jets dominated by instrumental effects or reconstruction failures~\cite{CMS-PAS-JME-16-003}.
Additionally, tight (single-lepton channel) or loose (SS dilepton and multi-lepton channels) leptons are removed from small-radius (large-radius) jets if their angular separation from the jet axis $\Delta R = \sqrt{\smash[b]{(\Delta\eta)^2+(\Delta\phi)^2}}< 0.4\,(0.8)$, i.e., if the leptons lie within the jet distance parameter of the jet axis.
Leptons are removed by subtracting the four-momentum of the matched lepton candidate from the four-momentum of the jet.
Jet energy corrections are applied after lepton removal.

Small-radius jets are tagged as coming from a bottom quark using the \textsc{DeepJet} algorithm~\cite{Bols:2020bkb}.
The working point used here provides 60--85\% efficiency for identifying bottom quark jets, varying with jet \pt, while misidentifying only 15--25\% of charm quark jets and 1--7\% of light-quark or gluon jets as bottom quark jets.
Corrections based on the jet \pt are applied to account for the differences between efficiencies in data and simulation~\cite{CMS-DP-2018-058}.
Small-radius jets with \pt $>$ 30\GeV and $\abs{\eta} < 2.4$ are considered in this search.

A large-radius jet may contain all the products of the hadronic decay of a top quark, or \PH, \PW, or \PZ boson that has been produced in the decay of a heavy VLQ and is therefore highly boosted.
The \textsc{DeepAK8}~\cite{JMETaggers} algorithm is used to identify the most probable parent particle of each large-radius jet as either a top or bottom quark, \PH, \PZ, or \PW boson, or light quark/gluon.
The identification is made by summing the \textsc{DeepAK8} scores for all individual decays of a specific massive particle, and determining which massive particle hypothesis has the largest score for each jet.  
In the 1.4\TeV VLQ signal simulation this tagging method has efficiencies of 54--74\% for identifying SM bosons, 76\% efficiency for identifying top quarks, 30\% efficiency for identifying bottom quarks, and 60\% efficiency for identifying light quarks or gluons.
The most significant sources of misidentification cause $\approx$25\% of \PZ bosons to be identified as \PW bosons, and 28\% of bottom quarks to be identified as Higgs bosons. Variations in the identification efficiency for each massive particle based on jet \pt or simulated VLQ mass range from 5 to 25\%, and are within the assigned uncertainties described in Section~\ref{sec:syst1lep}.
Large-radius jets are also characterized using the ratio between the 2-subjettiness and 1-subjettiness variables~\cite{Thaler:2010tr}, $\tau_{21}=\tau_{2}/\tau_{1}$, and the ``groomed'' jet mass.
The groomed jet mass is calculated after applying a modified mass-drop algorithm~\cite{Dasgupta:2013ihk,Butterworth:2008iy}, known as the ``soft-drop'' algorithm~\cite{SOFTDROP}, to large-radius jets using parameters $\beta=0$, $z_\text{cut}=0.1$, and $R_0 = 0.8$.
Large-radius jets with $\pt > 200\GeV$ and $\abs{\eta} < 2.4$ are considered in this search.

We define an observable \HT as the scalar \pt sum of all selected small-radius jets, \HTlep as the scalar sum of \HT and the \pt of all tight leptons, and \ST as the sum of \HTlep and \ptmiss.
These quantities provide a measure of the total event energy, which is typically larger for VLQ signal events than SM background events.

\section{Analysis strategy}
\label{sec:strategy}

The present search features three leptonic channels with sensitivity to different potential VLQ decays.

The single-lepton channel, described in Section~\ref{sec:1lep}, provides broad sensitivity to all \TTbar decays, as well as sensitivity to \PQB quark decays to $\PQt\PW$.
The decay of VLQ pairs to third-generation quarks and SM bosons produces two bottom or top quarks, and two \PW, \PZ, or Higgs bosons.
In the single-lepton final state, one of these top quarks or \PW bosons decays leptonically and produces the charged lepton and a neutrino, while the other three initial products decay hadronically and produce large-radius jets.
The parent particles of the large-radius jets can be identified using the \textsc{DeepAK8} algorithm. In this channel, VLQ candidates are reconstructed from the lepton, \ptmiss, and large-radius jets, and a multi-layer perceptron (MLP) neural network~\cite{almeida1997c1} is trained to identify events as \ttbar background, \wjet background, or VLQ signal events.
Events are categorized by lepton flavor, electron or muon, and then based on the particle identification of the VLQ candidates' decay products.

The SS dilepton channel, described in Section~\ref{sec:2lep}, is primarily sensitive to VLQ pair production with $\PQT \to \PQt\PH$ (with $\PH\to\PW\PW$) and $\PQB \to\PQt\PW$ decays.
With up to six \PW bosons produced (including those from the top quark decays), two same-sign \PW bosons can decay leptonically to produce two final-state leptons with the same electric charge. Events are categorized by lepton flavor combinations.

The multilepton channel, described in Section~\ref{sec:3lep}, is primarily sensitive to contributions from $\PQT \to\PQt\PZ$ and $\PQB \to\PQt\PW$ decays.
Leptonic decays of these \PZ or \PW bosons, combined with possible leptonic decays of the $\PW$ bosons from the decay of the top quarks, can produce three or more leptons---a rare final state in SM processes.
The high-energy signature of the VLQ signal in the \HTlep and \ST distributions is used to discriminate the signal from the background in the SS dilepton and multilepton channels, respectively.
Events are categorized by lepton flavor combinations.

Table~\ref{tab:strategy} summarizes the main event selection criteria used to form control regions (CRs) and signal regions (SRs) for the three channels, beyond the common selection criteria presented in Section~\ref{sec:reco}.
In the single-lepton channel all 2016--2018 data are analyzed according to the methods presented in this paper.
In the SS dilepton and multilepton channels, data from 2017--2018 are analyzed using the methods described here, while data from 2016 are reproduced from a previous search with the same analysis strategy~\cite{B2G-17-011}.
Template histograms from a variety of kinematic variables are taken from the SRs of all three channels, as well as some CRs in order to constrain uncertainties in the background estimation.
The template histograms are combined in a maximum likelihood fit, described in Section~\ref{sec:results}, to determine the presence of signal.

\begin{table}[htb]
  \topcaption{
    Summary of event selection criteria for the primary CRs and SRs in the three search channels.
    The label ``OSSF'' refers to opposite-sign charge, same-flavor lepton pairs, and the phrase ``max MLP'' refers to the largest score from the single-lepton MLP network.
  }
  \label{tab:strategy}
  \centering
  \begin{tabular}{l c @{$\quad$} c @{$\quad$} c}
    Channel                     & \multicolumn{3}{c}{Event selection}                                                   \\
                                & Overall                              & CR                  & SR                       \\[0.1em]
    \hline
    \\[-0.9em]
    \multirow{7}{*}{1$\ell$}    & 1 tight $\ell$                       & \NA                 & \NA                      \\
                                & $\pt(\ell) > 55$\GeV                 & \NA                 & \NA                      \\
                                & 0 other loose $\ell$, \pt $>$ 10\GeV & \NA                 & \NA                      \\
                                & $\ptmiss > 50\GeV$                   & \NA                 & \NA                      \\
                                & $\geq3$ large-radius jets            & \NA                 & \NA                      \\
                                & \NA                                  & max MLP not VLQ     & max MLP is VLQ           \\
                                & \NA                                  &                     & 2 VLQ candidates         \\[0.1em]
    \hline
    \\[-0.9em]
    \multirow{6}{*}{SS 2$\ell$} & 2 tight SS $\ell$                    & \NA                 & \NA                      \\
                                & $\pt(\ell) > 40$\GeV, 30\GeV         & \NA                 & \NA                      \\
                                & $\geq4$ small-radius jets            & \NA                 & \NA                      \\
                                & $M(\ell\ell) > 20$\GeV               & \NA                 & \NA                      \\
                                & $M(\Pe\Pe)$ outside 76--106\GeV      & \NA                 & \NA                      \\
                                & \NA                                  & $\HTlep < 1000\GeV$ & $\HTlep > 1000\GeV$      \\[0.1em]
    \hline
    \\[-0.9em]
    \multirow{6}{*}{3$\ell$}    & $\pt(\ell) > 30$\GeV                 & \NA                 & \NA                      \\
                                & $M$(OSSF $\ell\ell) > 20$\GeV        & \NA                 & \NA                      \\
                                & $\ptmiss > 20$\GeV                   & \NA                 & \NA                      \\      
                                & $\geq1$ $\PQb$-tagged jet            & \NA                 & \NA                      \\
                                & $\pt$($\PQb$ jet) $> 45$\GeV         & \NA                 & \NA                      \\
                                & \NA                                  & 3 loose $\ell$      & $\geq3$ tight $\ell$\GeV \\
                                & \NA                                  & 2 small-radius jets & $\geq3$ small-radius jets \\
  \end{tabular}
\end{table}

\section{Single-lepton channel}
\label{sec:1lep}

Events selected in the single-lepton channel must have exactly one electron or muon with $\pt > 55\GeV$ that passes the tight selection requirements described in the previous section.
No additional charged leptons passing the loose selection requirements with $\pt >10\GeV$ are permitted.
Since this channel targets signal events with a leptonic \PW boson decay, selected events must have $\ptmiss > 50\GeV$, which rejects multijet background events.
Additionally, at least three large-radius jets are required.
Large-radius jets are discarded if they lie within $\Delta R < 0.8$ from the charged lepton and the component of their momentum in the direction perpendicular to the lepton's momentum is less than 20\GeV.
These selection criteria are summarized in Table~\ref{tab:strategy}.

In the single-lepton channel the dominant SM background processes are \wjet production and \ttbar production.
To improve the modelling of the background estimate in this channel, discrepancies in the modeling of the \HT distribution in the {\wjet} samples are corrected for by applying a scaling function described in Ref.~\cite{B2G-17-011}.
Corrections to the modeling of the \ttbar simulation are derived in the \HT distribution of the \ttbar-enriched CR described in Section~\ref{sec:CSR} below.

The VLQ candidates are reconstructed by first identifying a ``leptonic particle'' candidate.
The \PW boson is reconstructed from the charged lepton and \ptvecmiss, applying a \PW boson mass constraint. The $z$-component of the missing momentum is taken to be the real part of the solution that results in a \PW boson mass nearest to 80.2\GeV.
If the minimum mass that can be formed by pairing this \PW boson with any small-radius jet is less than 150\GeV (a value chosen from studies performed on signal events) it is likely that this \PW boson is the decay product of an SM top quark.
In this case, the leptonic particle candidate is formed from the \PW boson and either a b-tagged small-radius jet within $\Delta R < 0.8$ or the small-radius jet that yielded the minimum mass pairing, if no such b-tagged small-radius jet is found (likely because of its misidentification as a light quark or gluon jet). Otherwise, the leptonic particle candidate is the \PW boson itself.

Two VLQ candidates can be formed in events that contain at least three large-radius jets that are separated from the leptonic particle candidate by $\Delta R > 0.8$, to ensure that the leptonic particle candidate and each jet represent unique decay products from the VLQ pair.
A parent particle hypothesis is provided for each jet from the \textsc{DeepAK8} tag.
Wherever possible, VLQ candidates are formed from the leptonic particle candidate and three large-radius jets according to the expected VLQ decay modes: {\PQb}\PW, {\PQt}\PZ, {\PQt}\PH, {\PQt}\PW, {\PQb}\PZ, or {\PQb}\PH.
Events with two VLQ candidates matching the expected decay modes are observed in simulation to have well-reconstructed VLQ mass peaks, particularly for hadronic VLQ candidates, and are used to form ``high-purity'' (HP) categories.
Events with VLQ candidates that do not match the expected decay modes are used to form ``low-purity'' (LP) categories.
In these events, the leptonic particle candidate and large-radius jets are paired so as to minimize the mass difference between the resulting VLQ candidates.

\subsection{Multilayer perceptron network}
\label{sec:MVA}

The HP event categories have very high background rejection rates, but also low signal acceptance rates. The majority of simulated signal events are reconstructed as LP events because of misidentification of the decay products. To maximize the potential sensitivity of the LP events, both the \TTbar and \BBbar analyses have an associated MLP, a type of fully connected neural network, with three output nodes, which is used to distinguish between either the \TTbar or \BBbar signal, the \wjet background, and the \ttbar background.
The MLPs are trained using events from \wjet and {\MGvATNLO} \ttbar background simulation samples, and either the \TTbar or \BBbar signal sample with a VLQ mass of 1\TeV. They provide strong classification performance across the entire VLQ mass range considered in this search. The {\MGvATNLO} and {\POWHEG} \ttbar simulations were compared for all input distributions to the MLP and only negligible differences were observed.

The ``training region'' consists of single-lepton events with at least three large-radius jets that either do not have reconstructed VLQ candidates, or were categorized as LP events.
Some of the LP events from each sample form a testing data set for evaluating network performance.
For the signal and \ttbar samples, which are not used elsewhere in this search, the training data sets contain as many of the remaining LP VLQ events as possible.
The \wjet events are restricted to those without VLQ candidates, to maintain separation from the SRs described below in Section~\ref{sec:CSR}.
Approximately equal-sized subsets of events from the three samples are provided for the training.

The MLP inputs include both event-level and jet-level observables that were chosen to maximize overall network accuracy and minimize the misidentification of \ttbar events as signal events. Event-level observables include $\HT$, $\ST$, \ptmiss, the minimum angular separation between the highest \pt large-radius jet and any other large-radius jet, and the numbers of small-radius jets, large-radius jets, and $\PQb$-tagged small-radius jets.
If a leptonically decaying top quark is reconstructed in the event, its \pt, mass, and the angular separation between the $\PW$ boson and the bottom quark are also included. Jet-level variables include the \pt, the \textsc{DeepAK8} light quark or gluon score, and the $N$-subjettiness ratio $\tau_2/\tau_1$ of the three highest \pt large-radius jets, as well as the softdrop mass of the highest \pt large-radius jet.

\begin{figure}[htb]
  \centering
  \includegraphics[width=0.49\textwidth]{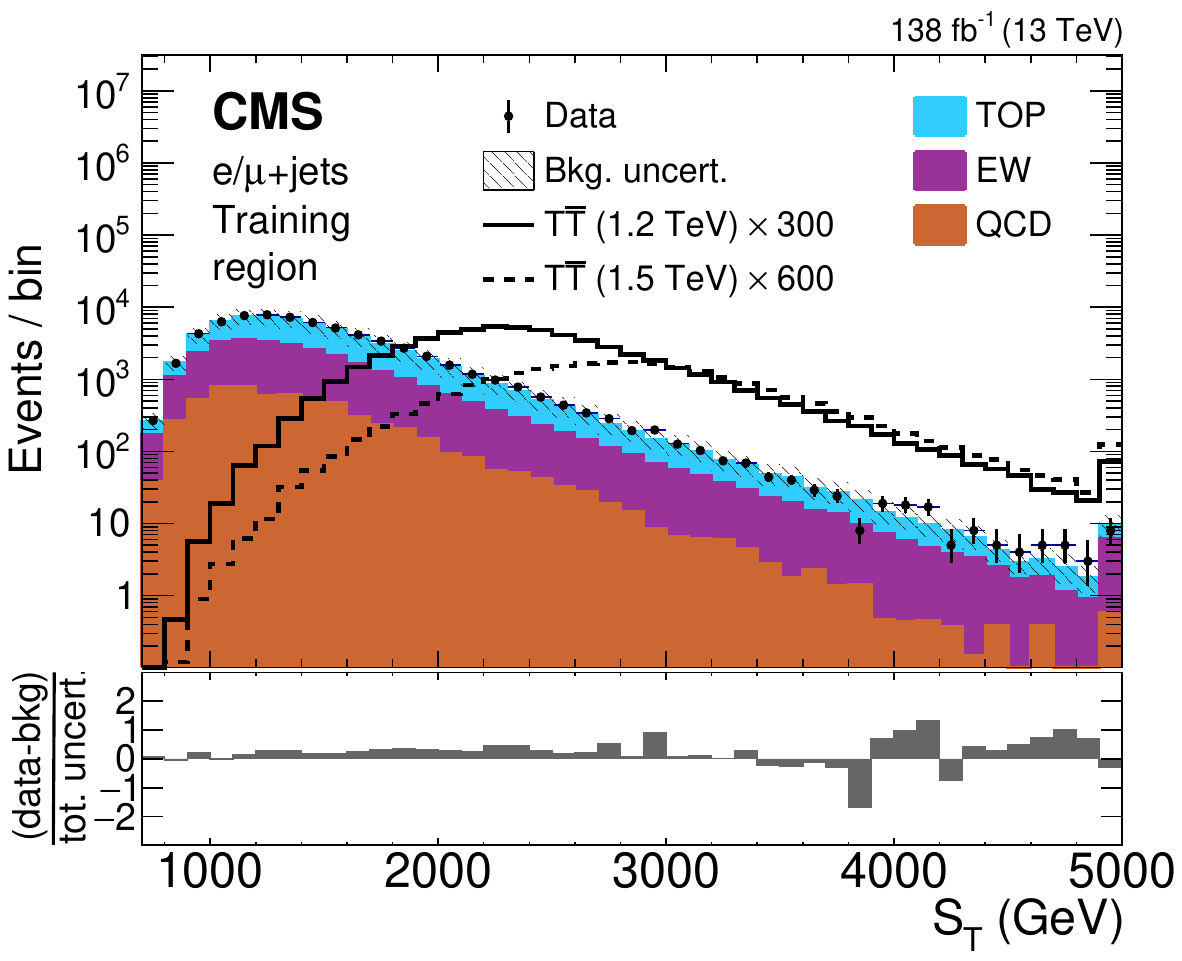}
  \includegraphics[width=0.49\textwidth]{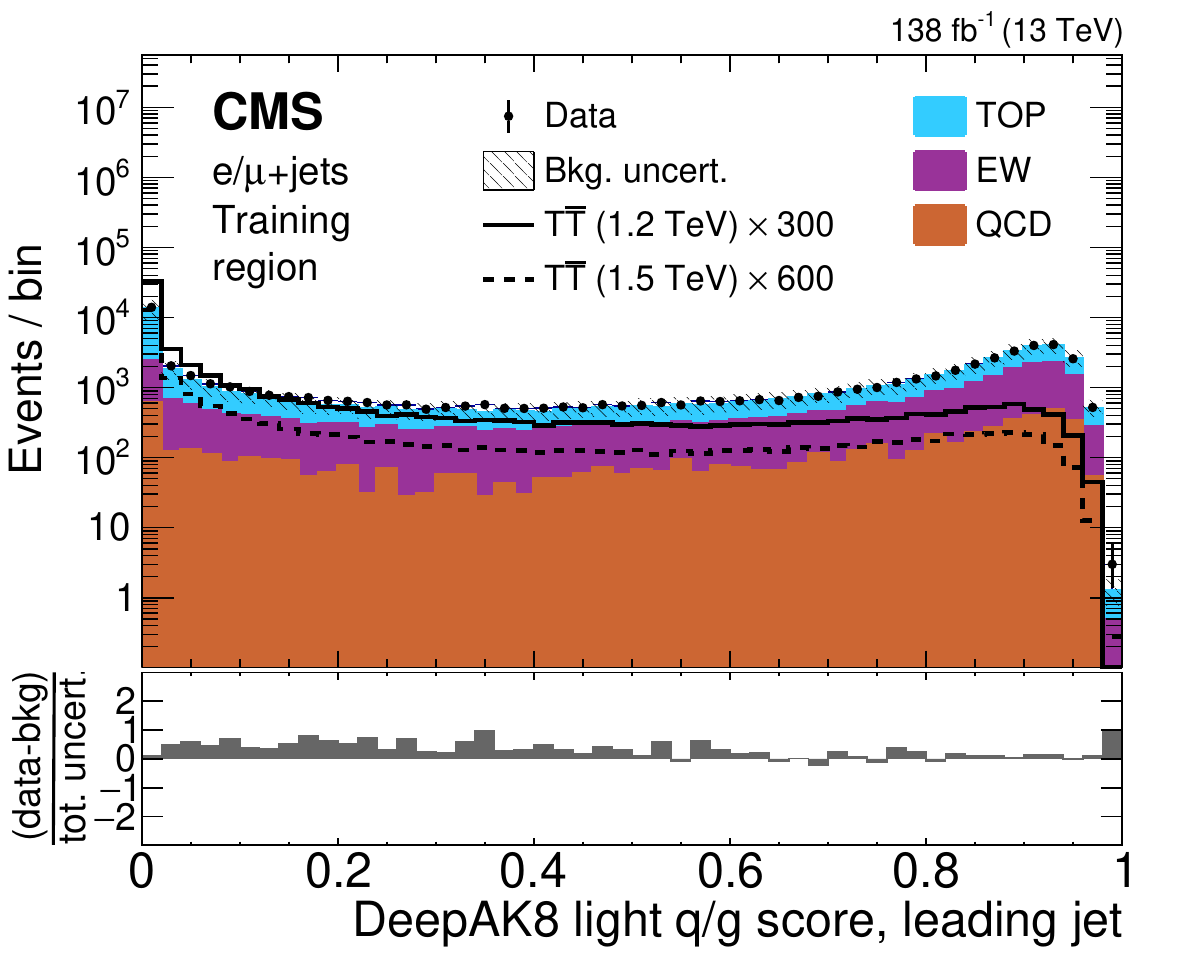}
  \caption{
    Example single-lepton channel MLP input distributions of $\ST$ (left) and the leading jet's \textsc{DeepAK8} light quark or gluon score (right) in the training region for the \TTbar MLP.
    The observed data are shown using black markers, predicted \TTbar signal with mass of 1.2 (1.5)\TeV in the singlet scenario using solid (dashed) lines, and backgrounds, using filled histograms.
    Statistical and systematic uncertainties in the background prediction before performing the fit to data are shown by the hatched region.
    The lower panels show the difference between the data and the background estimate as a multiple of the total uncertainty in both sources.
    The signal predictions have been scaled for visibility by the factors indicated in the figures.
  }
  \label{fig:inputs}
\end{figure}

Modeling of the \textsc{DeepAK8} light quark or gluon score is studied in a validation region containing events with only two large-radius jets, in which the predicted signal is negligible compared to the background.
The model is improved by a binned shape correction for the 2017 and 2018 simulations, taken from the data-to-background ratios in the light quark or gluon score distributions.
This ratio ranges from 1.2 in jets with lower scores to 0.7 in jets with very high scores.
The correction is applied to the simulation by weighting events by the product of the correction for the three leading large-radius jets, normalized so that the cross section of each simulated sample remains unchanged.

For jets produced by Higgs bosons, the \textsc{DeepAK8} light quark or gluon score distribution is calibrated using jets originating from $\Pg \to \bbbar$ fragmentation~\cite{ParticleNetDP}.
These jets are selected from samples of multijet data and background simulation using a boosted decision tree classifier so that their \textsc{DeepAK8} light quark or gluon score distribution resembles the distribution of \PH jets in the simulated 1.5\TeV \TTbar sample.
Corrections are derived by fitting the simulation to the data in several intervals of the \textsc{DeepAK8} light quark or gluon score.
Correction factors range from 1.2--1.4 in jets with high scores to 0.8--0.9 in jets with very low scores, depending on the data-taking period.
The corrections are applied only to events with jets originating from Higgs bosons.
All MLP input observables and pairs of input observables are then tested using the full training region to ensure that the background simulations accurately model the observed data.
Any bins expected to contain a significant signal component are removed from the test distributions.
Figure~\ref{fig:inputs} shows data and predictions in the training region for two example input variables, $\ST$ and the \textsc{DeepAK8} light quark or qluon score for the highest-\pt large-radius jet, after all the corrections described here are applied.
These observables show particularly strong separation between signal and background.

The MLP is implemented using the \textsc{scikit-learn} platform~\cite{SKL} and has three fully connected hidden layers with 10 nodes each.
It is trained using the ``Adam'' optimizer~\cite{ADAM} to minimize a cross-entropy loss function, with rectified linear unit activation functions between hidden layers and a softmax activation function for the output layer.
Training is halted when a validation sample, consisting of 10\% of the training events, indicates that the prediction accuracy has reached a plateau.
For the \TTbar- (\BBbar-)optimized MLP, the signal is identified with 92~(89)\% efficiency, while 14~(11)\% of \ttbar events and 2\% of \wjet events are misclassified as signal events.
The three outputs of the MLP are labelled as the \wjet node score, the \ttbar node score, and the VLQ node score. 

\subsection{Control and signal regions}
\label{sec:CSR}

\begin{figure}[htb]
  \centering
  \includegraphics[width=0.49\textwidth]{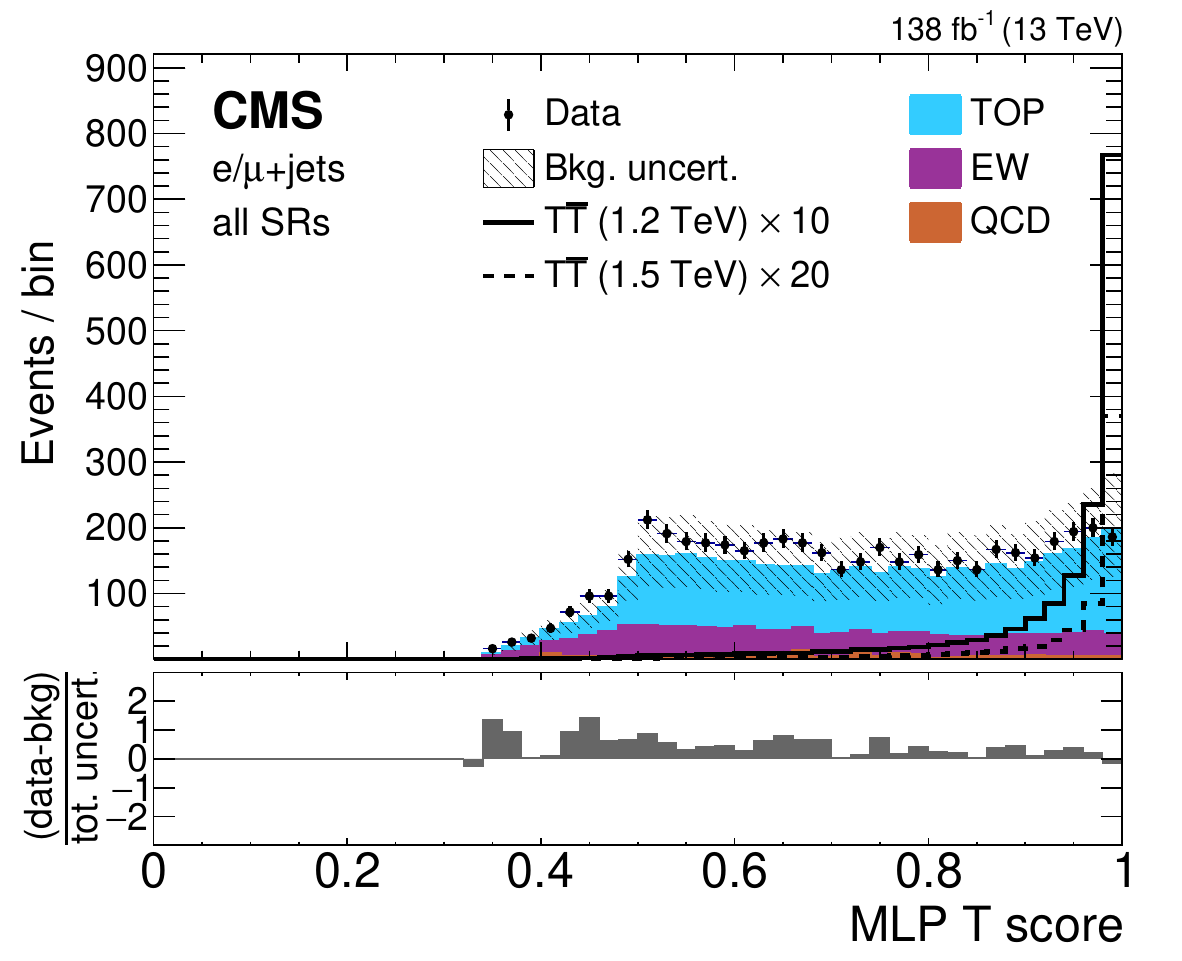}
  \includegraphics[width=0.49\textwidth]{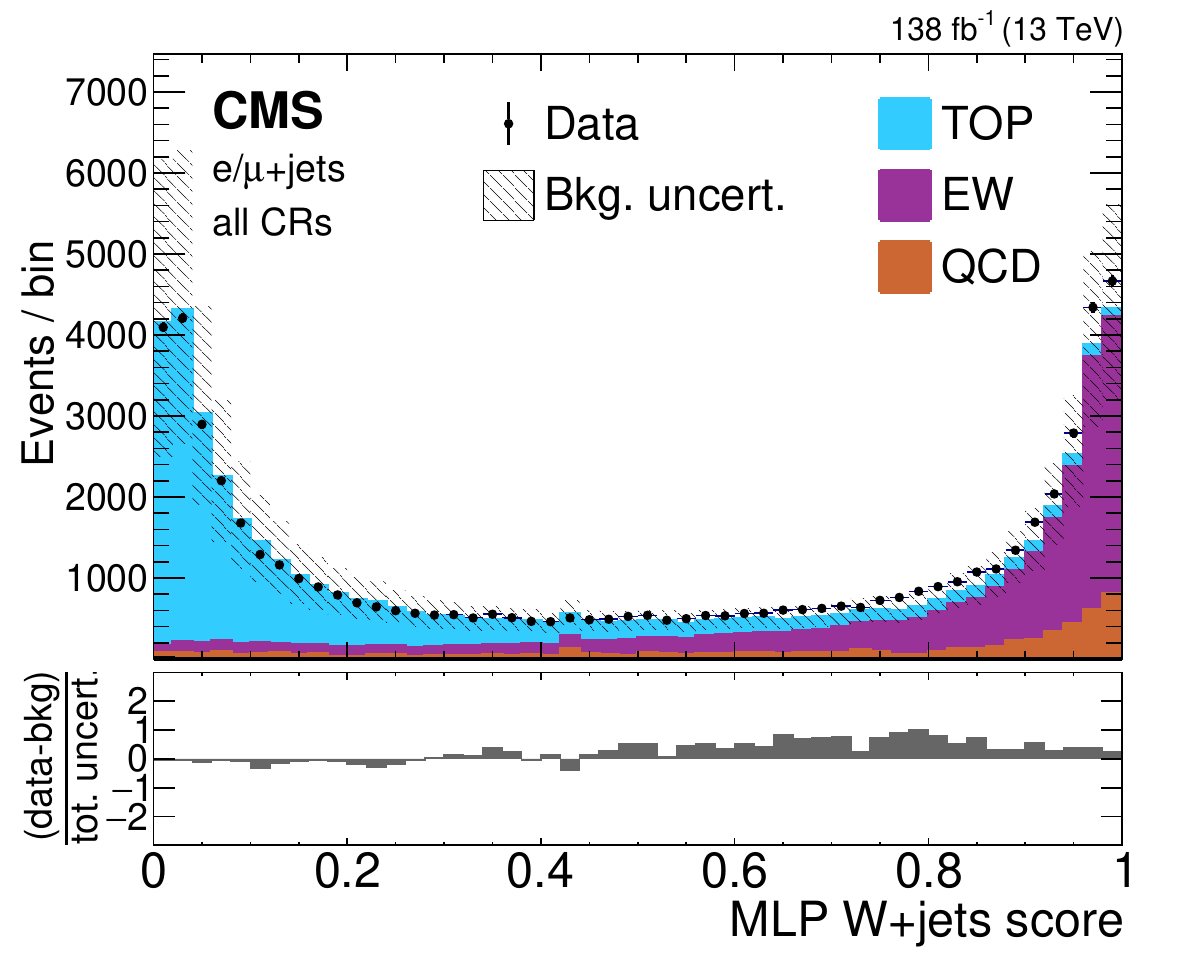}
  \caption{
    Example single-lepton channel \TTbar MLP output distributions of the \PQT quark score in the inclusive SR (left) and the \wjet score in the CRs (right).
    The observed data are shown using black markers, predicted \TTbar signal with mass of 1.2 (1.5)\TeV in the singlet scenario using solid (dashed) lines, and backgrounds, using filled histograms.
    Statistical and systematic uncertainties in the background estimate before performing the fit to data are shown by the hatched region.
    The lower panels show the difference between the data and the background estimate as a multiple of the total uncertainty in both sources.
    The signal predictions in the left distribution have been scaled for visibility by the factor indicated in the figure.
  }
  \label{fig:mlp}
\end{figure}

The MLP predictions provide a powerful separation between signal and background in the single-lepton channel, and are used to define an SR and CRs for each signal hypothesis (\TTbar or \BBbar).
The SR contains all HP VLQ events and LP VLQ events for which the VLQ node score was larger than either background score.
The CRs contain LP events in which one of the background node scores was larger than the VLQ node score.
Figure~\ref{fig:mlp} shows the strong distinction between the shape of the signal and the background in the VLQ node score distribution in the SR, as well as the separation between the \ttbar and \wjet background processes in the \wjet node score in the CRs.

\begin{figure}[hbtp]
  \centering
  \includegraphics[width=0.49\textwidth]{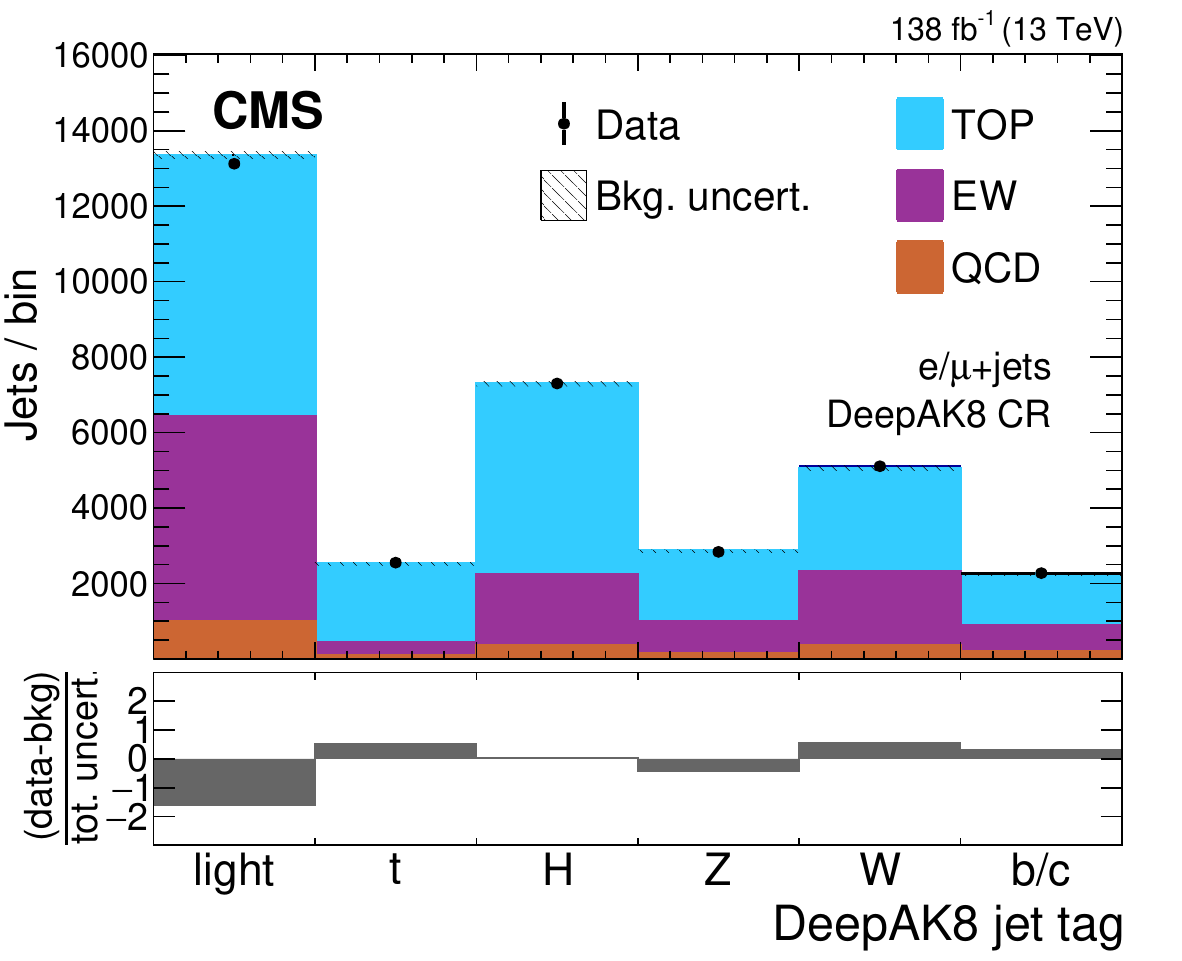}
  \includegraphics[width=0.49\textwidth]{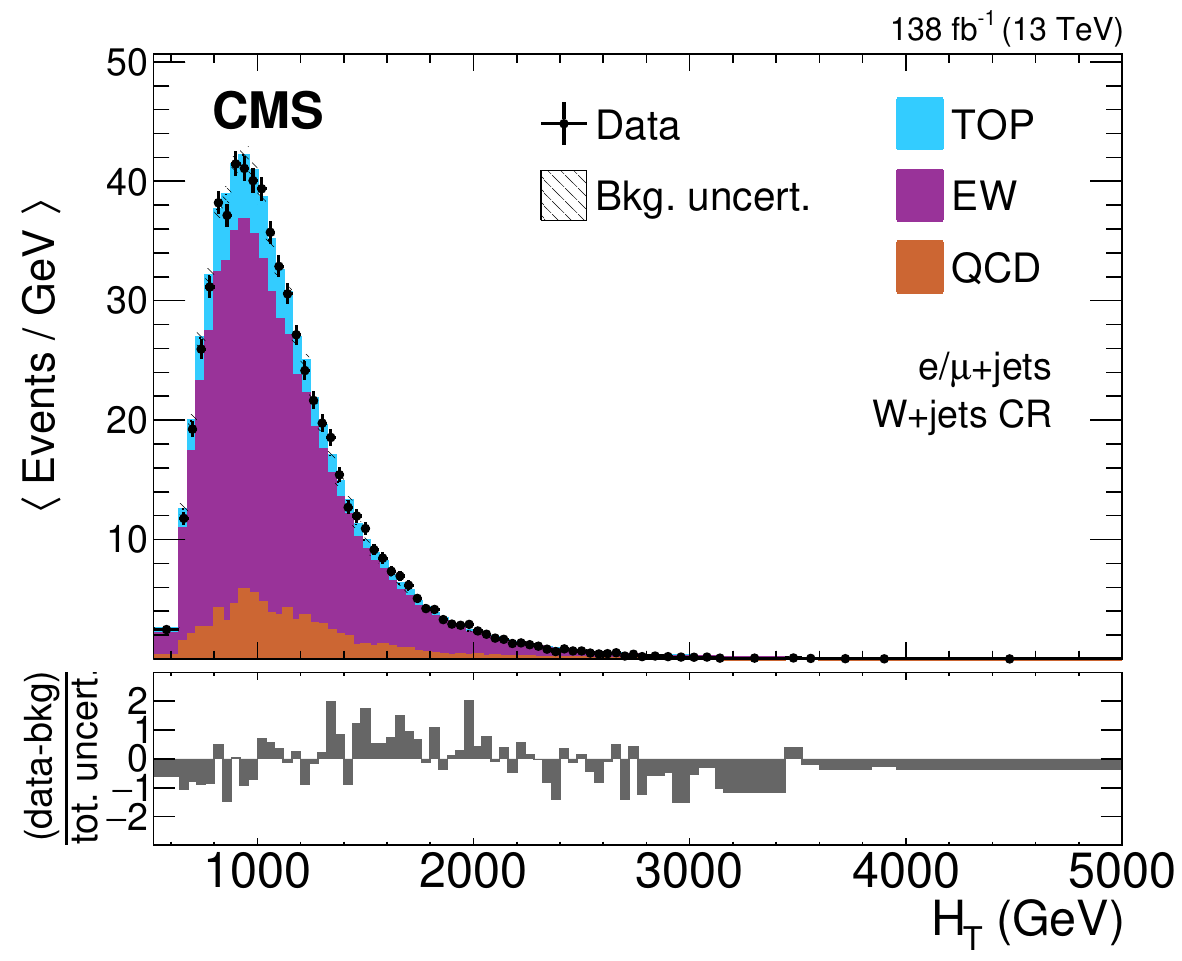}\\
  \includegraphics[width=0.49\textwidth]{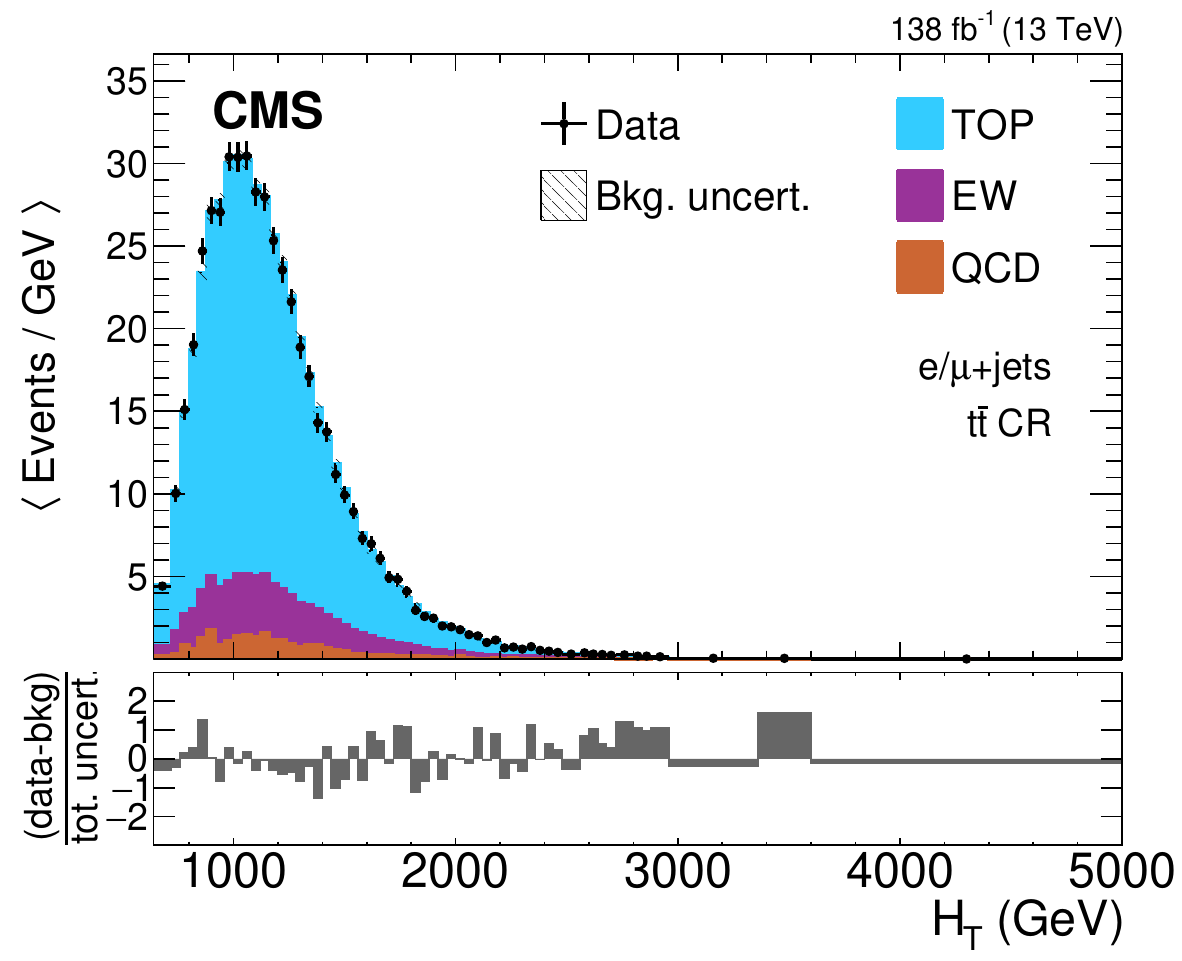}
  \caption{
    Single-lepton channel template histograms of the \textsc{DeepAK8} jet tags in the \textsc{DeepAK8} CR (upper left), and \HT in the \wjet (upper right) and \ttbar (lower) CRs for the \TTbar search.
    The observed data are shown using black markers and the post-fit background estimates, using filled histograms.
    Statistical and systematic uncertainties in the background estimate after performing the fit to data are shown by the hatched region.
    The lower panels show the difference between the data and the background estimate as a multiple of the total uncertainty in both sources.
    Electron and muon categories have been combined for illustration with their uncertainties added in quadrature.
  }
  \label{fig:CRCatTT}
\end{figure}

Three mutually exclusive CRs are formed in this channel. A \textsc{DeepAK8} CR is constructed from all CR events in which the highest-\pt large-radius jet was tagged as a massive particle, along with a randomly chosen half of all other CR events.
The remaining CR events are separated into either a \ttbar-enriched CR or a \wjet enriched CR, based on which of their MLP background node scores is larger.
Distributions from the three CRs are included in the fit to data (described in Section~\ref{sec:results}) because they relate to regions in which the background modeling can be constrained. 
In the \textsc{DeepAK8} CR the observable is the distribution of \textsc{DeepAK8} jet tags, which provides information about the efficiencies and misidentification rates for \textsc{DeepAK8} massive particle tagging.
In the \ttbar and \wjet CRs the observable is the \HT distribution, which is very sensitive to the overall energy scale of the events.
Figure~\ref{fig:CRCatTT} shows the data and simulation in the \TTbar CR categories after the fit, and the corresponding event yields are listed in Table~\ref{tab:CRCatTable}.

\begin{table}[htbp]
  \centering
  \topcaption{
    Numbers of predicted and observed CR events in 2016--2018 data (138\fbinv) for the \TTbar (upper section) and \BBbar (lower section) signal hypotheses in the single-lepton channel, after a background-only fit to data described in Section~\ref{sec:results}.
    Predicted numbers of signal events before the fit to data are included for comparison, using the singlet branching fraction scenario.
    Uncertainties include statistical and systematic components, and lepton flavor categories are combined for illustration with their uncertainties added in quadrature.
    Values in the \textsc{DeepAK8} CR represent the number of large-radius jets rather than the number of events.
  }
  \begin{tabular}{l r@{ }c@{ }l r@{ }c@{ }l r@{ }c@{ }l}
    Sample           & \multicolumn{3}{c}{\textsc{DeepAK8} CR} & \multicolumn{3}{c}{\wjet CR} & \multicolumn{3}{c}{\ttbar CR}                                                     \\[0.1em]
    \hline
    \\[-0.9em]
    \TTbar (1.2\TeV) & 8.21                                    & $\pm$                        & 0.95                          & 2.75    & $\pm$ & 0.31  & 6.40    & $\pm$ & 0.60  \\
    \TTbar (1.5\TeV) & 0.774                                   & $\pm$                        & 0.085                         & 0.486   & $\pm$ & 0.052 & 0.694   & $\pm$ & 0.066 \\[0.3em]
    TOP              & 20\,070                                 & $\pm$                        & 190                           & 3\,227  & $\pm$ & 61    & 16\,490 & $\pm$ & 160   \\
    EW               & 11\,120                                 & $\pm$                        & 210                           & 20\,700 & $\pm$ & 280   & 2\,782  & $\pm$ & 71    \\
    QCD              & 2\,170                                  & $\pm$                        & 190                           & 3\,280  & $\pm$ & 260   & 975     & $\pm$ & 85    \\[0.5em]
    Total bkgd.      & 33\,360                                 & $\pm$                        & 190                           & 27\,210 & $\pm$ & 150   & 20\,240 & $\pm$ & 130   \\
    Data             & \multicolumn{3}{c}{33207}               & \multicolumn{3}{c}{27189}    & \multicolumn{3}{c}{20190}                                                         \\
    Data/bkgd.       & 0.996                                   & $\pm$                        & 0.008                         & 0.999   & $\pm$ & 0.008 & 0.997   & $\pm$ & 0.010 \\[0.8em]

    Sample           & \multicolumn{3}{c}{\textsc{DeepAK8} CR} & \multicolumn{3}{c}{\wjet CR} & \multicolumn{3}{c}{\ttbar CR}                                                     \\[0.1em]
    \hline
    \\[-0.9em]
    \BBbar (1.2\TeV) & 12.0                                    & $\pm$                        & 1.5                           & 1.52    & $\pm$ & 0.16  & 8.21    & $\pm$ & 0.83  \\
    \BBbar (1.5\TeV) & 1.42                                    & $\pm$                        & 0.17                          & 0.302   & $\pm$ & 0.033 & 1.04    & $\pm$ & 0.11  \\[0.3em]
    TOP              & 20\,800                                 & $\pm$                        & 230                           & 2\,956  & $\pm$ & 88    & 16\,950 & $\pm$ & 170   \\
    EW               & 11\,540                                 & $\pm$                        & 230                           & 21\,010 & $\pm$ & 270   & 2\,965  & $\pm$ & 79    \\
    QCD              & 2\,210                                  & $\pm$                        & 200                           & 3\,260  & $\pm$ & 240   & 975     & $\pm$ & 88    \\[0.5em]
    Total bkgd.      & 34\,550                                 & $\pm$                        & 200                           & 27\,230 & $\pm$ & 150   & 20\,890 & $\pm$ & 130   \\
    Data             & \multicolumn{3}{c}{34476}               & \multicolumn{3}{c}{27221}    & \multicolumn{3}{c}{20815}                                                         \\
    Data/bkgd.       & 0.998                                   & $\pm$                        & 0.008                         & 1.000   & $\pm$ & 0.008 & 0.996   & $\pm$ & 0.009 \\
  \end{tabular}
  \label{tab:CRCatTable}
\end{table}

\section{Same-sign dilepton channel}
\label{sec:2lep}

Events with exactly two tight leptons with the same sign of electric charge are selected.
They must have satisfied a dilepton trigger and the leading (subleading) lepton must have $\pt >40\,(30)\GeV$.
Additionally, the events are required to contain at least four small-radius jets.
Events are categorized based on the flavors of the two leptons: ${\Pe}\Pe$, ${\Pe}\PGm$, or ${\PGm}\PGm$.

To reject low mass dilepton resonances, it is required that the invariant mass of the SS lepton pair is greater than 20\GeV.
To veto $\PZ\to\Pe\Pe$ decays with charge misidentification it is required that the invariant mass does not lie in the range 76--106\GeV, and further required that neither of the leptons forms a pair within this same mass range with any same-flavor loose lepton in the event.
The SR consists of selected events with \HTlep above 1000\GeV, and a CR is formed using events with \HTlep below 1000\GeV.
These selection criteria are summarized in Table~\ref{tab:strategy}.

\subsection{Background modeling}
\label{sec:mmethod}

Three categories of background are considered: prompt, nonprompt, and charge misidentification.
Prompt background refers to SM processes with SS dilepton final states and is estimated using simulation.
Such SM processes include $\PV\PV$, $\PV\PV\PV$, $\ttbar\PV$, $\ttbar\PH$, and $\ttbar\ttbar$ production.
Nonprompt background refers to events with nonprompt leptons passing the tight lepton identification and/or jets misidentified as leptons.

Processes producing a pair of oppositely charged prompt leptons can also contribute to the background when the sign of the charge of one of the leptons has been misidentified.
Because of the design of the CMS muon system, the charge misidentification rate for muons is negligible for muons with \pt below the \TeVns scale~\cite{CMS:2019ied}, so this is only significant in the ${\Pe}\Pe$ and ${\Pe}\PGm$ categories.
The charge misidentification rate for electrons is measured in observed $\PZ \to \Pe\Pe$ events  using a likelihood fit to the $\pt$-$\abs{\eta}$ distribution of the lepton candidates.
The fit compares SS and opposite-sign $\Pe\Pe$ pairs which have a dilepton mass compatible with the \PZ mass.
To simplify the calculation and ensure a sufficient sample size in each bin, we split events into three regions defined by \pt ($<$100, 100--200, $>$200\GeV) and five regions defined by $\abs{\eta}$  ($<$0.4, 0.4--0.8, 0.8--1.44, 1.57--2.0, 2.0--2.4).
The charge misidentification rates derived vary between 0.005\% and 5\%, increasing with both \pt and $\abs{\eta}$.
The background from charge misidentification is then estimated by weighting opposite-sign dilepton events that pass all other event selection requirements by the charge misidentification rate per electron.
For dielectron events, the cases in which either of the electrons is misidentified are considered.

Nonprompt background is estimated from events with at least one loose lepton using ``prompt rates'' and ``nonprompt rates'', following the ``matrix'' method explained in Ref.~\cite{Wong:2808538}.
The prompt rate, the probability of prompt loose leptons to pass the tight categorization, is measured using the tag-and-probe method in Drell--Yan events in data.
The events used for this measurement are required to pass dilepton triggers and have a tight lepton with $\pt>30\GeV$ and $\abs{\eta}< 2.4$, as well as a second ``probe'' lepton that satisfies the loose requirements.
The prompt rate is the fraction of probe leptons passing the tight requirements.
The $\eta$-averaged prompt rate of muons is $0.928\pm0.015$ and $0.931\pm0.011$ in 2017 and 2018 data, respectively.
For electrons, the prompt rate depends on \pt and varies between 0.78 and 0.83 in both years.

The nonprompt rate is the probability of a nonprompt lepton or jet to pass the tight lepton identification.
As a result of changes in trigger thresholds with respect to 2016, the method for evaluating nonprompt rates used in Ref.~\cite{B2G-17-011} has been superseded
for 2017--2018 data by evaluating the lepton \pt distributions in the multilepton channel CR, as described in Section~\ref{sec:multiback}.
The measured constant electron and muon nonprompt rates are applied to tight-loose and loose-loose dilepton events.
The systematic uncertainties in the nonprompt rates in the SS dilepton channel are obtained by scanning nonprompt rate values and finding the values that provide the best fit between data and predicted background in the \HTlep distributions in the region $\HTlep<1000\GeV$, where possible signal contributions are negligible compared to the background. The corresponding event yields in the CR are listed in Table~\ref{tab:2lepCR}.

\begin{table}[htbp]
  \centering
  \topcaption{
    Numbers of predicted and observed CR events in 2017--2018 data (101\fbinv) in the SS dilepton channel, after a background-only fit to data.
    Predicted numbers of signal events before the fit to data are included for comparison, using the singlet branching fraction scenario.
    Uncertainties include statistical and systematic components.
    Predictions for 2017 and 2018 are combined with their uncertainties added in quadrature.
  }
  \begin{tabular}{l r@{ }c@{ }l r@{ }c@{ }l r@{ }c@{ }l}
    Sample                                 & \multicolumn{3}{c}{$\Pe\Pe$} & \multicolumn{3}{c}{$\Pe\PGm$} & \multicolumn{3}{c}{$\PGm\PGm$}                                                     \\[0.1em]
    \hline
    \\[-0.9em]
    \TTbar (1.2\TeV)                       & 0.0395                       & $\pm$                         & 0.0036                         & 0.1074 & $\pm$ & 0.0065 & 0.0678 & $\pm$ & 0.0043 \\
    \TTbar (1.5\TeV)                       & 0.0024                       & $\pm$                         & 0.0002                         & 0.0085 & $\pm$ & 0.0006 & 0.0054 & $\pm$ & 0.0004 \\[0.3em]
    \BBbar (1.2\TeV)                       & 0.0979                       & $\pm$                         & 0.0056                         & 0.237  & $\pm$ & 0.014  & 0.157  & $\pm$ & 0.012  \\
    \BBbar (1.5\TeV)                       & 0.0053                       & $\pm$                         & 0.0005                         & 0.0163 & $\pm$ & 0.0014 & 0.0115 & $\pm$ & 0.0010 \\[0.3em]
    $\ttbar\PV$+$\ttbar\PH$+$\ttbar\ttbar$ & 94                           & $\pm$                         & 13                             & 269    & $\pm$ & 38     & 161    & $\pm$ & 22     \\
    $\PV\PV$(\PV)                          & 29.1                         & $\pm$                         & 9.4                            & 72     & $\pm$ & 26     & 40     & $\pm$ & 16     \\
    Charge misid.                          & 121                          & $\pm$                         & 46                             & 114    & $\pm$ & 44     & 0.0    & $\pm$ & 0.0    \\
    Nonprompt                              & 265                          & $\pm$                         & 46                             & 798    & $\pm$ & 58     & 555    & $\pm$ & 27     \\[0.5em]
    Total bkgd.                            & 509                          & $\pm$                         & 35                             & 1254   & $\pm$ & 42     & 757    & $\pm$ & 25     \\
    Data                                   & \multicolumn{3}{c}{507}      & \multicolumn{3}{c}{1276}      & \multicolumn{3}{c}{734}                                                            \\
    Data/bkgd.                             & 0.995                        & $\pm$                         & 0.081                          & 1.018  & $\pm$ & 0.045  & 0.970  & $\pm$ & 0.048  \\
  \end{tabular}
  \label{tab:2lepCR}
\end{table}

\section{Multilepton channel}
\label{sec:3lep}

In the multilepton channel events with three or more tight leptons are selected for the SR.
The leptons must have $\pt > 30\GeV$, and the event must have passed a dilepton trigger.
Events are categorized based on the flavors of the three leading leptons: \eee, \eem, \emm, and \mmm.
Events are required to have $\ptmiss> 20\GeV$, since at least one neutrino is expected in the final state, and at least three small-radius jets among which at least one is \PQb tagged, as expected from top quark decays.
To reduce background from low-mass resonance decays to leptons, it is required that the mass of any opposite-sign same-flavor lepton pairs in the event must be greater than 20\GeV.
To optimize signal discrimination in the SR, it is further required that each lepton must have a $\vec{\pt}$ component orthogonal to the closest jet that is greater than 8\GeV, and there must be at least one \PQb-tagged jet with $\pt > 45\, (50)\GeV$ in 2017 (2018).

Events are also selected for a mutually exclusive CR.
Events in this CR are required to have exactly three leptons that pass the loose requirements and exactly two small-radius jets.
The CR is used to estimate the nonprompt background (as described in the next section), with the assumption that the nonprompt background contributes at the same level to the CR and the SR.
This assumption is cross-checked in another mutually exclusive region with exactly three leptons that pass the loose requirements and exactly one small-radius jet.
These selection criteria are summarized in Table~\ref{tab:strategy}.

\subsection{Background modeling}
\label{sec:multiback}

Two categories of background are considered: prompt and nonprompt.
Prompt background, estimated from simulation, refers to SM processes that can produce multilepton final states, including $\PV\PV$, $\PV\PV\PV$, and $\ttbar\PV$ production.
Nonprompt background is again estimated by the Matrix method, but extended to three leptons.

\begin{figure}
  \centering
  \includegraphics[width=0.48\linewidth]{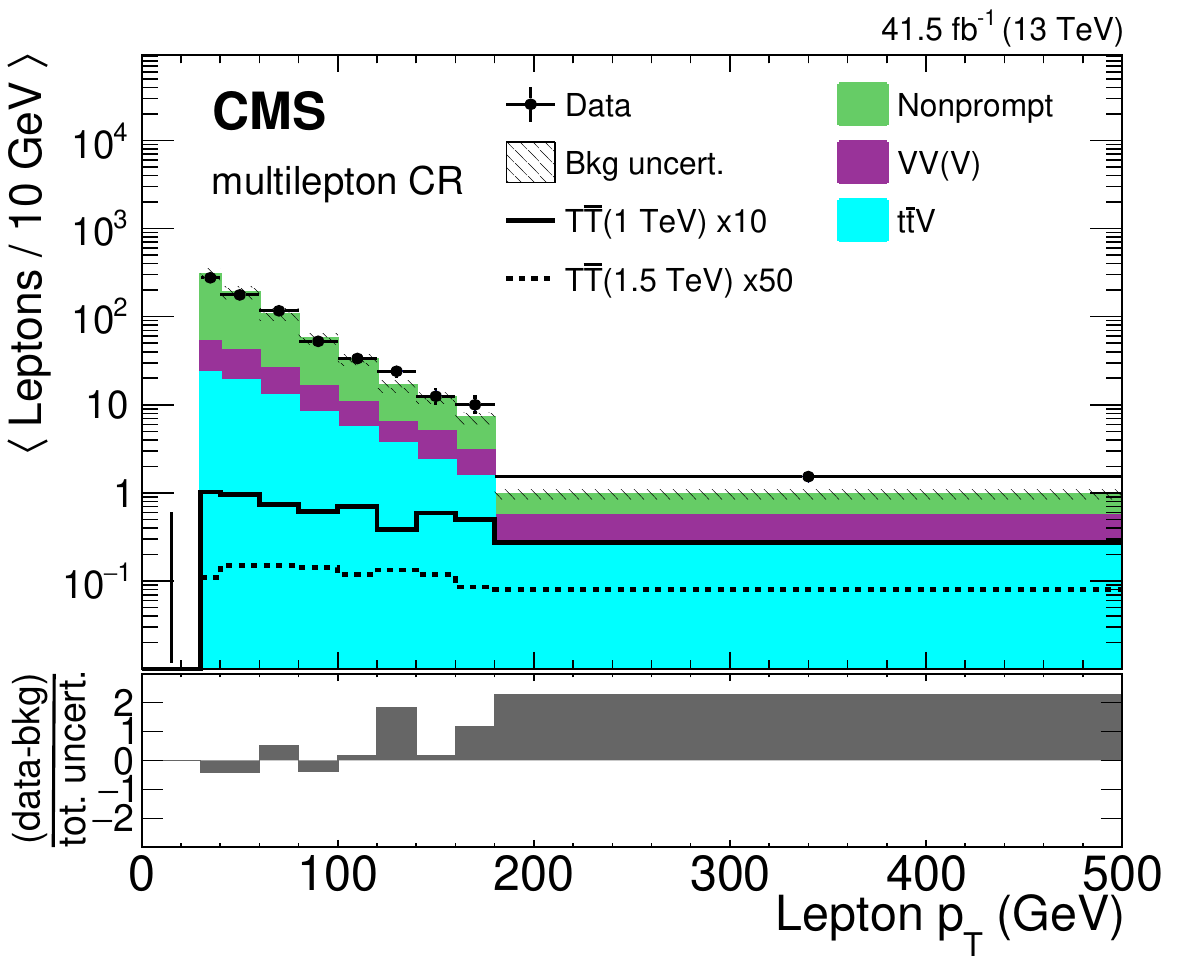}
  \includegraphics[width=0.48\linewidth]{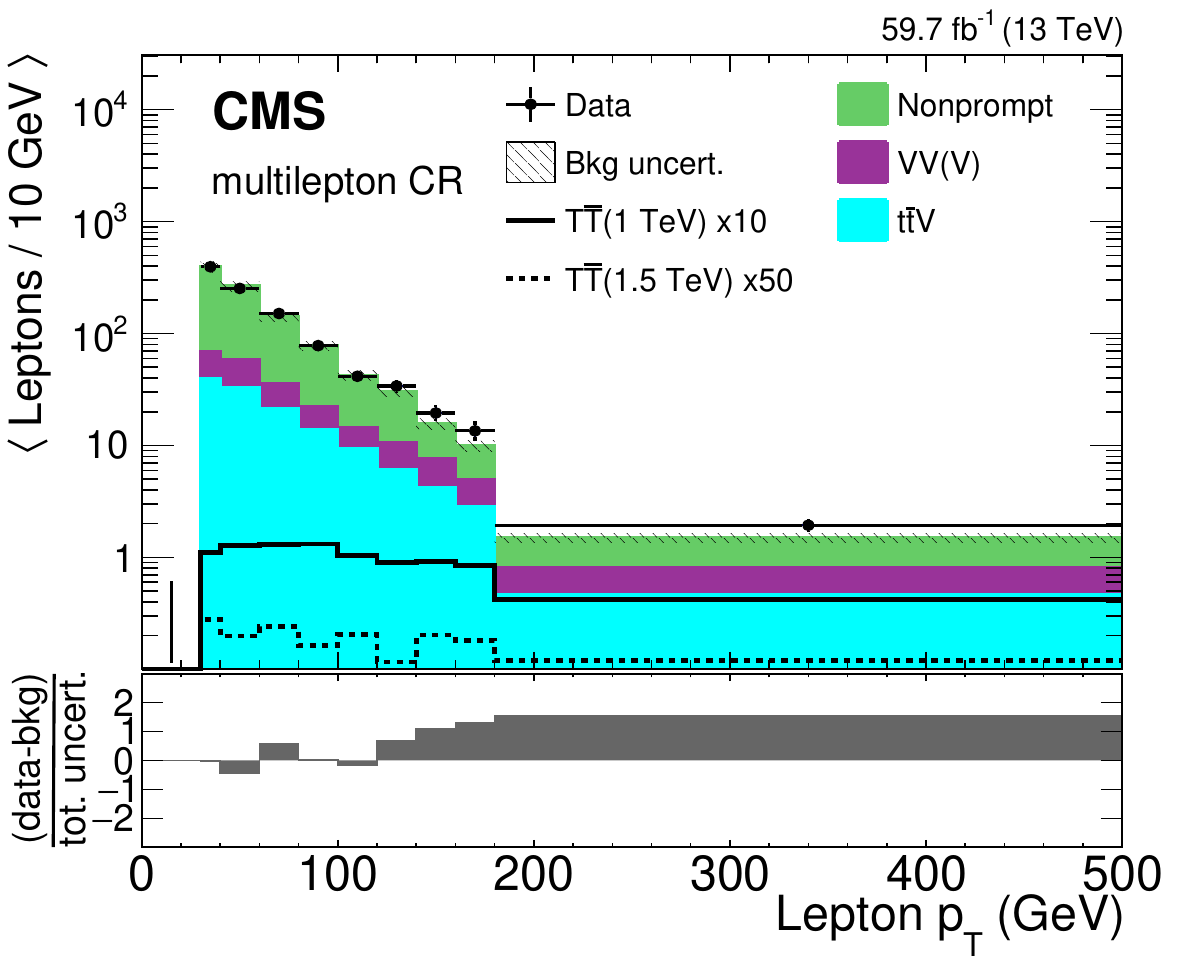}
  \caption{
    Distribution of lepton \pt in 2017 (left) and 2018 (right) in the multilepton channel nonprompt lepton CR for all flavor categories, evaluated with the best fit nonprompt rates.
    The \pt values of the three leptons in each event are included in the histogram.
    The uncertainty shown is the quadratic sum of the statistical and systematic components.
    The lower panel shows the difference between the data and the background estimate as a multiple of the total uncertainty in both sources.
    The signal predictions have been scaled for visibility by the factors indicated in the figures.
  }
  \label{fig:CR2_lepPt}
\end{figure}

The prompt rates applied are identical to those used in the SS dilepton channel.
Nonprompt rates are extracted by fitting the predicted lepton \pt distributions in the multilepton CR, shown in Fig. \ref{fig:CR2_lepPt}, through the minimization of the $\chi^2$ between the data and the total estimated background.
The electron and muon nonprompt rates ($f_\Pe$ and $f_\PGm$) are varied independently in steps of 0.01 from 0.0 to 0.5, producing a 2D $\chi^2$ distribution.
The $\chi^2$  distribution is then converted to a 2D Gaussian probability distribution using the relation $P(f_\PGm, f_\Pe)\propto \exp(-\chi^2/2)$.
Uncertainties are determined by marginalizing the nonprompt rates individually, i.e., summing the distribution along one axis producing a 1D Gaussian distribution for each nonprompt rate, with the width of the Gaussian assigned as the uncertainty.
This ignores correlations between lepton flavors, which are considered by other systematic uncertainties.

\begin{table}[htbp]
  \centering
  \topcaption{
    Numbers of predicted and observed CR events in 2017--2018 data (101\fbinv) in the multilepton channel, after a background-only fit to data.
    Predicted numbers of signal events before the fit to data are included for comparison, using the singlet branching fraction scenario.
    Uncertainties include statistical and systematic components.
    Predictions for 2017 and 2018 are combined with their uncertainties added in quadrature.
  }
  \begin{tabular}{l r@{ }c@{ }l r@{ }c@{ }l r@{ }c@{ }l r@{ }c@{ }l}
    Sample           & \multicolumn{3}{c}{\eee} & \multicolumn{3}{c}{\eem} & \multicolumn{3}{c}{\emm} & \multicolumn{3}{c}{\mmm}                                                                      \\[0.1em]
    \hline
    \\[-0.9em]
    \TTbar (1.2\TeV) & 0.0085                   & $\pm$                    & 0.0034                   & 0.0173                   & $\pm$ & 0.0033 & 0.0088 & $\pm$ & 0.0028 & 0.0085 & $\pm$ & 0.0017 \\
    \TTbar (1.5\TeV) & 0.0007                   & $\pm$                    & 0.0003                   & 0.0026                   & $\pm$ & 0.0007 & 0.0016 & $\pm$ & 0.0006 & 0.0019 & $\pm$ & 0.0004 \\[0.3em]
    \BBbar (1.2\TeV) & 0.0163                   & $\pm$                    & 0.0054                   & 0.0394                   & $\pm$ & 0.0083 & 0.0342 & $\pm$ & 0.0061 & 0.0187 & $\pm$ & 0.0054 \\
    \BBbar (1.5\TeV) & 0.0020                   & $\pm$                    & 0.0007                   & 0.0059                   & $\pm$ & 0.0009 & 0.0055 & $\pm$ & 0.0017 & 0.0025 & $\pm$ & 0.0006 \\[0.3em]
    $\ttbar\PV$      & 5.76                     & $\pm$                    & 0.69                     & 12.1                     & $\pm$ & 1.3    & 14.4   & $\pm$ & 1.6    & 8.8    & $\pm$ & 1.1    \\
    $\PV\PV$(\PV)    & 13.3                     & $\pm$                    & 1.2                      & 18.8                     & $\pm$ & 1.4    & 19.9   & $\pm$ & 1.5    & 23.2   & $\pm$ & 1.9    \\
    Nonprompt        & 32.8                     & $\pm$                    & 5.2                      & 93.7                     & $\pm$ & 7.0    & 105.0  & $\pm$ & 6.9    & 70.0   & $\pm$ & 7.6    \\[0.5em]
    Total bkgd.      & 51.8                     & $\pm$                    & 5.2                      & 124.7                    & $\pm$ & 7.1    & 139.4  & $\pm$ & 7.1    & 102.1  & $\pm$ & 7.8    \\
    Data             & \multicolumn{3}{c}{53}   & \multicolumn{3}{c}{140}  & \multicolumn{3}{c}{158}  & \multicolumn{3}{c}{98}                                                                        \\
    Data/bkgd.       & 1.02                     & $\pm$                    & 0.17                     & 1.12                     & $\pm$ & 0.11   & 1.13   & $\pm$ & 0.11   & 0.96   & $\pm$ & 0.12   \\
  \end{tabular}
  \label{tab:3lepCR}
\end{table}

For 2017 (2018) data, the minimum $\chi^2$ is found at $f_\Pe=0.10 (0.08) \pm 0.01$ and $f_\PGm=0.15 (0.16) \pm 0.01$.
The differences in yields between the data and the total background relative to the estimated nonprompt background are 23\% (\eee), 16\% (\eem), 2.7\% (\emm), and 11\% (\mmm) in 2017; and 24\% (\eee), $-5.2$\% (\eem), $-0.4$\% (\emm) and 7.7\% (\mmm) in 2018. These differences are assigned as uncertainties.

A cross-check of the nonprompt rate measurement is performed using a sample of simulated decays of top quark pairs in which both top quarks decay leptonically.
Events are required to contain at least three leptons, of which two are prompt and one is nonprompt.
The number of true nonprompt events in which the nonprompt lepton passes the tight identification is compared to the prediction made using the numbers of nonprompt leptons and the nonprompt rates, 
The discrepancies between predicted and observed events in this sample range from 1.3\% to 7.5\% across the 2017 and 2018 lepton flavor categories, and are incorporated in the nonprompt rate systematic uncertainties. The discrepancies are studied across the leptons' \pt distribution and no significant trends are observed.

In addition, to check that the nonprompt rates measured in the CR are applicable to the SR, a $\chi^2$ minimization is performed on the \ttbar simulated samples, comparing the true nonprompt distributions to the predicted distributions in both the CR and the SR.
The measured $f_\Pe$ is the same in both regions in 2017 and 2018, while $f_\PGm$ deviates by 0.02 and 0.01 in the 2017 and 2018 samples, respectively.

\section{Systematic uncertainties}
\label{sec:syst}

Systematic uncertainties can affect both the normalization and the shape of the predicted background and signal distributions,
and are summarized in Table~\ref{table:Systematics}.
\begin{table}[hp]
  \centering
  \topcaption{
    Summary of systematic uncertainties for the various analysis channels, grouped according to the channel.
    The second column shows uncertainties for the single-lepton channel, evaluated with 2016--2018 data combined.
    The third and fourth columns show uncertainties for the other channels, evaluated using 2017 or 2018 data.
    Uncertainties that are common across multiple channels appear in the first relevant column, followed by the label ``same'' in other columns.
    Ranges indicate values across different lepton flavor categories, and functional forms describe the quantities on which the uncertainty's numerical value depends.
    The final column shows which predictions are affected by each uncertainty: ``MC'' denotes all simulation (including signal), ``OS'' denotes charge misidentification background, and ``NP'' denotes nonprompt lepton background.
  }
  \label{table:Systematics}
  \cmsTable{
    \begin{tabular}{lcccc}
      \multirow{2}{*}{Uncertainty}                     & \multicolumn{3}{c}{Value or formula} & Affected                                                \\
                                                       & 1$\ell$                              & 2--3$\ell$ 2017         & 2--3$\ell$ 2018     & Samples \\
      \hline
      Integrated luminosity                            & 1.6\%                                & 2.3\%                   & 2.5\%               & MC      \\
      \Pe reconstr. and iso.                           & 3.1\%                                & 1.4--4.2\%              & 1.8--5.4\%          & MC      \\[0.5em]
      \Pe identification                               & $\pm\sigma$(\pt,$\eta$)              & same                    & same                & MC      \\
      \PGm ident. and iso.                             & 4.3\%                                & 2.5--7.5\%              & 2.5--7.5\%          & MC      \\
      Pileup                                           & $\sigma_\mathrm{inel}\pm$4.6\%       & same                    & same                & MC      \\
      L1 trigger timing                                & $\pm\sigma$(\pt,$\eta$)              & same                    & same                & MC      \\
      Jet energy scale                                 & $\pm\sigma$(\pt,$\eta$)              & same                    & same                & MC      \\
      Jet energy res.                                  & $\pm\sigma$($\eta$)                  & same                    & same                & MC      \\
      1, 3$\ell$ \PQb tag: {\PQb}/{\PQc}               & $\pm\sigma$(\pt)                     & same                    & same                & MC      \\
      1, 3$\ell$ \PQb tag: light                       & $\pm\sigma$(\pt)                     & same                    & same                & MC      \\
      Ren./fact. scales                                & env($\times2,\times0.5$)             & same                    & same                & MC      \\
      PDF                                              & $\pm$RMS (Hessian)                   & same                    & same                & MC      \\[0.5em]
      $\Pe$ trigger                                    & $\pm\sigma$(\pt,$\eta$)              & \NA                     & \NA                 & MC      \\
      $\PGm$ trigger                                   & $\pm\sigma$(\pt,$\eta$)              & \NA                     & \NA                 & MC      \\
      {\HT} {\PW}+jets corr.                           & env(upper, lower fits)               & \NA                     & \NA                 & \wjet   \\
      {\HT} \ttbar corr.                               & $\pm\sigma$(\HT)                     & \NA                     & \NA                 & \ttbar  \\
      \textsc{DeepAK8}: light {\Pq}/{\Pg}              & $+0$, $-\sigma$(score)               & \NA                     & \NA                 & MC      \\
      \textsc{DeepAK8}: light {\Pq}/{\Pg} ({\PH} jets) & $\pm\sigma$(score)                   & \NA                     & \NA                 & MC      \\
      \textsc{DeepAK8}: heavy eff.                     & $\pm$25\%/jet/flavor                 & \NA                     & \NA                 & MC      \\
      \textsc{DeepAK8}: heavy misid.                   & $\pm$25\%/jet/flavor                 & \NA                     & \NA                 & MC      \\[0.5em]
      $\Pe\Pe$ trigger                                 & \NA                                  & $\pm\sigma$(\pt,$\eta$) & same                & MC      \\
      $\Pe\PGm$ trigger                                & \NA                                  & $\pm\sigma$(\pt,$\eta$) & same                & MC      \\
      $\PGm\PGm$ trigger                               & \NA                                  & $\pm\sigma$(\pt,$\eta$) & same                & MC      \\
      Charge misid.                                    & \NA                                  & 34.5\%                  & 60.0\%              & OS      \\
      2$\ell$ $\Pe/\PGm$ prompt rate                   & \NA                                  & $\pm\sigma$(\pt,$\eta$) & same                & NP      \\
      2$\ell$ $\Pe$ NP rate                            & \NA                                  & NP rate$+$0.07          & NP rate$+$0.08      & NP      \\
      2$\ell$ $\PGm$ NP rate                           & \NA                                  & NP rate$-$0.02          & same                & NP      \\[0.5em]
      3$\ell$ $\Pe$ prompt rate                        & \NA                                  & 1.0--6.6\%              & 0.9--7.3\%          & NP      \\
      3$\ell$ $\PGm$ prompt rate                       & \NA                                  & 1.0--6.6\%              & 1.2--5.5\%          & NP      \\
      3$\ell$ $\Pe$ NP rate                            & \NA                                  & NP rate$\pm$0.010       & same                & NP      \\
      3$\ell$ $\PGm$ NP rate                           & \NA                                  & NP rate$\pm$0.022       & NP rate$\pm$0.014   & NP      \\
      3$\ell$ NP bkgd., by flavor                      & \NA                                  & 3.9--23.0\%             & 1.4--24.6\%         & NP      \\
      3$\ell$ $\PGm$ NP rate in $\eta$($\PGm$)         & \NA                                  & 10.0--38.7\%            & 1.0--1.4\%          & NP
    \end{tabular}
  }
\end{table}

\subsection{Common uncertainties}

Several uncertainties are found in common throughout the search and are correlated across the three analysis channels.
The effects of all shape-based uncertainties are evaluated by varying inputs to the analysis by their respective uncertainties.
\begin{itemize}
  \item Integrated luminosity: the integrated luminosities for the years 2016, 2017, and 2018 have 1.2--2.5\% individual uncertainties, and the total luminosity has an uncertainty of 1.6\%~\cite{LUM16,LUM17,LUM18}.
        Components of these uncertainties are correlated between the 2016--2018 and 2017--2018 data periods.
  \item Muon identification and isolation scale factors: uncertainties in these corrections are 2\% per muon for identification and 1.5\% per muon for isolation.
        In the single-lepton channel the uncertainties for the three data periods are combined in quadrature.
  \item Electron reconstruction and isolation: uncertainties in these corrections are 1\% per electron for reconstruction and 1.5\% per electron for isolation.
        In the single-lepton channel the uncertainties for the three data periods are combined in quadrature.
  \item Electron identification: the uncertainty in the correction is applied as a two-dimensional function of \pt and $\eta$.
  \item Pileup correction: the uncertainty in the pileup weighting for simulation is evaluated by varying the total inelastic cross section ($\sigma_\mathrm{inel}$) of 69.2\unit{mb} by $\pm 4.6$\%~\cite{CMS:2018mlc}.
  \item L1 trigger timing: the uncertainty in this correction is applied as a two-dimensional function of $\pt$ and $\eta$ to data from 2016 (in the single-lepton channel) and 2017.
  \item Jet energy scale and resolution: the uncertainties in these corrections affect both small-radius and large-radius jet momenta, and are propagated to the \ptmiss distribution and all observables calculated from jets.
        In the single-lepton channel, the LOWESS algorithm~\cite{LOWESS1,LOWESS2} is used to smooth the resulting shifted histograms.
  \item \textsc{DeepJet} \PQb tagging: uncertainties in these corrections are applied separately for bottom and charm quark tagging, and for light quark or gluon misidentification~\cite{CMS-DP-2018-058}.
  \item Scale uncertainties: the uncertainty in the choice of the renormalization ($\mu_R$) and factorization ($\mu_F$) scales in the simulation is used to estimate the effect of not including higher-order matrix elements. Scale uncertainties are treated independently for each group of simulated physics processes described in Section~\ref{sec:samples}.
        The uncertainty is computed in each bin of the final observable by varying $\mu_R$ up and down by a factor of two, and also varying $\mu_F$ up and down by a factor of two, including symmetric shifts of both scales, and forming an envelope from the seven resulting distributions.
        In Table~\ref{table:Systematics} this envelope is summarized as ``env($\times 2, \times 0.5$)''.
        For the signal, the normalization impact of this uncertainty is scaled down to reflect only the effect of the analysis selection.
  \item PDFs: the PDF uncertainty is correlated across all samples that use the same PDF set. Specifically, uncertainties in the PDF sets applied in the simulation are treated separately for 2016 and 2017--2018 to account for the update from NNPDF~3.0 to NNPDF~3.1 PDF sets.
        Additionally, the 2017--2018 uncertainty is treated separately for signal, to which the PDF4LHC15 PDF set was applied.
        In the 2016 simulation, the NNPDF~3.0 uncertainty is computed bin-by-bin in the final observable from 100 PDF replicas using a quantile method to identify the RMS~\cite{PDF4LHC}, summarized as ``$\pm$RMS(replicas)'' in Table~\ref{table:Systematics}.
        In the 2017--2018 simulation, Hessian uncertainties are summed in quadrature to form a total uncertainty in each bin (``$\pm$RMS(Hessian)'').
        For the signal, the normalization impact of the PDF uncertainty is also scaled down to reflect only the effect of the analysis selection.
\end{itemize}

The uncertainties in electron identification, jet energy scale, and jet energy resolution are uncorrelated across the 2016, 2017, and 2018 data periods.
Other common uncertainties are correlated between run periods, unless described otherwise above.

\subsection{Single-lepton uncertainties}\label{sec:syst1lep}

Uncertainties affecting only the single-lepton channel are:
\begin{itemize}
  \item Single-lepton triggers: the uncertainty in these corrections to the simulation are applied as 2D functions of \pt and $\eta$, independently for each lepton flavor.
  \item \HT distribution correction for the {\PW}+jets background: the uncertainty in this correction is formed by repeating the fits with all points shifted up or down by their statistical uncertainties to form an envelope-like uncertainty.
        In Table~\ref{table:Systematics} this procedure is summarized as ``env(upper, lower fits)''.
  \item \HT distribution correction for the \ttbar background: the uncertainty in this correction is calculated using the fit parameter covariance matrix and depends on the value of \HT in each event.
  \item \textsc{DeepAK8} light quark or gluon score corrections: for jets not associated with a generator-level \PH, a single-sided uncertainty applied to 2017--2018 simulation is formed by turning on and off the correction.
        The uncertainty in the \PH jet correction is evaluated by shifting the correction factors by their uncertainty.
        For all jets the value of this uncertainty varies, based on the \textsc{DeepAK8} light quark or gluon score in each event.
  \item \textsc{DeepAK8} heavy-particle tagging and misidentification: independent uncertainties in the efficiency for the \textsc{DeepAK8} algorithm to identify correctly or incorrectly each massive particle within a jet are formed in order to perform an in situ correction of any differences in efficiency between data and simulation.
        To form a large input uncertainty for correct and incorrect identification of each flavor, the \textsc{DeepAK8} tag of each large-radius jet is compared to its ``true'' parent particle, which is determined by spatial matching between the jets and generated particles.
        For each jet, the relevant flavor's correct or incorrect identification uncertainty is incremented by a large value (25\%).
        The misidentification uncertainties are constrained significantly in the fit for all jet flavors, and the correct identification uncertainties are constrained for all flavors except \PH and \PZ bosons, which are very rare in the background events.
\end{itemize}

Of these uncertainties, the trigger scale factor uncertainties, \textsc{DeepAK8} light quark or gluon score uncertainty for \PH jets, and \textsc{DeepAK8} heavy-particle misidentification uncertainties are treated independently for each data period.
The dominant uncertainties in the single-lepton background predictions are the renormalization and factorization scale uncertainties, and the signal predictions are most sensitive to the \textsc{DeepAK8} heavy-particle misidentification uncertainties.

\subsection{SS dilepton uncertainties}

Uncertainties affecting the SS dilepton channel are:
\begin{itemize}
  \item Dilepton triggers: uncertainties in the trigger scale factors for each lepton flavor combination are applied as 2D functions of lepton \pt and $\eta$.
  \item Charge misidentification: the uncertainty in the charge misidentification background has two contributions.
        The first is the statistical uncertainty in the minimization procedure.
        The second arises from systematic differences in event topology between the Drell--Yan process used to measure the misidentification rate and the \ttbar-like SR.
        The latter is determined from simulation by counting the number of tight electrons with $\pt > 30\GeV$ and an incorrectly reconstructed charge.
        Comparing the rates measured in the Drell--Yan and \ttbar samples, the uncertainties in the charge misidentification yields are 31\% and 59\% for 2017 and 2018 data, respectively.
  \item Prompt rates: uncertainties in the nonprompt background estimate due to the prompt rate measurement are evaluated by varying the prompt rates by their measurement uncertainty separately for each flavor.
  \item Nonprompt rates: uncertainties in the nonprompt background estimate due to the nonprompt rates are estimated by recalculating the nonprompt background in the $\HTlep < 1000\GeV$ region, with the differences between the best fitted values in $\HTlep < 1000\GeV$ and those in the multilepton CR taken as one-sided uncertainties affecting the shape of the nonprompt background estimate.
\end{itemize}

With the exception of the trigger scale factor uncertainties, all of the SS dilepton uncertainties are treated independently for each data period, including the uncertainties affecting 2016 data that are unchanged from the previously published result.
The dominant background uncertainties are those affecting the nonprompt background estimation.

\subsection{Multilepton uncertainties}

Uncertainties affecting the multilepton channel are:
\begin{itemize}
  \item Dilepton triggers: uncertainties in the trigger scale factors for each lepton flavor combination are applied as 2D functions of lepton \pt and $\eta$.
  \item Prompt rates: the prompt rates are measured in dilepton events so the corresponding uncertainties must cover the possible effect of different event topologies, as well as the measurement uncertainty itself.
        Since the measured values are not too far from unity and as using a prompt rate of unity does not significantly affect the nonprompt rates, uncertainties are conservatively estimated by comparing the nonprompt background event yields when applying the measured prompt rates and unity prompt rates, separately for each lepton flavor.
  \item Nonprompt rates ($\Pe/\PGm$): the nonprompt rates for each lepton flavor are varied by the sum in quadrature of the measurement error from marginalization and the deviation in the nonprompt rates obtained from fitting simulated \ttbar events in the SR and the CR.
        This variation of nonprompt rates from the best fit values presented in Section~\ref{sec:multiback} results in shape uncertainties in the nonprompt background estimate.
  \item Nonprompt background (by flavor): a normalization uncertainty is assigned to the nonprompt background estimate in each of the four lepton flavor categories to account for remaining discrepancies observed during cross-checks.
        This uncertainty is the quadrature sum of two sources: the differences between the data and the predicted background in the CR, and the difference between the true nonprompt and the predicted number of events in the cross-check using simulated \ttbar events.  
  \item Nonprompt rate ($\eta(\PGm)$): the SS dilepton channel measurement in the 2016 analysis observed $\eta$-dependence in the muon nonprompt rate~\cite{B2G-17-011}: this effect is studied in 2017--2018 data and included as an uncertainty. 
\end{itemize}
With the exception of the trigger scale factor uncertainties, all of these multilepton uncertainties are treated independently for each data period, including the uncertainties affecting 2016 data that are unchanged from the previously published result.
As in the SS dilepton channel, the dominant background uncertainties are those affecting the nonprompt background estimation.

\section{Results}
\label{sec:results}

The possible presence of a signal is determined by simultaneously fitting template histograms, shown below, from a variety of discriminating variables in all three channels. 
In the single-lepton channel, the \HT and jet tag CR distributions are included in the fit to constrain uncertainties in the background modeling. In the SR, the VLQ score from the MLP is used to form template histograms for both HP events, in which both VLQ candidates contain the expected particle labels, and for LP events, which have at least one VLQ Candidate without the expected particle labels.
The SR data are subdivided into 24 (\TTbar) or 18 (\BBbar) exclusive categories based both on the lepton flavor and the set of \textsc{DeepAK8} jet tags observed.
The categorization according to \textsc{DeepAK8} jet tags, applied to electron and muon events separately, is described in Table~\ref{tab:cats}.
\begin{table}[htbp]
  \centering
  \topcaption{
    Category labels and definitions for the SRs of the single-lepton channel \TTbar analysis. Electron and muon events are analyzed separately in all categories.
    The VLQ candidate tag describes the pairings formed from the leptonic particle candidate and three large-radius jets.
    The hadronic VLQ candidate is reconstructed from two large-radius jets, and a VLQ candidate tag of ``other'' indicates that the hadronic VLQ candidate did not consist of $\PQb\PW$-, $\PQt\PZ$-, or $\PQt\PH$-tagged jets.
    In the \BBbar analysis, the VLQ candidate tags considered are $\PQt\PW$, $\PQb\PZ$, and $\PQb\PH$. Categories 4, 6, and 7 are not included in the \BBbar analysis.
  }
  \begin{tabular}{c c c c@{$\quad$}c@{$\quad$}c}
    \centering
    SR label & Purity & VLQ candidate tags                                     & \multicolumn{3}{c}{{\sc DeepAK8} jet tags}                       \\
             &        &                                                        & $\PQt$                                     & $\PH$    & $\PZ$    \\[0.1em]
    \hline
    \\[-0.9em]
    1        & HP     & $\PQb\PW\PQb\PW$                                       &                                            &          &          \\
    2        & HP     & $\PQt\PZ\PQb\PW$                                       &                                            &          &          \\
    3        & HP     & $\PQt\PH\PQb\PW$                                       &                                            &          &          \\
    4        & HP     & $\PQt\PZ\PQt\PZ$ + $\PQt\PZ\PQt\PH$ + $\PQt\PH\PQt\PH$ &                                            &          &          \\[0.5em]
    5        & LP     & hadronic $\PQb\PW$                                     &                                            &          &          \\
    6        & LP     & hadronic $\PQt\PZ$                                     &                                            &          &          \\
    7        & LP     & hadronic $\PQt\PH$                                     &                                            &          &          \\[0.5em]
    8        & LP     & other                                                  & $\geq 2$                                   &          &          \\
    9        & LP     & other                                                  & 1                                          & 0        &          \\
    10       & LP     & other                                                  & 0-1                                        & $\geq 2$ &          \\
    11       & LP     & other                                                  & 0-1                                        & 1        &          \\
    12       & LP     & other                                                  & 0                                          & 0        & $\geq 1$ \\
  \end{tabular}
  \label{tab:cats}
\end{table}

\begin{figure}[htbp]
  \centering
  \includegraphics[width=0.32\textwidth]{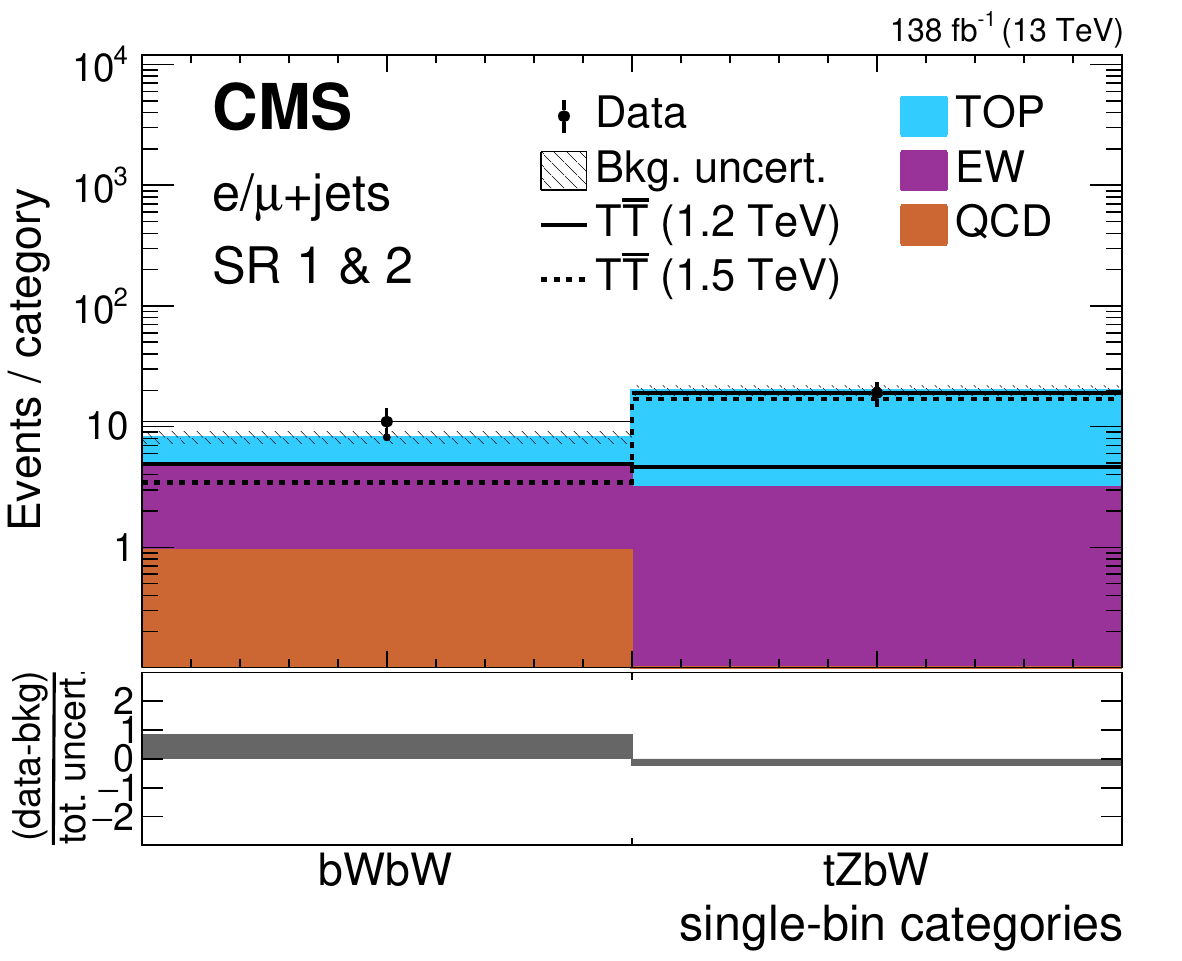}
  \includegraphics[width=0.32\textwidth]{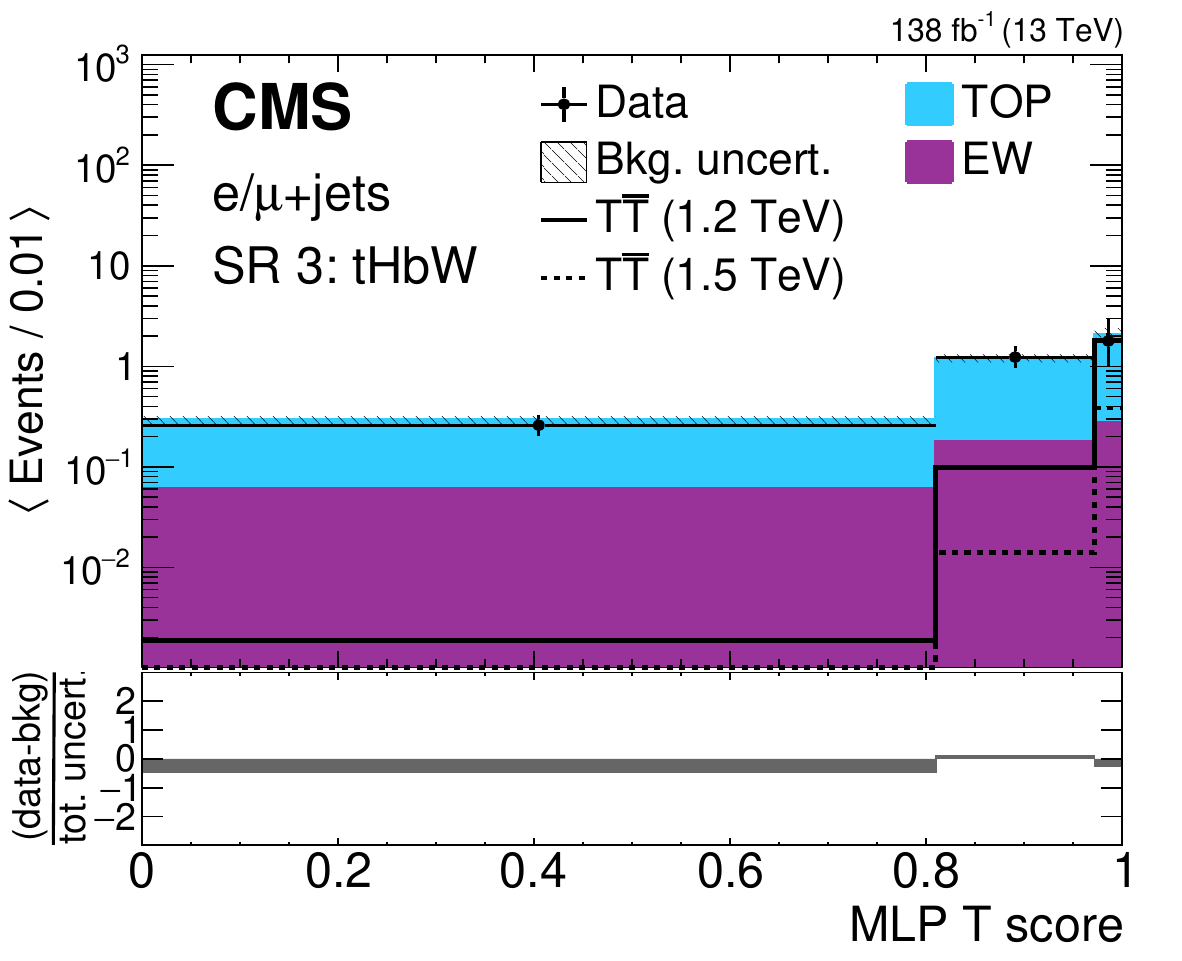}
  \includegraphics[width=0.32\textwidth]{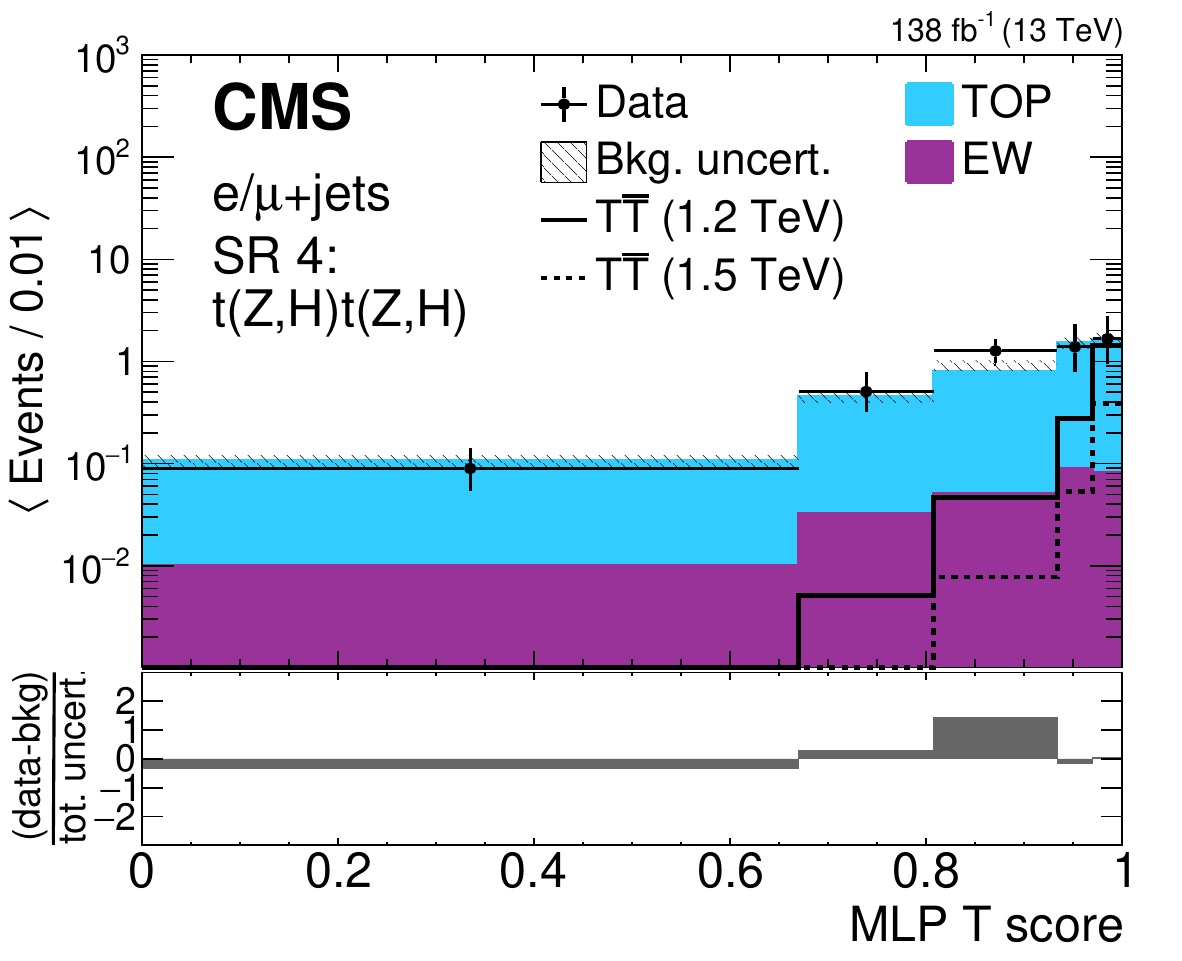}\\
  \includegraphics[width=0.32\textwidth]{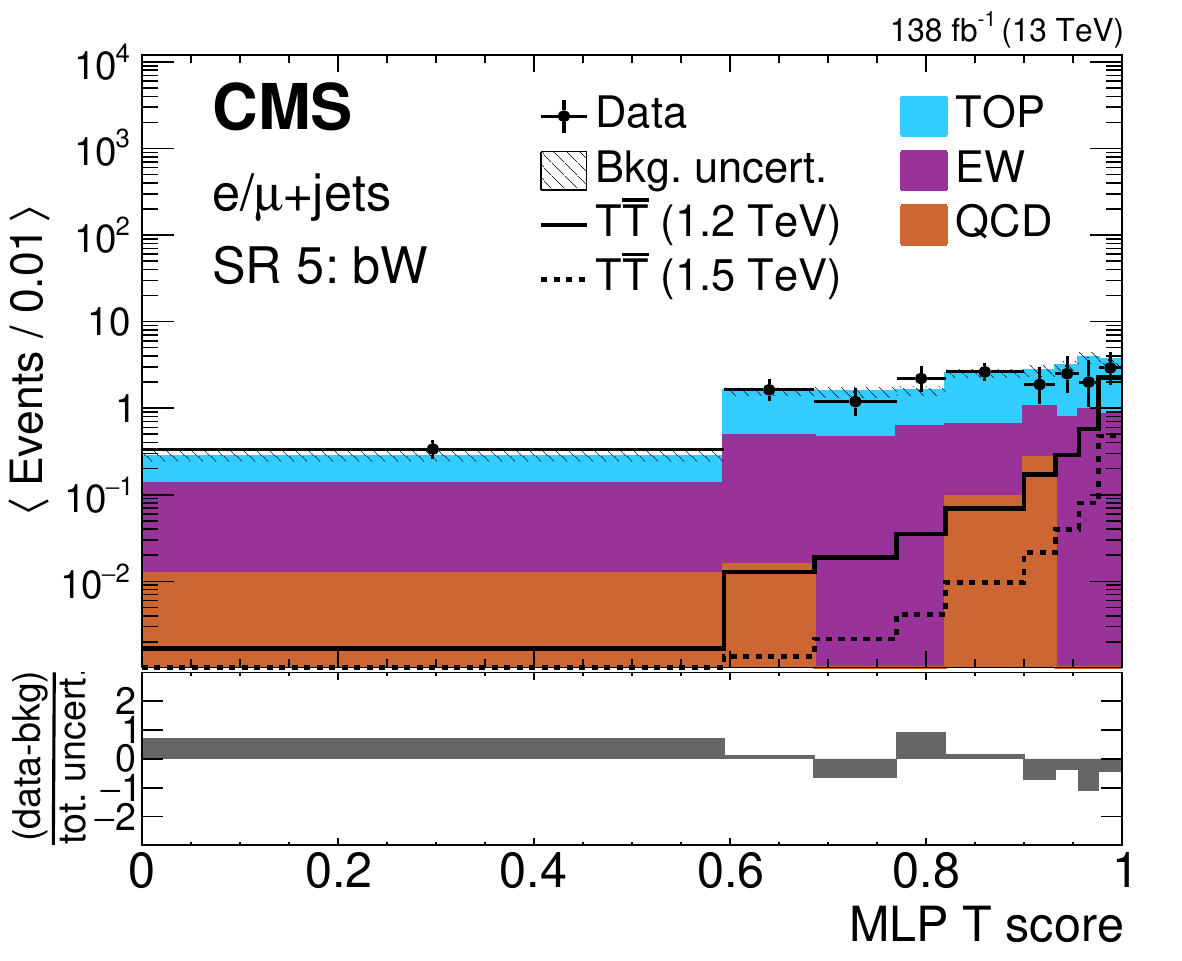}
  \includegraphics[width=0.32\textwidth]{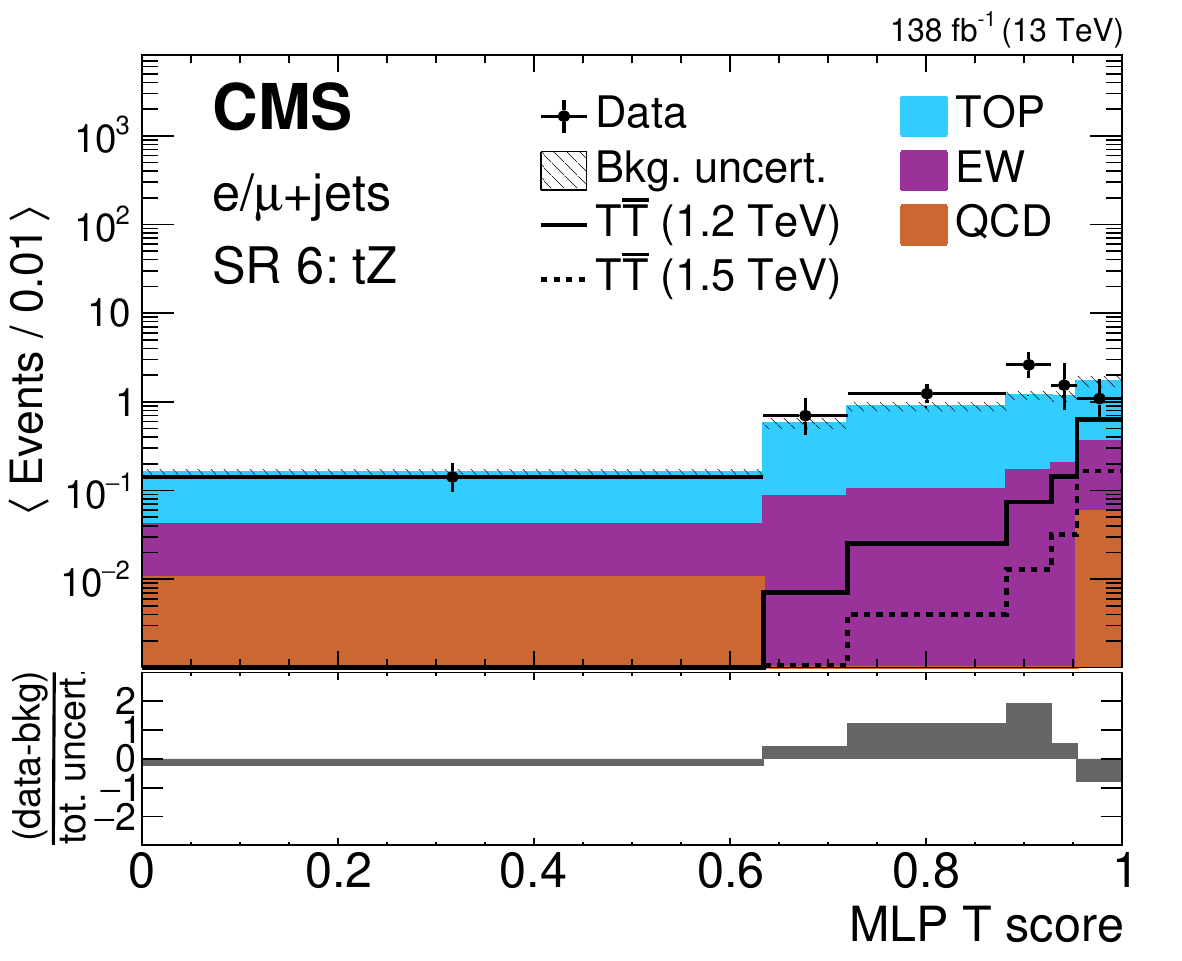}
  \includegraphics[width=0.32\textwidth]{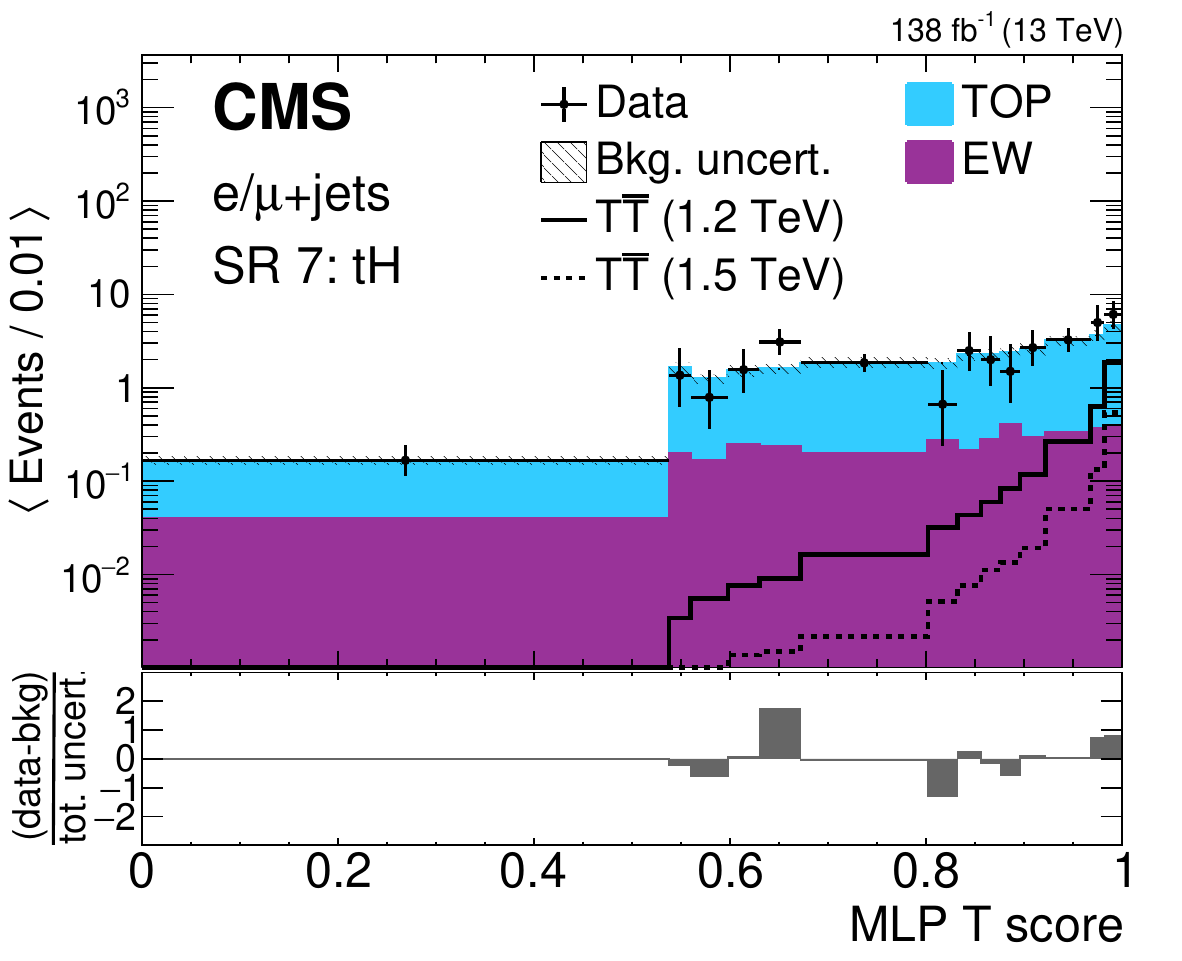}\\
  \includegraphics[width=0.32\textwidth]{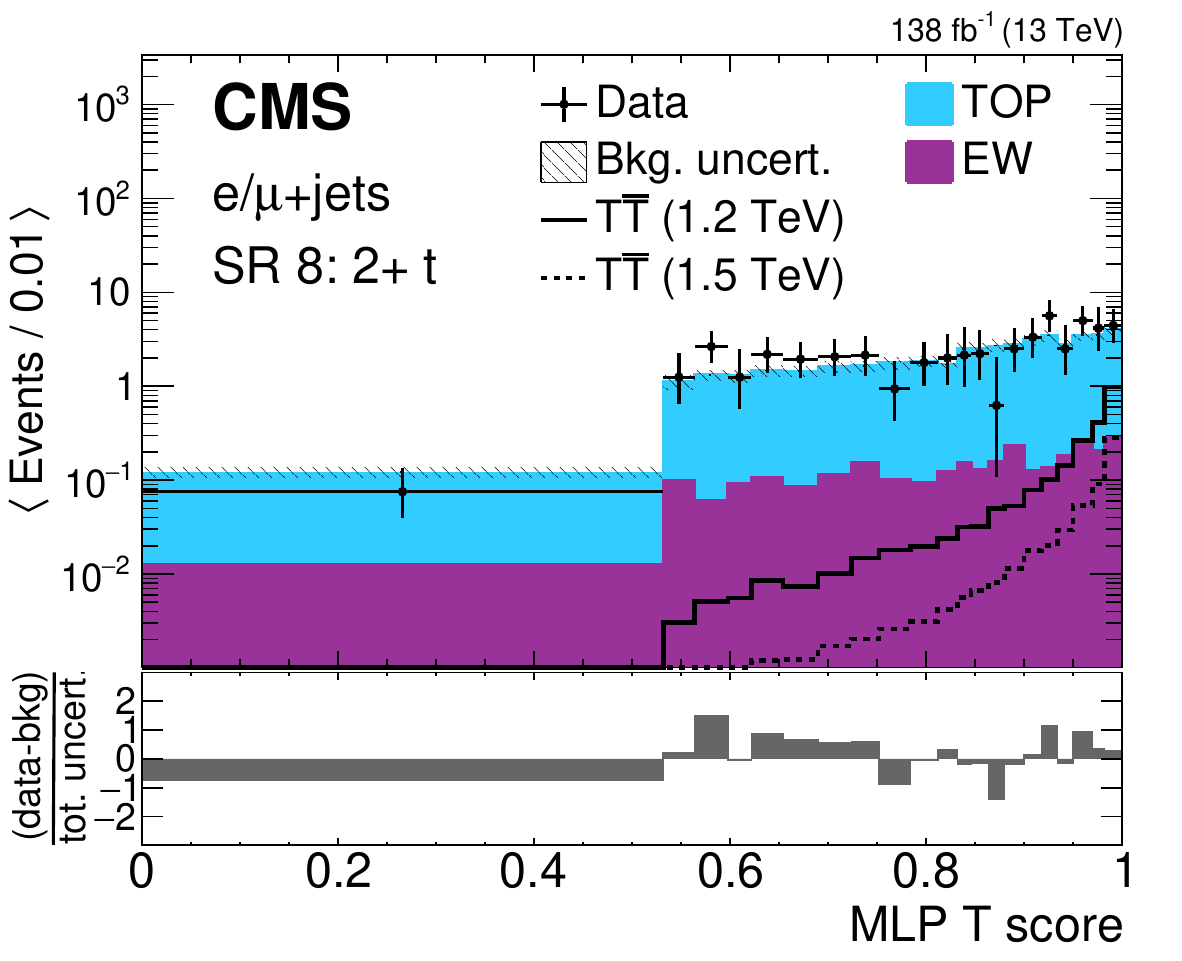}
  \includegraphics[width=0.32\textwidth]{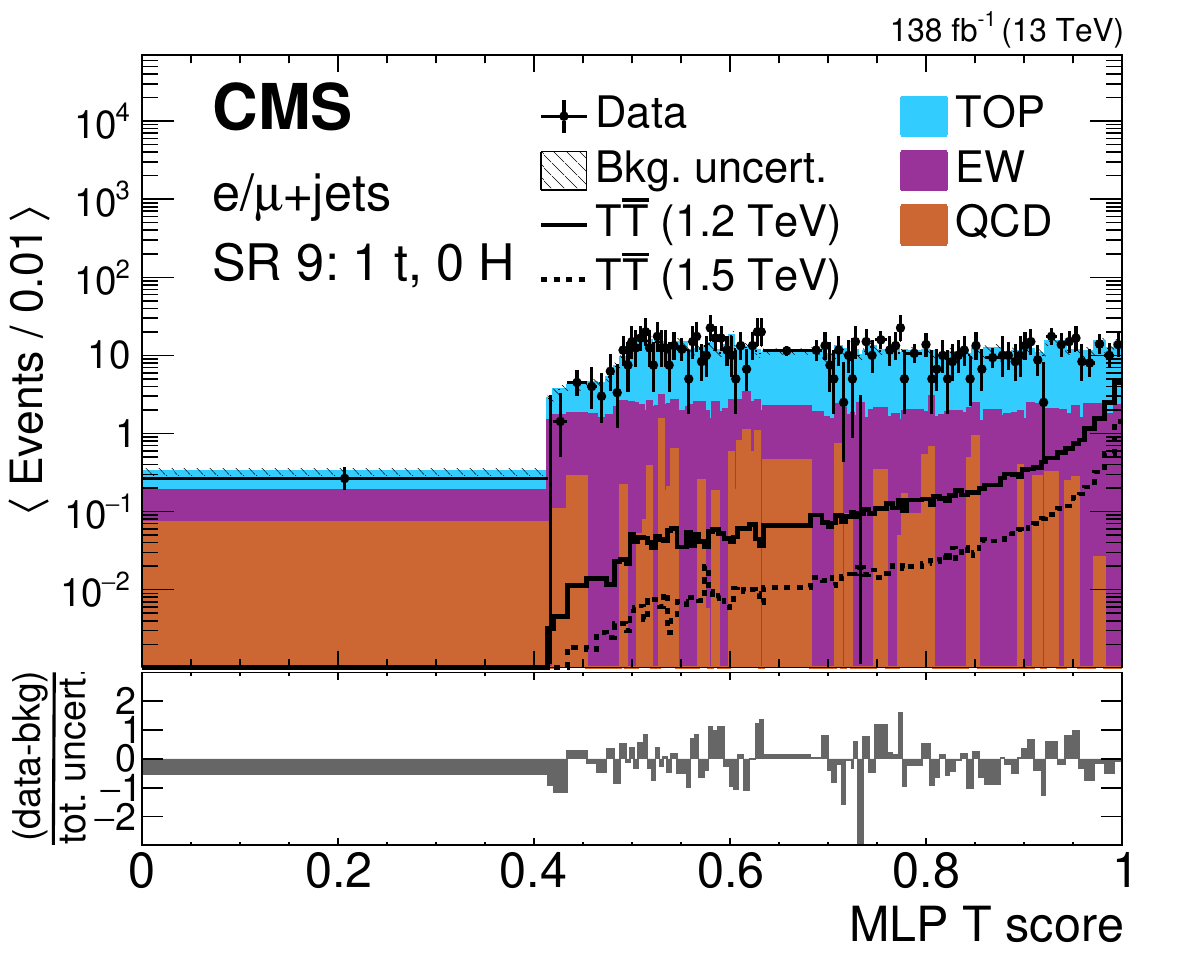}
  \includegraphics[width=0.32\textwidth]{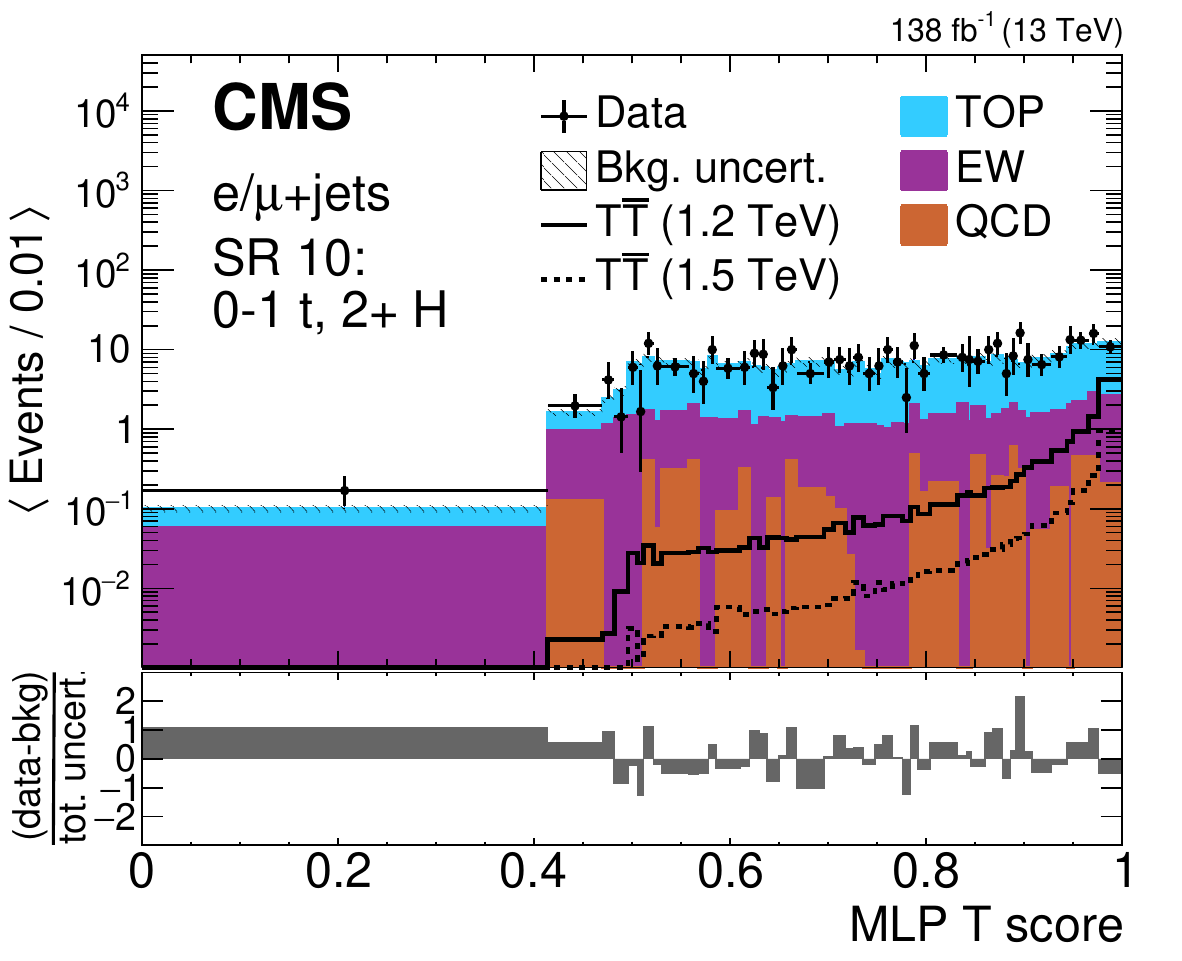}\\
  \includegraphics[width=0.32\textwidth]{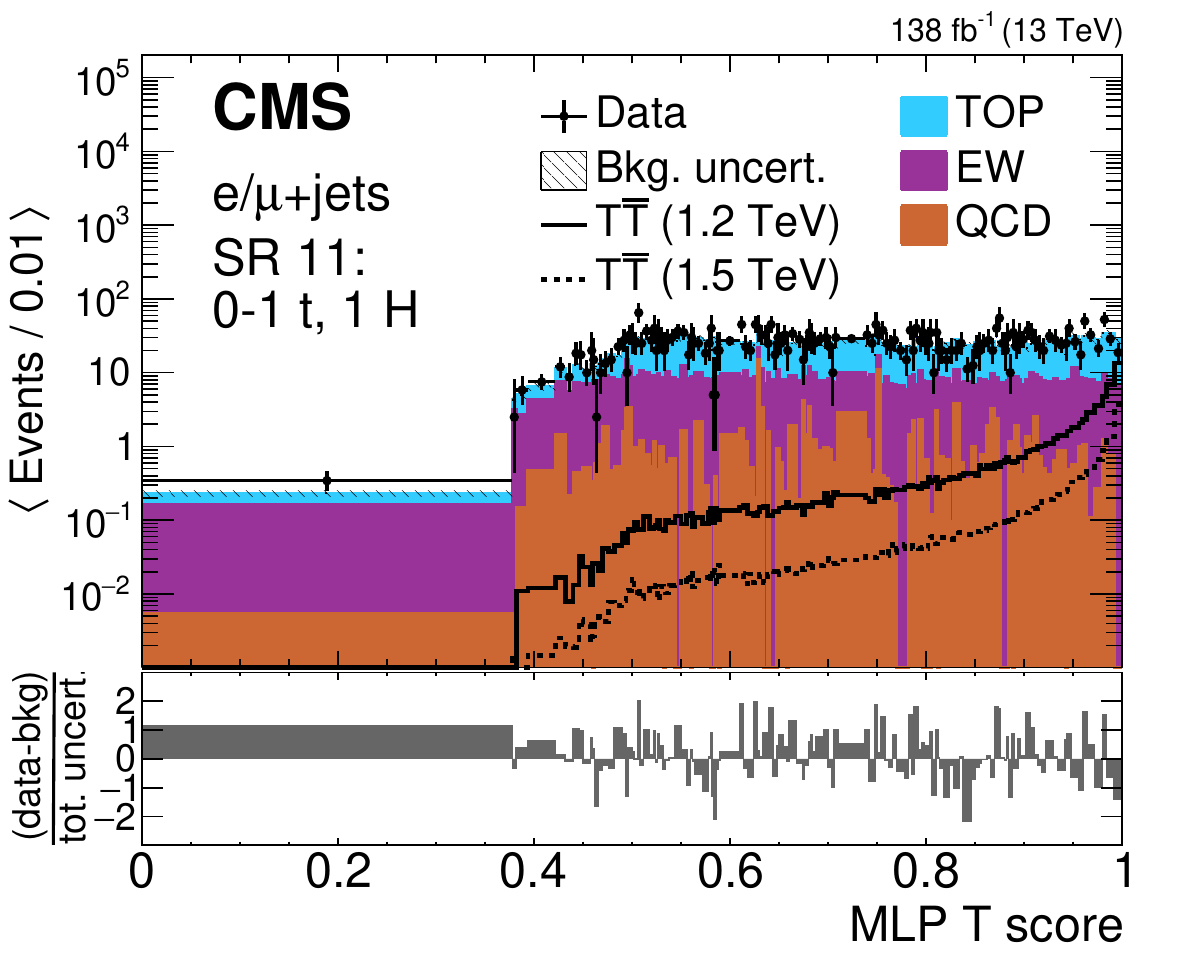}
  \includegraphics[width=0.32\textwidth]{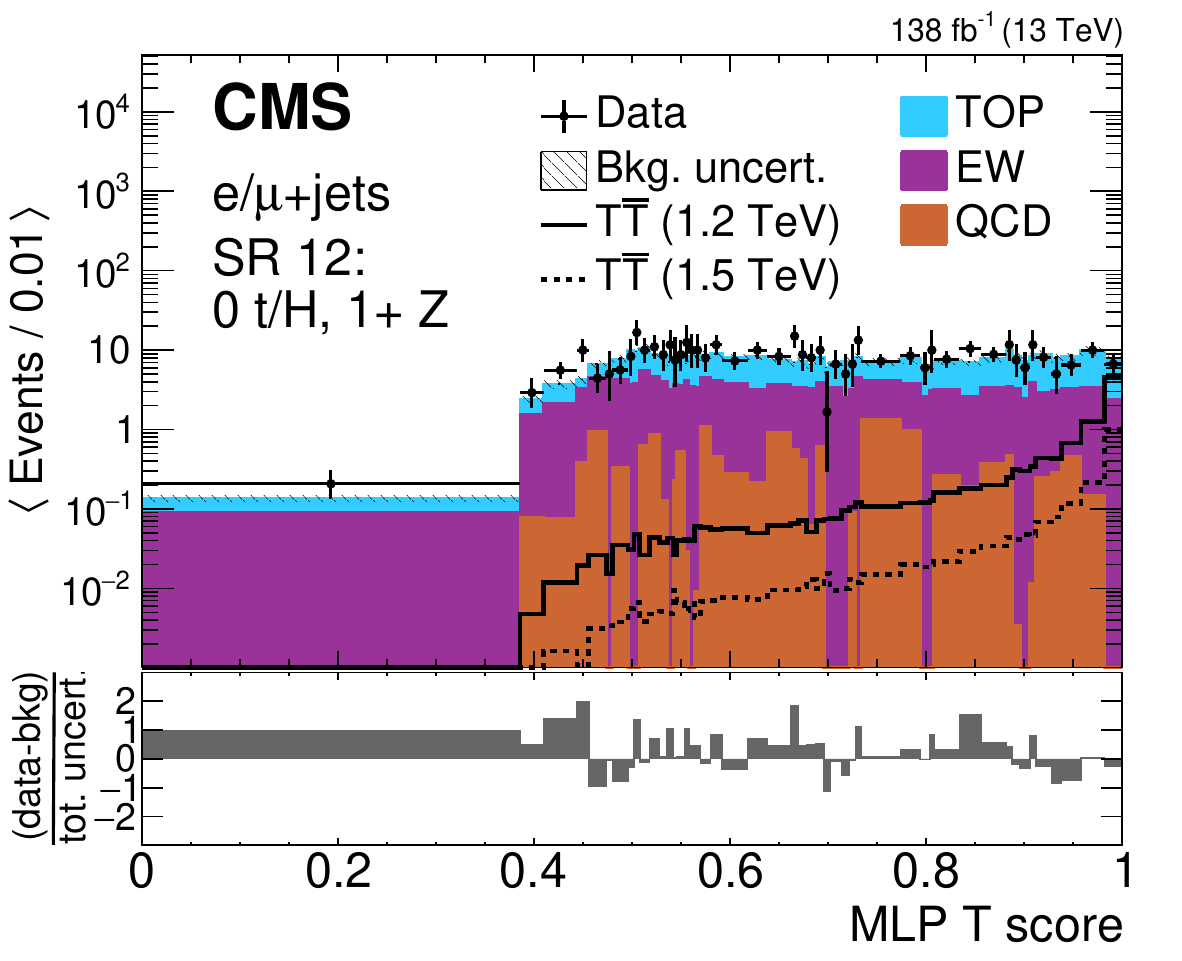}
  \caption{
    Template histograms of the VLQ score in single-lepton SRs 1 and 2 combined, and SRs 3--12 (left-to-right, upper-to-lower).
    The observed data are shown using black markers, the predicted \TTbar signal for a mass of 1.2 (1.5)\TeV in the singlet scenario using solid (dashed) lines, and the post-fit background estimates, using filled histograms.
    Statistical and systematic uncertainties in the background estimate after performing the fit to data are shown by the hatched region.
    The lower panels show the difference between the data and the background estimate as a multiple of the total uncertainty in both sources.
    Electron and muon categories have been combined for illustration with their uncertainties added in quadrature.
  }
  \label{fig:templatesTTinvalid}
\end{figure}

Figure~\ref{fig:templatesTTinvalid} shows the template histograms for each single-lepton SR category listed in Table~\ref{tab:cats} for the \TTbar analysis, with lepton flavor categories combined for illustration.
The histograms are binned such that the total background in each bin has a statistical uncertainty smaller than 20\%. Tables~\ref{tab:SRLTT}--\ref{tab:SRLBB} list the numbers of events selected for each category in both the \TTbar and \BBbar analyses, the latter of which is not shown in the figures.

\begin{figure}[hbtp]
  \centering
  \includegraphics[width=0.49\linewidth]{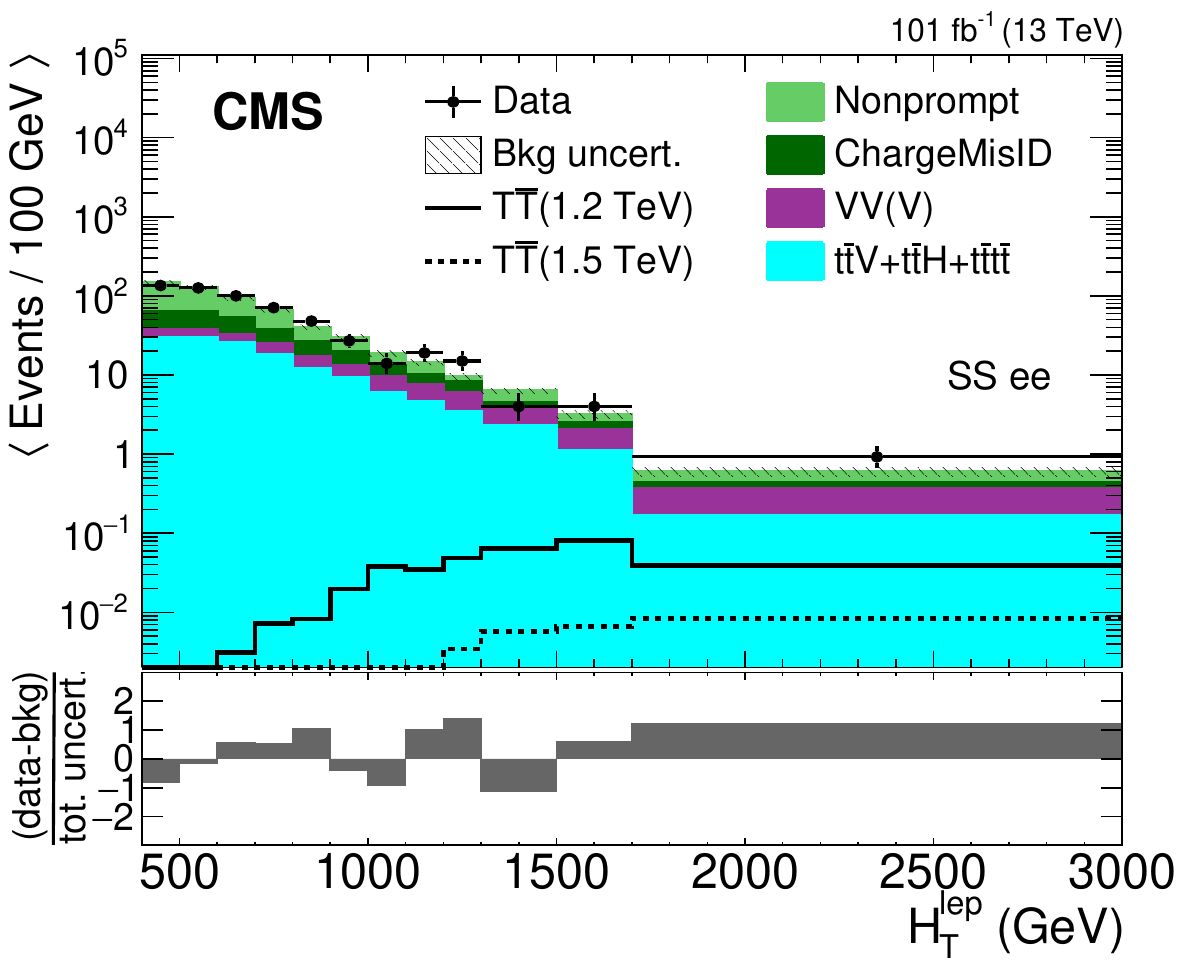}
  \includegraphics[width=0.49\linewidth]{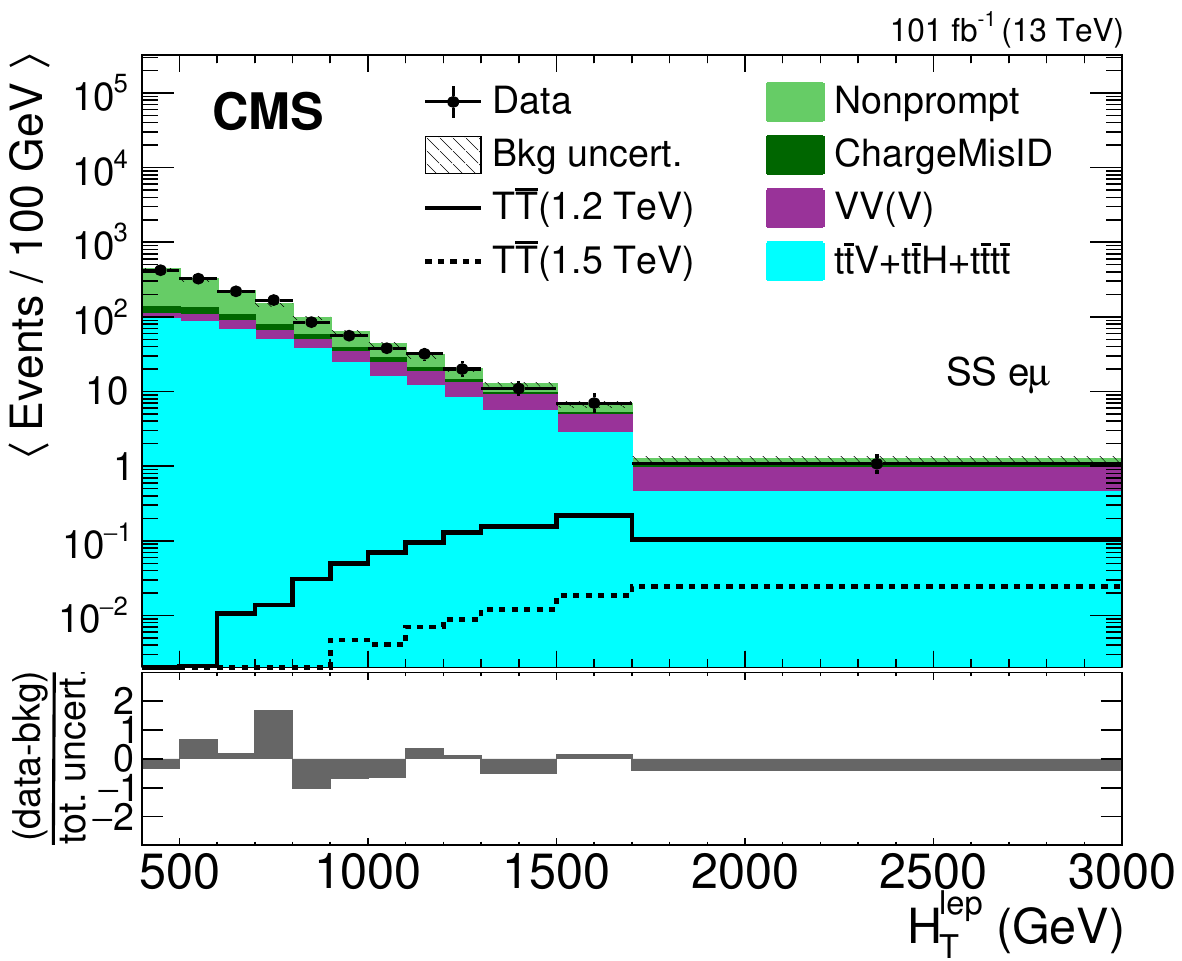}
  \includegraphics[width=0.49\linewidth]{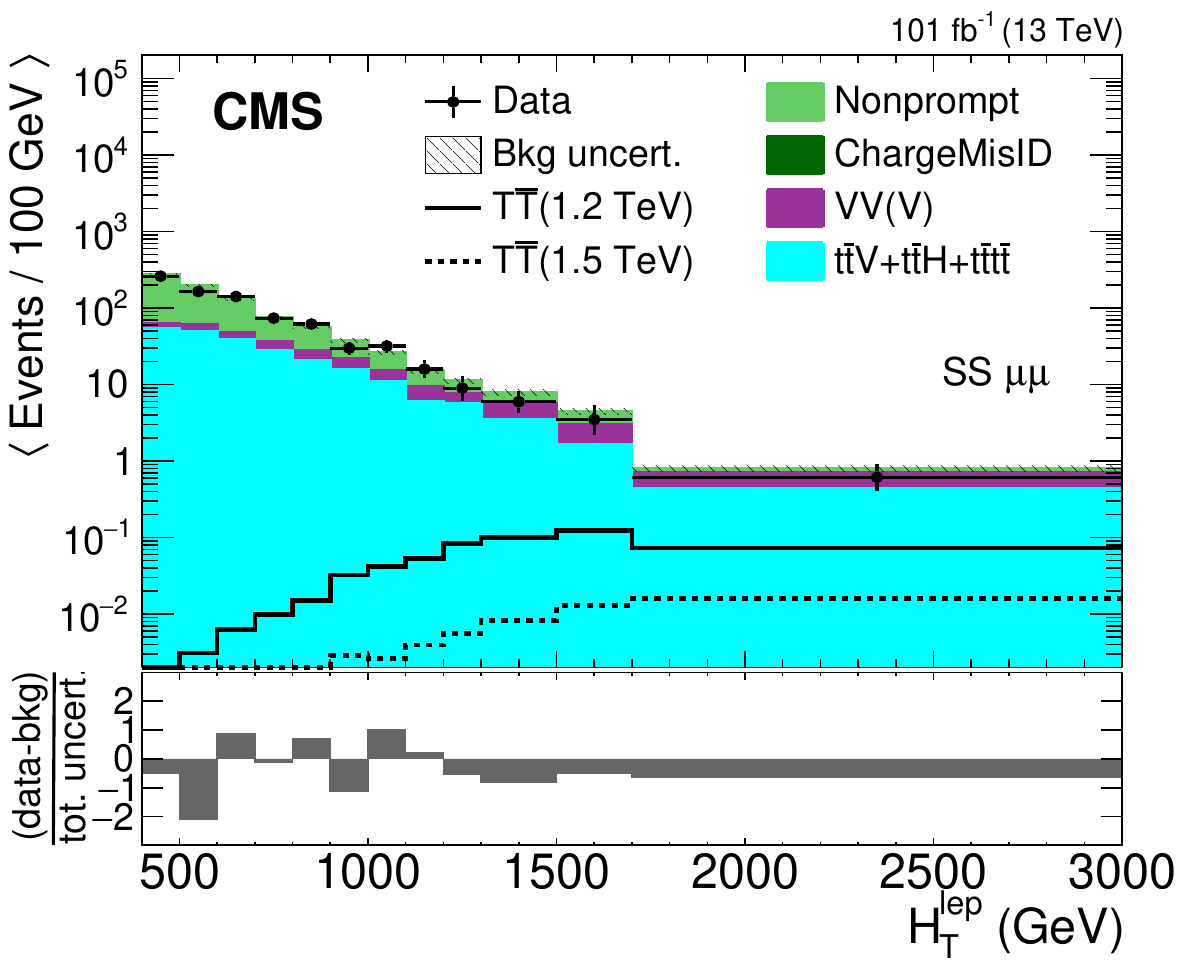}
  \caption{
    Template histograms of \HTlep in the same-sign dilepton signal region for $\Pe\Pe$ (upper left), $\Pe\PGm$ (upper right), and $\PGm\PGm$ categories (lower).
    The observed data from 2017--2018 (combined for illustration) are shown using black markers, the predicted \TTbar signal for a mass of 1.2 (1.5)\TeV in the singlet scenario using solid (dashed) lines, and the post-fit background estimates, using filled histograms.
    Statistical and systematic uncertainties in the background estimate after performing the fit to data are shown by the hatched region.
    The lower panels show the difference between the data and the background estimate as a multiple of the total uncertainty in both sources.
  }
  \label{fig:SS2l1718_HT_final}
\end{figure}

\begin{figure}[hbtp]
  \centering
  \includegraphics[width=0.49\linewidth]{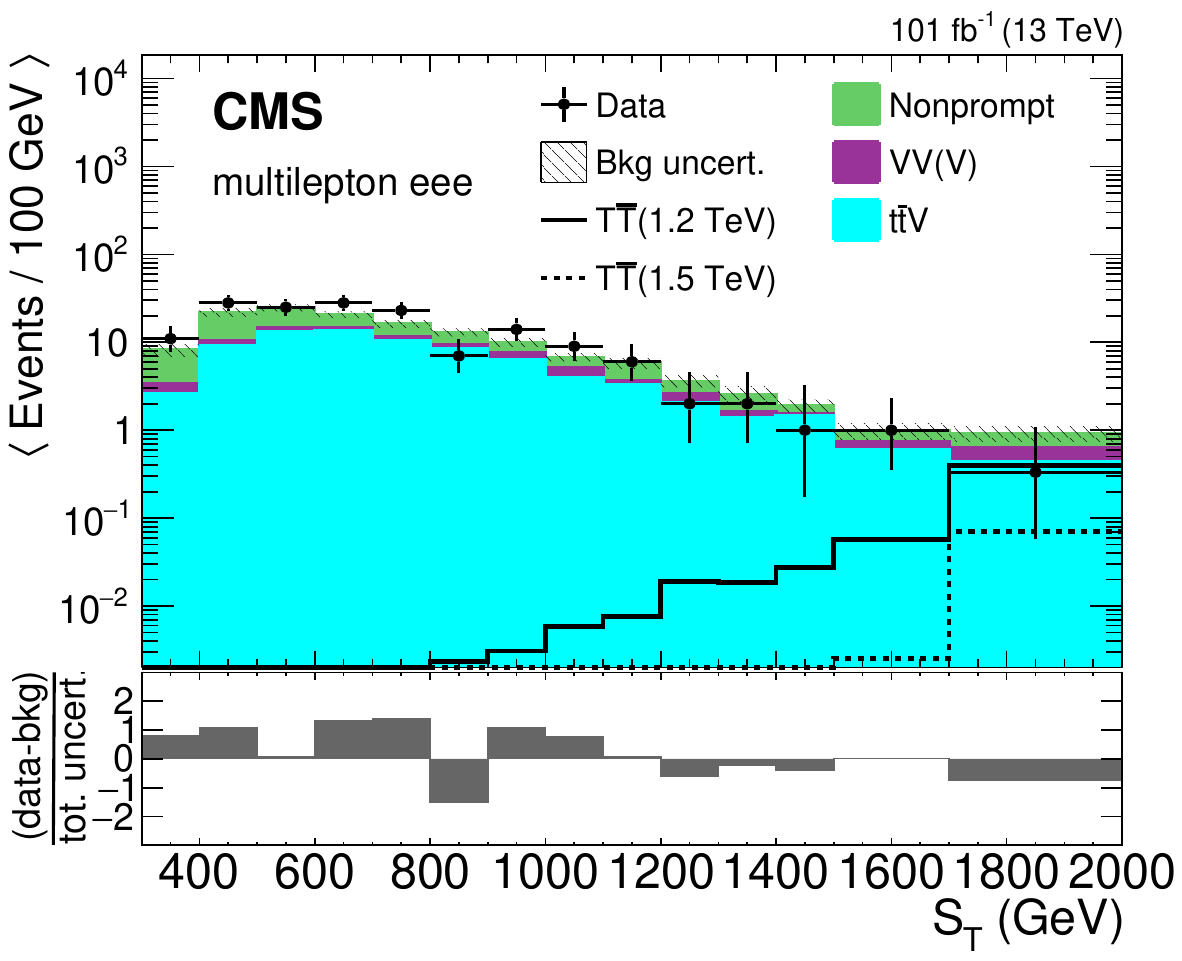}
  \includegraphics[width=0.49\linewidth]{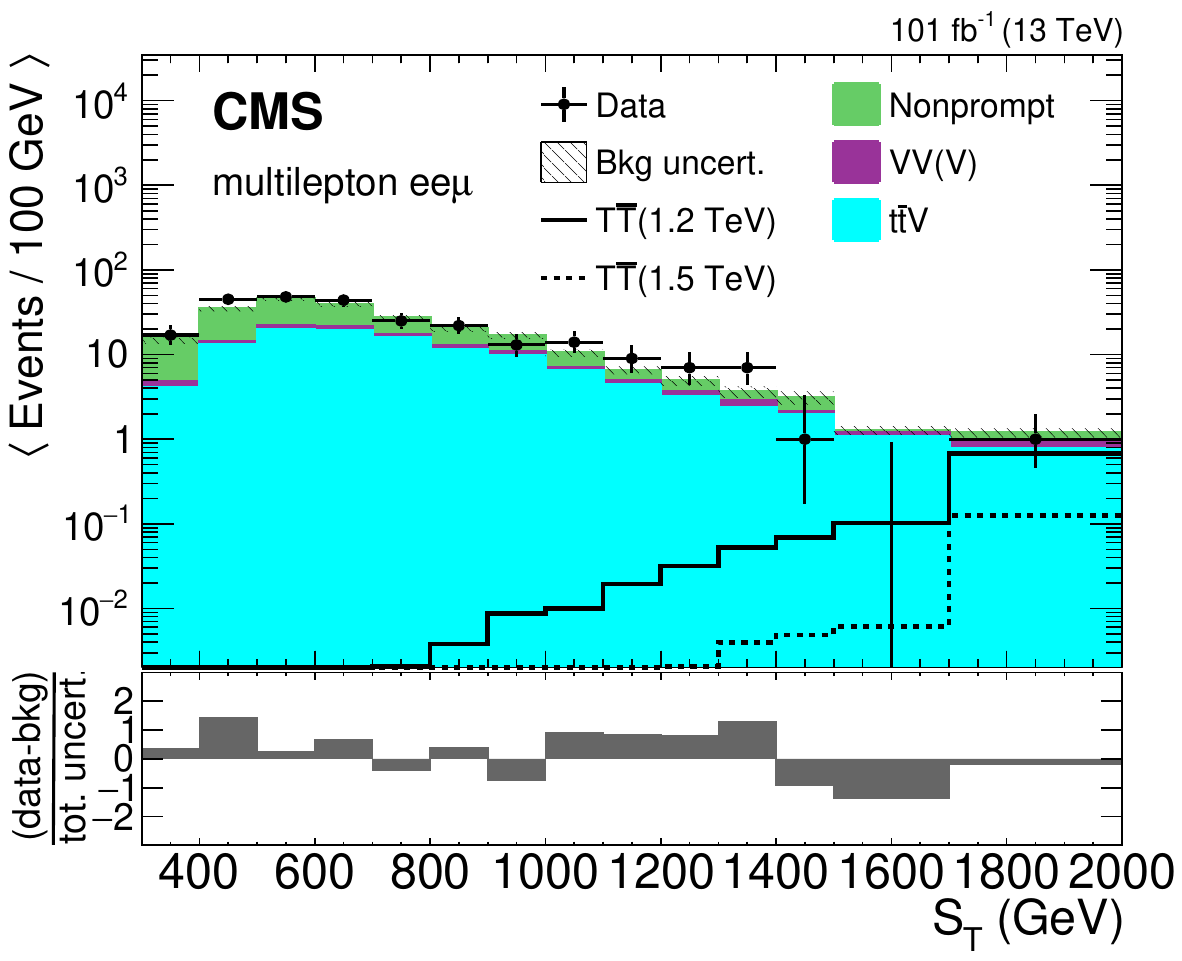}\\
  \includegraphics[width=0.49\linewidth]{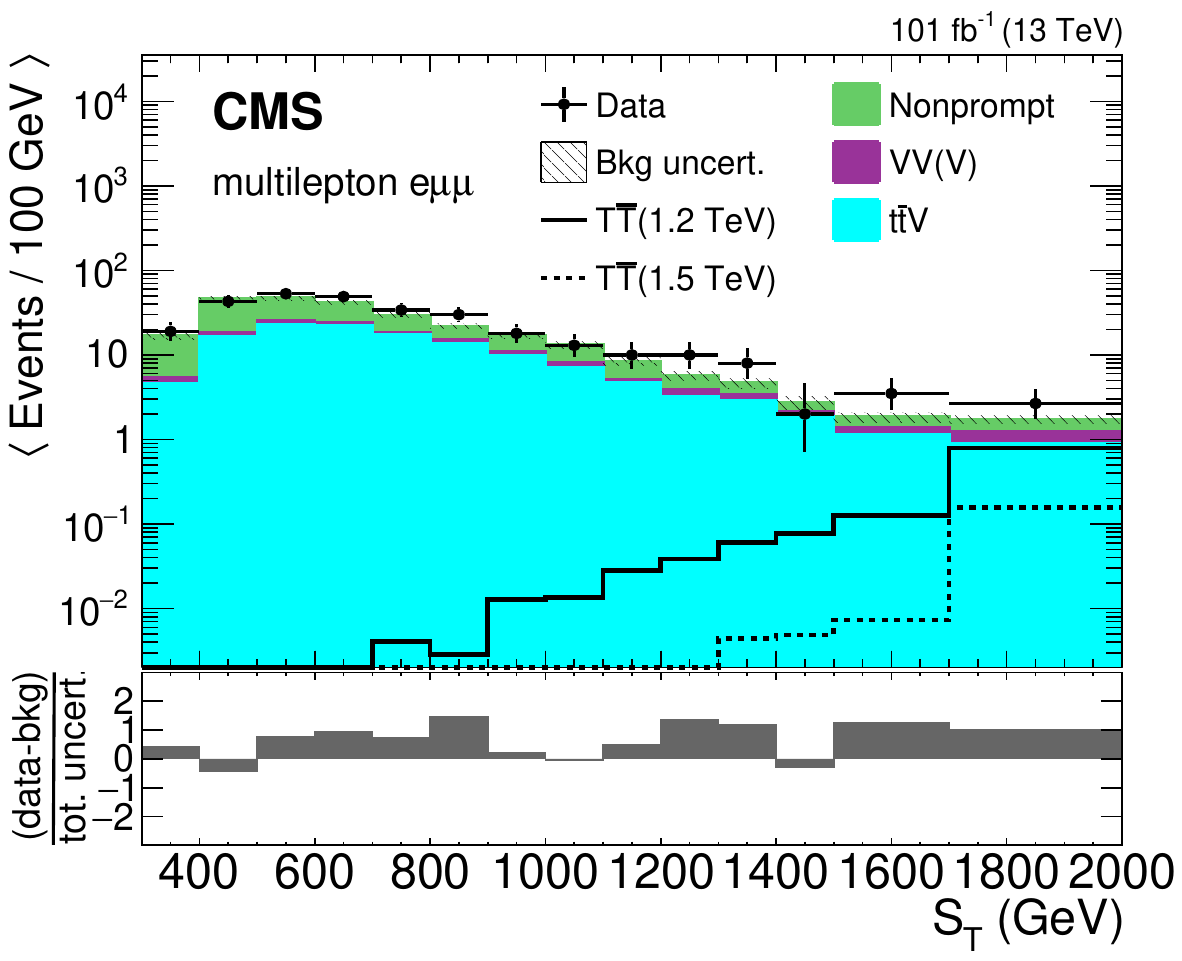}
  \includegraphics[width=0.49\linewidth]{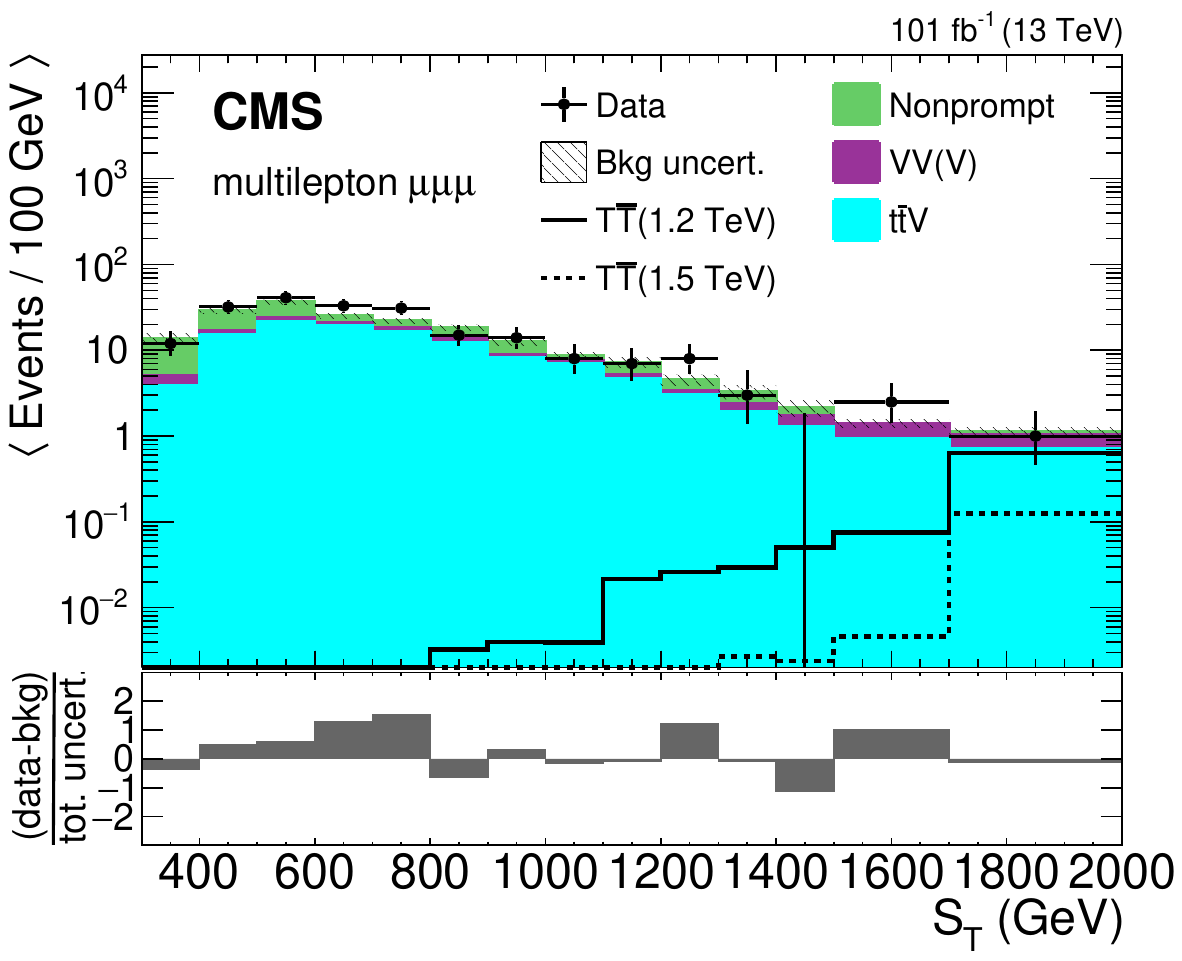}
  \caption{Template histograms of \ST in the multilepton signal region for \eee, \eem, \emm, and \mmm~categories (left-to-right, upper-to-lower).
    The observed data from 2017--2018 (combined for illustration) are shown using black markers, the predicted \TTbar signal for a mass of 1.2 (1.5)\TeV in the singlet scenario using solid (dashed) lines, and the post-fit background estimates, using filled histograms.
    Statistical and systematic uncertainties in the background estimate after performing the fit to data are shown by the hatched region.
    The lower panels show the difference between the data and the background estimate as a multiple of the total uncertainty in both sources.
  }
  \label{fig:3L1718_SRFinal_ST}
\end{figure}

In the SS dilepton channel, the {\HTlep} distribution is used to form template histograms in the three lepton flavor categories for 2017 and 2018 data, while results from 2016 data are included as a counting experiment.
In the multilepton channel, the \ST distribution is fitted in the four lepton flavor categories for all data-taking periods.
In both of these channels the template histograms from 2016 data are reproduced from Ref.~\cite{B2G-17-011}.
Figure~\ref{fig:SS2l1718_HT_final} shows the {\HTlep} templates for 2017 and 2018 data in the SS dilepton channel, in which the total background in each bin has a statistical uncertainty smaller than 30\%.
Figure~\ref{fig:3L1718_SRFinal_ST} shows the \ST templates for 2017 and 2018 data in the multilepton channel, with histograms binned to match Ref.~\cite{B2G-17-011}. Tables~\ref{tab:2lep}--\ref{tab:3lep} list the numbers of events selected for each category of the SS dilepton and multilepton channels.
Event yields for 2016 data in these two channels are reproduced from Ref.~\cite{B2G-17-011}.
No significant excess of data over the SM background estimate is observed in any channel.

\begin{table}[htbp]
  \centering
  \topcaption{
    Numbers of predicted and observed events in 2016--2018 data (138\fbinv) in the \TTbar SR categories considered in the single-lepton channel, after a background-only fit to data.
    Electron and muon categories have been combined for illustration.
    Predicted numbers of signal events before the fit to data are included for comparison, using the singlet branching fraction scenario.
    Uncertainties include statistical and systematic components, with the uncertainties in the electron and muon categories added in quadrature.
  }
  \begin{tabular}{lr@{ }c@{ }lr@{ }c@{ }lr@{ }c@{ }lr@{ }c@{ }l}
    Sample           & \multicolumn{3}{c}{\TTbar SR 1} & \multicolumn{3}{c}{\TTbar SR 2}  & \multicolumn{3}{c}{\TTbar SR 3}  & \multicolumn{3}{c}{\TTbar SR 4}                                                                                     \\[0.1em]
    \hline
    \\[-0.9em]
    \TTbar (1.2\TeV) & 4.9                             & $\pm$                            & 1.8                              & 4.6                              & $\pm$                    & 1.1   & 6.9   & $\pm$ & 1.5   & 6.0   & $\pm$ & 1.2   \\
    \TTbar (1.5\TeV) & 0.80                            & $\pm$                            & 0.30                             & 0.95                             & $\pm$                    & 0.22  & 1.33  & $\pm$ & 0.30  & 1.47  & $\pm$ & 0.32  \\[0.3em]
    TOP              & 3.44                            & $\pm$                            & 0.60                             & 16.9                             & $\pm$                    & 1.9   & 40.5  & $\pm$ & 2.9   & 31.6  & $\pm$ & 1.9   \\
    EW               & 3.80                            & $\pm$                            & 0.22                             & 3.10                             & $\pm$                    & 0.15  & 8.54  & $\pm$ & 0.35  & 2.34  & $\pm$ & 0.20  \\
    QCD              & 0.937                           & $\pm$                            & 0                                & \multicolumn{3}{c}{$<1$}         & \multicolumn{3}{c}{$<1$} & 1.52  & $\pm$ & 0.88                                  \\[0.5em]
    Total bkgd.      & 8.2                             & $\pm$                            & 1.0                              & 20.0                             & $\pm$                    & 1.9   & 40.9  & $\pm$ & 3.0   & 43.6  & $\pm$ & 2.1   \\
    Data             & \multicolumn{3}{c}{11}          & \multicolumn{3}{c}{19}           & \multicolumn{3}{c}{46}           & \multicolumn{3}{c}{39}                                                                                              \\
    Data/bkgd.       & 1.35                            & $\pm$                            & 0.44                             & 0.95                             & $\pm$                    & 0.24  & 1.12  & $\pm$ & 0.19  & 0.89  & $\pm$ & 0.15  \\[0.8em]
    Sample           & \multicolumn{3}{c}{\TTbar SR 5} & \multicolumn{3}{c}{\TTbar SR 6}  & \multicolumn{3}{c}{\TTbar SR 7}  & \multicolumn{3}{c}{\TTbar SR 8}                                                                                     \\[0.1em]
    \hline
    \\[-0.9em]
    \TTbar (1.2\TeV) & 8.9                             & $\pm$                            & 2.2                              & 4.13                             & $\pm$                    & 0.78  & 6.6   & $\pm$ & 1.1   & 3.95  & $\pm$ & 0.85  \\
    \TTbar (1.5\TeV) & 1.61                            & $\pm$                            & 0.39                             & 0.99                             & $\pm$                    & 0.19  & 1.57  & $\pm$ & 0.28  & 0.94  & $\pm$ & 0.20  \\[0.3em]
    TOP              & 70.8                            & $\pm$                            & 5.0                              & 37.4                             & $\pm$                    & 2.2   & 94.0  & $\pm$ & 4.4   & 96.6  & $\pm$ & 3.8   \\
    EW               & 31.2                            & $\pm$                            & 1.4                              & 7.03                             & $\pm$                    & 0.53  & 13.70 & $\pm$ & 0.91  & 6.86  & $\pm$ & 0.81  \\
    QCD              & 2.50                            & $\pm$                            & 0.64                             & 0.93                             & $\pm$                    & 0.57  & 0.71  & $\pm$ & 0.11  & 0.34  & $\pm$ & 0.22  \\[0.5em]
    Total bkgd.      & 104.5                           & $\pm$                            & 5.0                              & 45.3                             & $\pm$                    & 2.5   & 108.4 & $\pm$ & 4.3   & 103.8 & $\pm$ & 4.2   \\
    Data             & \multicolumn{3}{c}{100}         & \multicolumn{3}{c}{56}           & \multicolumn{3}{c}{112}          & \multicolumn{3}{c}{114}                                                                                             \\
    Data/bkgd.       & 0.96                            & $\pm$                            & 0.11                             & 1.24                             & $\pm$                    & 0.18  & 1.03  & $\pm$ & 0.11  & 1.10  & $\pm$ & 0.11  \\[0.8em]
    Sample           & \multicolumn{3}{c}{\TTbar SR 9} & \multicolumn{3}{c}{\TTbar SR 10} & \multicolumn{3}{c}{\TTbar SR 11} & \multicolumn{3}{c}{\TTbar SR 12}                                                                                    \\[0.1em]
    \hline
    \\[-0.9em]
    \TTbar (1.2\TeV) & 18.4                            & $\pm$                            & 3.4                              & 42.1                             & $\pm$                    & 5.5   & 16.4  & $\pm$ & 2.5   & 18.3  & $\pm$ & 3.1   \\
    \TTbar (1.5\TeV) & 3.64                            & $\pm$                            & 0.68                             & 8.4                              & $\pm$                    & 1.1   & 3.60  & $\pm$ & 0.54  & 3.44  & $\pm$ & 0.56  \\[0.3em]
    TOP              & 303.8                           & $\pm$                            & 7.8                              & 920                              & $\pm$                    & 15    & 525   & $\pm$ & 15    & 243.5 & $\pm$ & 7.1   \\
    EW               & 81.9                            & $\pm$                            & 3.0                              & 457                              & $\pm$                    & 10    & 112.8 & $\pm$ & 6.4   & 187.6 & $\pm$ & 4.9   \\
    QCD              & 9.3                             & $\pm$                            & 1.9                              & 63.5                             & $\pm$                    & 9.2   & 13.6  & $\pm$ & 4.8   & 23.3  & $\pm$ & 4.1   \\[0.5em]
    Total bkgd.      & 395.0                           & $\pm$                            & 9.5                              & 1440                             & $\pm$                    & 19    & 652   & $\pm$ & 13    & 454   & $\pm$ & 11    \\
    Data             & \multicolumn{3}{c}{416}         & \multicolumn{3}{c}{1525}         & \multicolumn{3}{c}{626}          & \multicolumn{3}{c}{499}                                                                                             \\
    Data/bkgd.       & 1.053                           & $\pm$                            & 0.057                            & 1.059                            & $\pm$                    & 0.030 & 0.961 & $\pm$ & 0.043 & 1.098 & $\pm$ & 0.056 \\
  \end{tabular}
  \label{tab:SRLTT}
\end{table}

\begin{table}[htbp]
  \centering
  \topcaption{
    Numbers of predicted and observed events in 2016--2018 data (138\fbinv) in the \BBbar SR categories considered in the single-lepton channel, after a background-only fit to data.
    Electron and muon categories have been combined for illustration.
    Predicted numbers of signal events before the fit to data are included for comparison, using the singlet branching fraction scenario.
    Uncertainties include statistical and systematic components, with the uncertainties in the electron and muon categories added in quadrature.
  }
  \begin{tabular}{l r@{ }c@{ }l r@{ }c@{ }l r@{ }c@{ }l r@{ }c@{ }l r@{ }c@{ }l}
    Sample           & \multicolumn{3}{c}{\BBbar SR 1} & \multicolumn{3}{c}{\BBbar SR 2} & \multicolumn{3}{c}{\BBbar SR 3}  & \multicolumn{3}{c}{\BBbar SR 5}                                                                                                                     \\[0.1em]
    \hline
    \\[-0.9em]
    \BBbar (1.2\TeV) & 5.0                             & $\pm$                           & 1.5                              & 4.4                              & $\pm$                            & 1.1   & 6.4   & $\pm$ & 1.6   & 10.7  & $\pm$ & 2.3   &       &       &       \\
    \BBbar (1.5\TeV) & 1.37                            & $\pm$                           & 0.44                             & 0.88                             & $\pm$                            & 0.23  & 1.27  & $\pm$ & 0.32  & 2.46  & $\pm$ & 0.54  &       &       &       \\[0.3em]
    TOP              & 13.5                            & $\pm$                           & 1.2                              & 17.7                             & $\pm$                            & 2.1   & 36.7  & $\pm$ & 3.1   & 75.4  & $\pm$ & 3.8   &       &       &       \\
    EW               & 2.72                            & $\pm$                           & 0.66                             & 3.23                             & $\pm$                            & 0.17  & 9.03  & $\pm$ & 0.45  & 19.9  & $\pm$ & 1.9   &       &       &       \\
    QCD              & \multicolumn{3}{c}{$<1$}        & 0.25                            & $\pm$                            & 0.14                             & \multicolumn{3}{c}{$<1$}         & 1.51  & $\pm$ & 0.62  &       &       &                                       \\[0.5em]
    Total bkgd.      & 16.2                            & $\pm$                           & 1.5                              & 21.2                             & $\pm$                            & 2.1   & 45.8  & $\pm$ & 3.1   & 96.8  & $\pm$ & 3.7   &       &       &       \\
    Data             & \multicolumn{3}{c}{22}          & \multicolumn{3}{c}{19}          & \multicolumn{3}{c}{43}           & \multicolumn{3}{c}{89}           &                                  &       &                                                                       \\
    Data/bkgd.       & 1.36                            & $\pm$                           & 0.32                             & 0.90                             & $\pm$                            & 0.23  & 0.94  & $\pm$ & 0.16  & 0.92  & $\pm$ & 0.10  &       &       &       \\[0.8em]
    Sample           & \multicolumn{3}{c}{\BBbar SR 8} & \multicolumn{3}{c}{\BBbar SR 9} & \multicolumn{3}{c}{\BBbar SR 10} & \multicolumn{3}{c}{\BBbar SR 11} & \multicolumn{3}{c}{\BBbar SR 12}                                                                                 \\[0.1em]
    \hline
    \\[-0.9em]
    \BBbar (1.2\TeV) & 6.1                             & $\pm$                           & 1.4                              & 13.9                             & $\pm$                            & 3.0   & 37.8  & $\pm$ & 5.4   & 17.5  & $\pm$ & 2.8   & 11.2  & $\pm$ & 1.9   \\
    \BBbar (1.5\TeV) & 1.44                            & $\pm$                           & 0.35                             & 2.81                             & $\pm$                            & 0.62  & 7.8   & $\pm$ & 1.1   & 3.8   & $\pm$ & 0.6   & 2.07  & $\pm$ & 0.35  \\[0.3em]
    TOP              & 80.7                            & $\pm$                           & 3.8                              & 255.5                            & $\pm$                            & 8.0   & 793   & $\pm$ & 15    & 394   & $\pm$ & 16    & 200.2 & $\pm$ & 6.6   \\
    EW               & 8.0                             & $\pm$                           & 1.2                              & 81.4                             & $\pm$                            & 3.3   & 435   & $\pm$ & 11    & 116   & $\pm$ & 8.5   & 165.3 & $\pm$ & 5.1   \\
    QCD              & 0.88                            & $\pm$                           & 0.37                             & 9.0                              & $\pm$                            & 1.5   & 65.3  & $\pm$ & 9.4   & 10.8  & $\pm$ & 2.6   & 18.5  & $\pm$ & 4.4   \\[0.5em]
    Total bkgd.      & 89.6                            & $\pm$                           & 4.4                              & 345.9                            & $\pm$                            & 9.6   & 1294  & $\pm$ & 20    & 521   & $\pm$ & 11    & 384   & $\pm$ & 10    \\
    Data             & \multicolumn{3}{c}{90}          & \multicolumn{3}{c}{357}         & \multicolumn{3}{c}{1378}         & \multicolumn{3}{c}{482}          & \multicolumn{3}{c}{434}                                                                                          \\
    Data/bkgd.       & 1.00                            & $\pm$                           & 0.12                             & 1.032                            & $\pm$                            & 0.062 & 1.065 & $\pm$ & 0.033 & 0.926 & $\pm$ & 0.046 & 1.130 & $\pm$ & 0.062 \\
  \end{tabular}
  \label{tab:SRLBB}
\end{table}

\begin{table}[htbp]
  \centering
  \topcaption{
    Numbers of predicted and observed SR events in 2017--2018 data (101\fbinv) in the SS dilepton channel, after a background-only fit to data.
    Predicted numbers of signal events before the fit to data are included for comparison, using the singlet branching fraction scenario.
    Uncertainties include statistical and systematic components.
    Predictions for 2017 and 2018 are combined for illustration with their uncertainties added in quadrature.
  }
  \begin{tabular}{l r@{ }c@{ }l r@{ }c@{ }l r@{ }c@{ }l}
    Sample                                 & \multicolumn{3}{c}{{\Pe}\Pe} & \multicolumn{3}{c}{{\Pe}{\PGm}} & \multicolumn{3}{c}{{\PGm}\PGm}                                                 \\[0.1em]
    \hline
    \\[-0.9em]
    \TTbar (1.2\TeV)                       & 0.968                        & $\pm$                           & 0.053                          & 2.52  & $\pm$ & 0.12  & 1.651 & $\pm$ & 0.092 \\
    \TTbar (1.5\TeV)                       & 0.142                        & $\pm$                           & 0.008                          & 0.404 & $\pm$ & 0.023 & 0.270 & $\pm$ & 0.015 \\[0.3em]
    \BBbar (1.2\TeV)                       & 2.24                         & $\pm$                           & 0.12                           & 5.85  & $\pm$ & 0.28  & 3.63  & $\pm$ & 0.20  \\
    \BBbar (1.5\TeV)                       & 0.357                        & $\pm$                           & 0.021                          & 0.961 & $\pm$ & 0.070 & 0.588 & $\pm$ & 0.032 \\[0.3em]
    $\ttbar\PV$+$\ttbar\PH$+$\ttbar\ttbar$ & 148                          & $\pm$                           & 12                             & 407   & $\pm$ & 32    & 246   & $\pm$ & 19    \\
    $\PV\PV$(\PV)                          & 53.7                         & $\pm$                           & 4.6                            & 129   & $\pm$ & 11    & 72.6  & $\pm$ & 6.6   \\
    Charge misid.                          & 114                          & $\pm$                           & 44                             & 107   & $\pm$ & 42    & 0.00  & $\pm$ & 0.00  \\
    Nonprompt                              & 259                          & $\pm$                           & 41                             & 766   & $\pm$ & 51    & 539   & $\pm$ & 29    \\[0.5em]
    Total bkgd.                            & 575                          & $\pm$                           & 31                             & 1409  & $\pm$ & 37    & 857   & $\pm$ & 25    \\
    Data                                   & \multicolumn{3}{c}{582}      & \multicolumn{3}{c}{1415}        & \multicolumn{3}{c}{819}                                                        \\
    Data/bkgd.                             & 1.012                        & $\pm$                           & 0.069                          & 1.004 & $\pm$ & 0.037 & 0.956 & $\pm$ & 0.044 \\
  \end{tabular}
  \label{tab:2lep}
\end{table}

\begin{table}[htbp]
  \centering
  \topcaption{
    Numbers of predicted and observed SR events in 2017--2018 data (101\fbinv) in the multilepton channel, after a background-only fit to data.
    Predicted numbers of signal events before the fit to data are included for comparison, using the singlet branching fraction scenario.
    Uncertainties include statistical and systematic components.
    Predictions for 2017 and 2018 are combined for illustration with their uncertainties added in quadrature.
  }
  \begin{tabular}{l r@{ }c@{ }l r@{ }c@{ }l r@{ }c@{ }l r@{ }c@{ }l}
    Sample           & \multicolumn{3}{c}{\eee} & \multicolumn{3}{c}{\eem} & \multicolumn{3}{c}{\emm} & \multicolumn{3}{c}{\mmm}                                                                 \\[0.1em]
    \hline
    \\[-0.9em]
    \TTbar (1.2\TeV) & 1.39                     & $\pm$                    & 0.10                     & 2.43                     & $\pm$ & 0.15  & 2.86  & $\pm$ & 0.17  & 2.19  & $\pm$ & 0.15  \\
    \TTbar (1.5\TeV) & 0.220                    & $\pm$                    & 0.018                    & 0.404                    & $\pm$ & 0.026 & 0.497 & $\pm$ & 0.033 & 0.396 & $\pm$ & 0.027 \\[0.3em]
    \BBbar (1.2\TeV) & 1.15                     & $\pm$                    & 0.08                     & 2.86                     & $\pm$ & 0.17  & 3.32  & $\pm$ & 0.19  & 1.99  & $\pm$ & 0.13  \\
    \BBbar (1.5\TeV) & 0.198                    & $\pm$                    & 0.015                    & 0.472                    & $\pm$ & 0.028 & 0.554 & $\pm$ & 0.032 & 0.354 & $\pm$ & 0.027 \\[0.3em]
    $\ttbar\PV$      & 78.2                     & $\pm$                    & 4.3                      & 116.5                    & $\pm$ & 6.1   & 132.4 & $\pm$ & 6.9   & 119.3 & $\pm$ & 5.8   \\
    $\PV\PV$(\PV)    & 11.51                    & $\pm$                    & 0.67                     & 12.87                    & $\pm$ & 0.68  & 14.69 & $\pm$ & 0.80  & 15.19 & $\pm$ & 0.77  \\
    Nonprompt        & 50.1                     & $\pm$                    & 8.2                      & 105.3                    & $\pm$ & 9.8   & 117.8 & $\pm$ & 9.8   & 55.9  & $\pm$ & 9.0   \\[0.5em]
    Total bkgd.      & 139.8                    & $\pm$                    & 8.1                      & 234.7                    & $\pm$ & 9.1   & 265   & $\pm$ & 10    & 190.3 & $\pm$ & 9.3   \\
    Data             & \multicolumn{3}{c}{159}  & \multicolumn{3}{c}{255}  & \multicolumn{3}{c}{304}  & \multicolumn{3}{c}{212}                                                                  \\
    Data/bkgd.       & 1.14                     & $\pm$                    & 0.11                     & 1.086                    & $\pm$ & 0.080 & 1.148 & $\pm$ & 0.079 & 1.114 & $\pm$ & 0.094 \\
  \end{tabular}
  \label{tab:3lep}
\end{table}

Upper limits at 95\% \CL on the production cross sections of \TTbar and \BBbar pairs are set using a binned maximum likelihood fit to all categories for a total of 54 (48) templates in the \TTbar (\BBbar) analysis. The searches for each VLQ flavor are independent, with only one flavor considered in the signal templates.
Uncertainties due to limited event counts in simulated samples are included as Poisson-distributed nuisance parameters using the Barlow--Beeston method~\cite{BBLITE1,BBLITE2}.
Systematic uncertainties listed in Table~\ref{table:Systematics} are included as lognormal-distributed nuisance parameters if they affect only the normalization of the background or signal predictions.
Systematic uncertainties affecting the shapes of the template distributions are treated using template morphing with Gaussian probability distributions~\cite{BBLITE2}. The fit model is validated by studying the results of fits to pseudo-data generated from the background estimate, with and without known rates of \TTbar or \BBbar signal injected. 
In the \TTbar (\BBbar) analysis the fit to data has 1361 (1219) parameters and a goodness-of-fit measure from a saturated $\chi^2$ model of 1379 (1207).
Expected limits at 95\% \CL are calculated using a profile likelihood test statistic in the asymptotic approximation~\cite{Cowan}, with the \CLs method~\cite{CLS1,CLS2}.

\begin{figure}[htbp]
  \centering
  \includegraphics[width=0.49\textwidth]{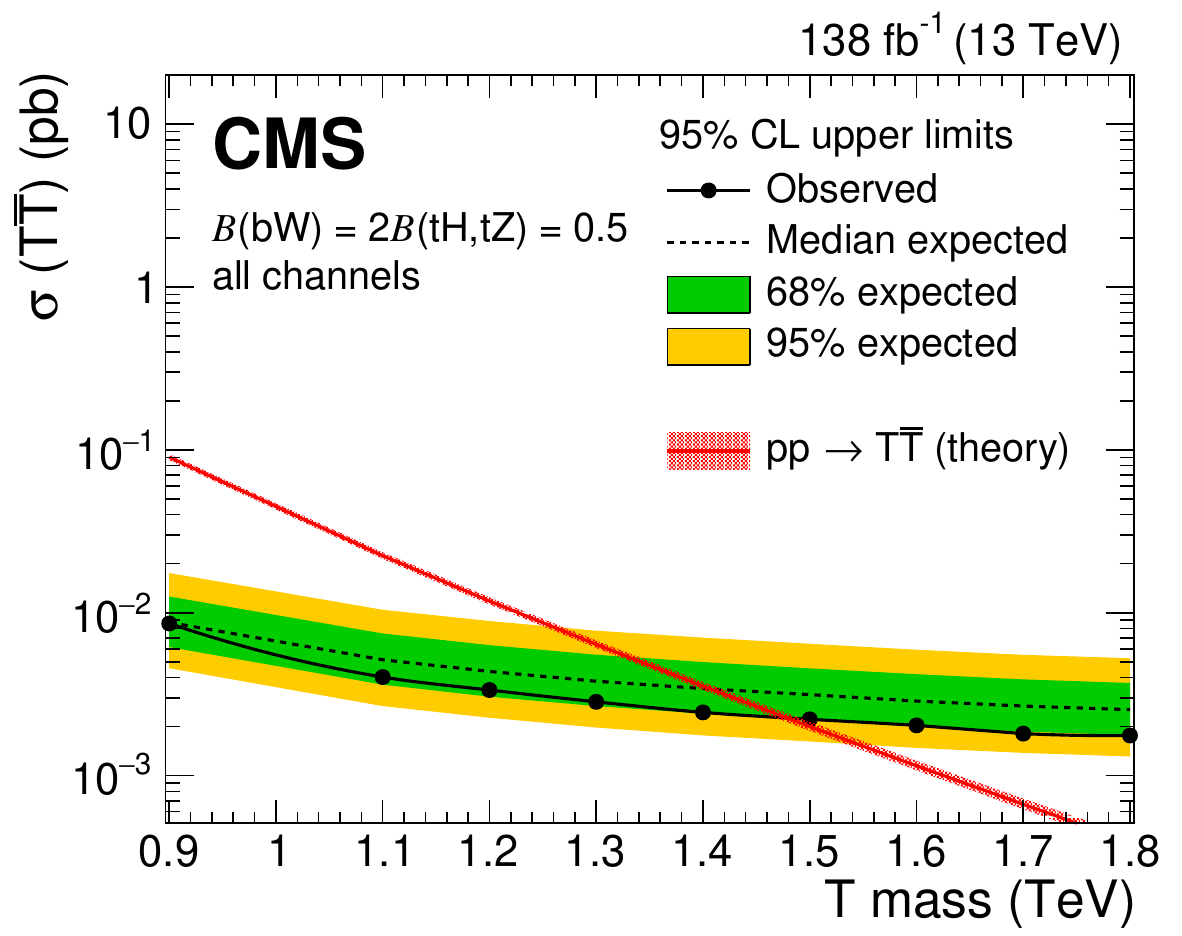}
  \includegraphics[width=0.49\textwidth]{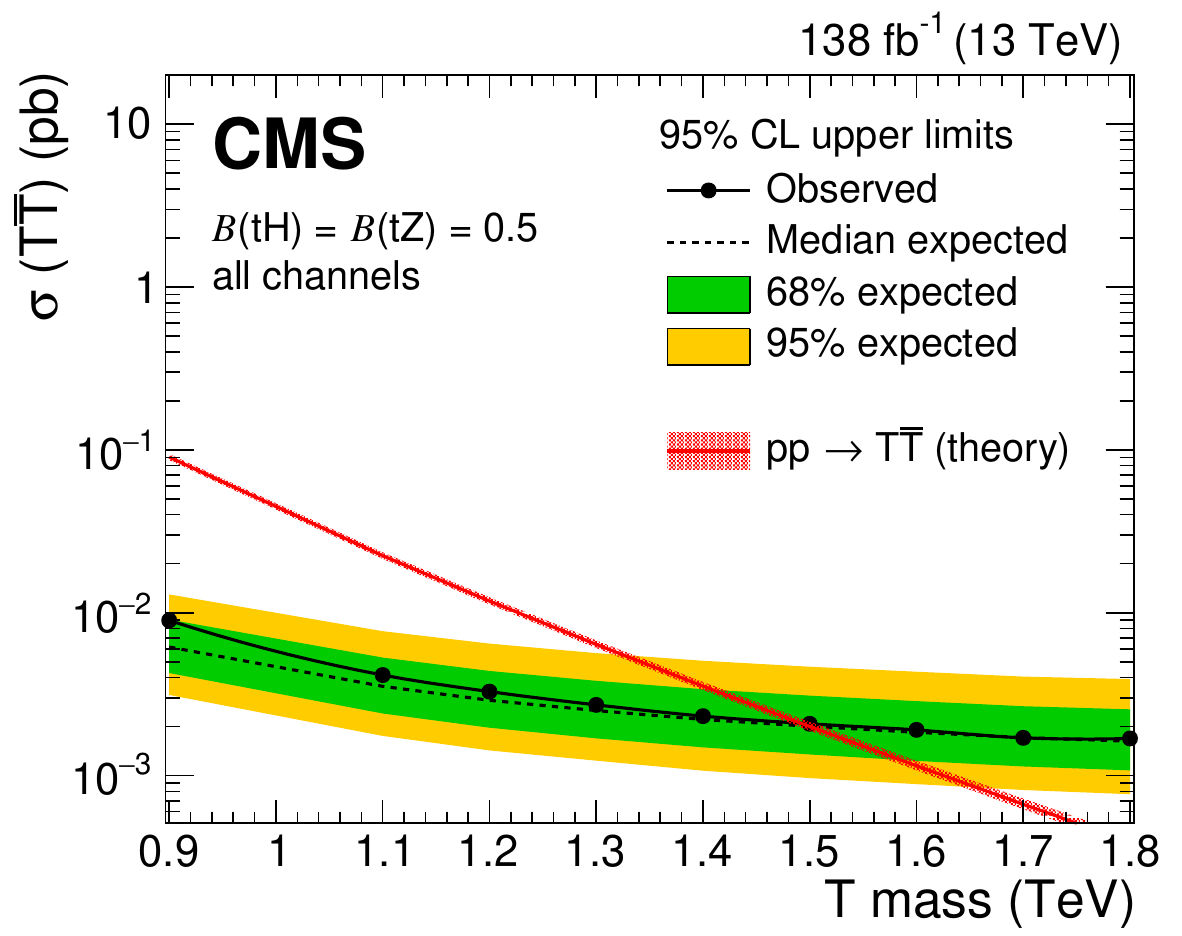}\\
  \includegraphics[width=0.49\textwidth]{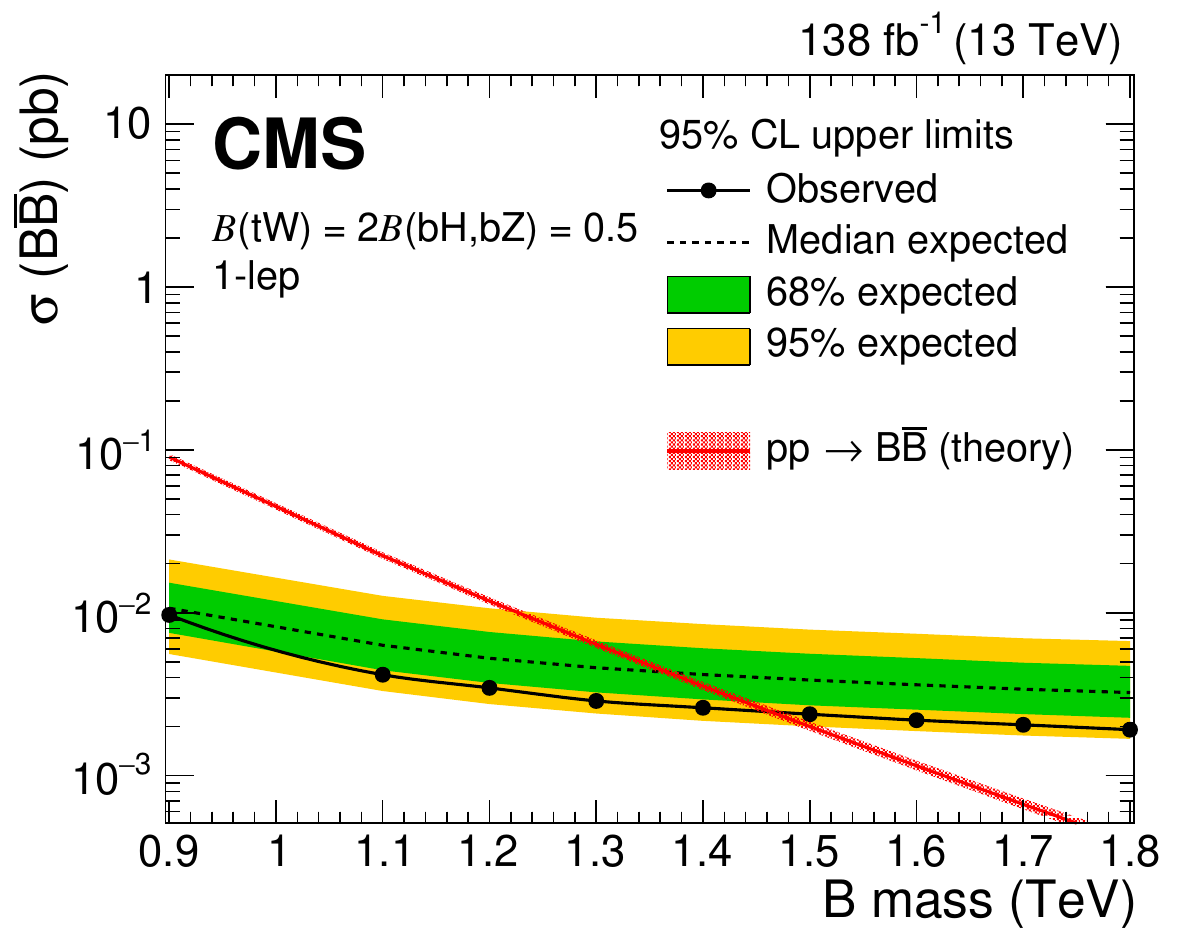}
  \includegraphics[width=0.49\textwidth]{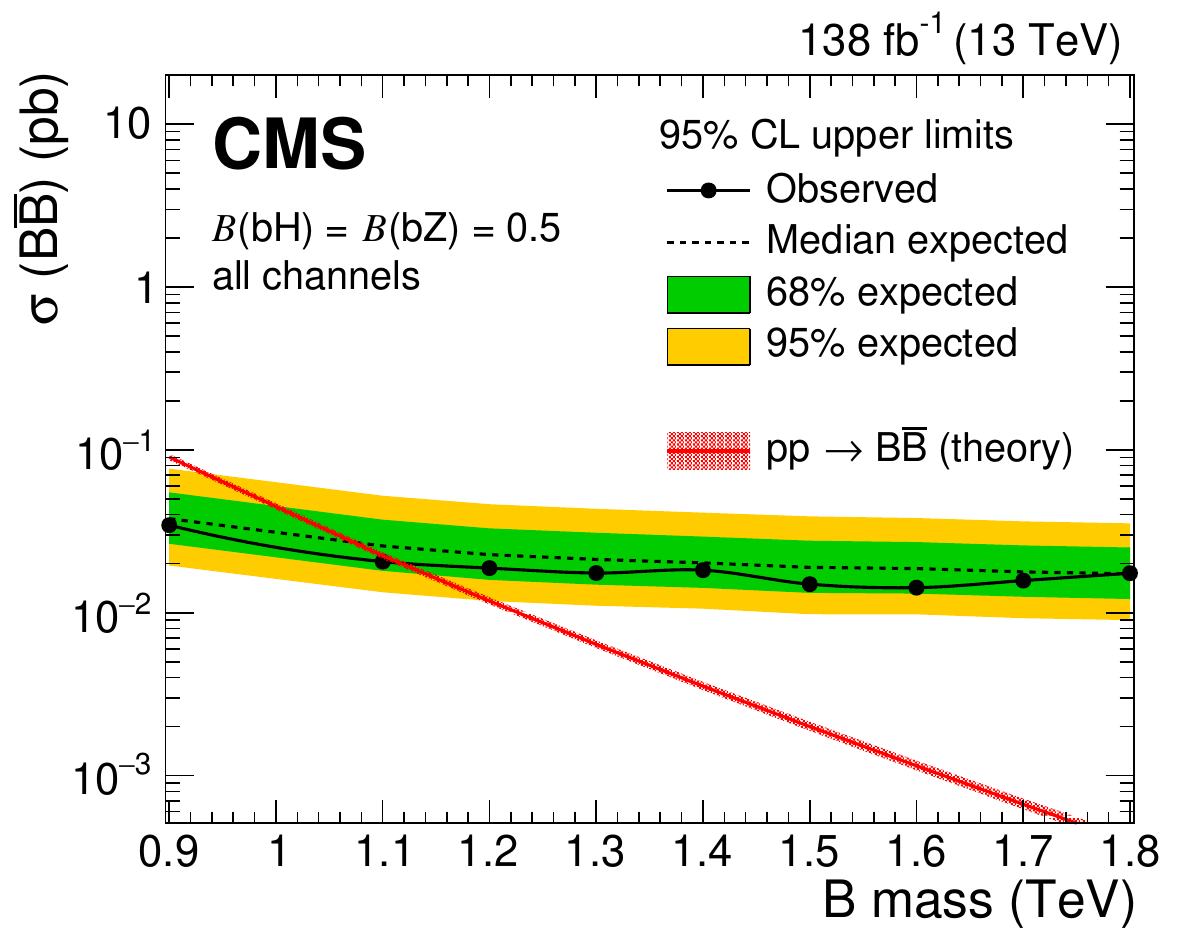}
  \caption{
    Observed (solid lines) and expected (dashed lines) 95\% \CL upper limits on \TTbar (upper) and \BBbar (lower) production cross sections for the singlet (left) and doublet (right) hypotheses, from the combined fit to all channels.
    Predicted cross sections are shown by the red line surrounded by a band representing energy scale and PDF uncertainties in the calculation.
  }
  \label{fig:123leplimits}
\end{figure}

Limits on the \TTbar and \BBbar production cross sections for both benchmark branching fraction scenarios are shown in Figure~\ref{fig:123leplimits}, where the band around the theory prediction indicates the scale and PDF uncertainties expected at NNLO. 
The corresponding mass limits are listed in Table~\ref{tab:123lepLimits}, along with 95\% \CL limits for scenarios with 100\% branching fractions to \PH, \PW, or \PZ bosons. The maximum likelihood fit produces a slight deficit of data in the signal-enriched regions of single-lepton SRs 2, 3, 5, and 9, contributing to an observed limit that is stronger than the expected limit in branching fractions with significant \PW boson contributions.

\begin{figure}[htbp]
  \centering
  \includegraphics[width=0.48\textwidth]{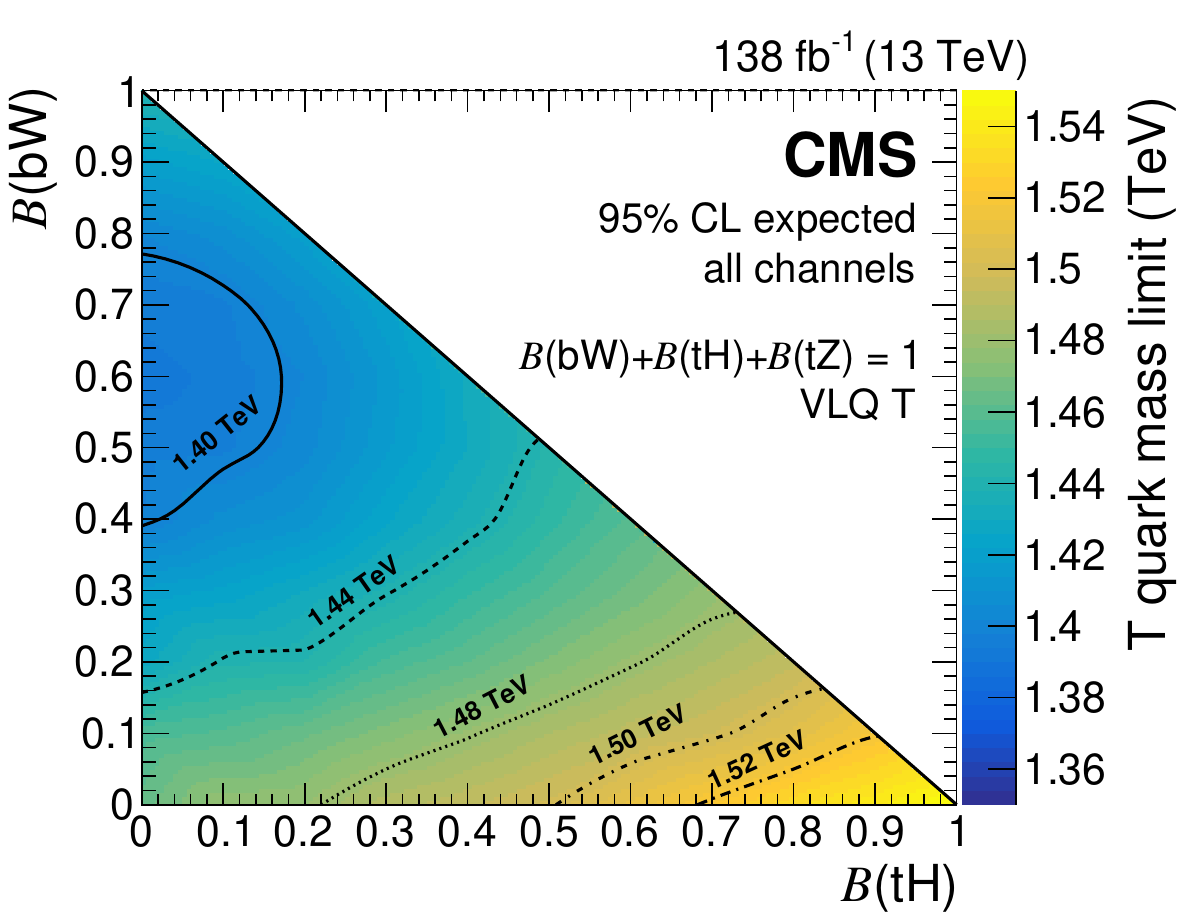}
  \includegraphics[width=0.48\textwidth]{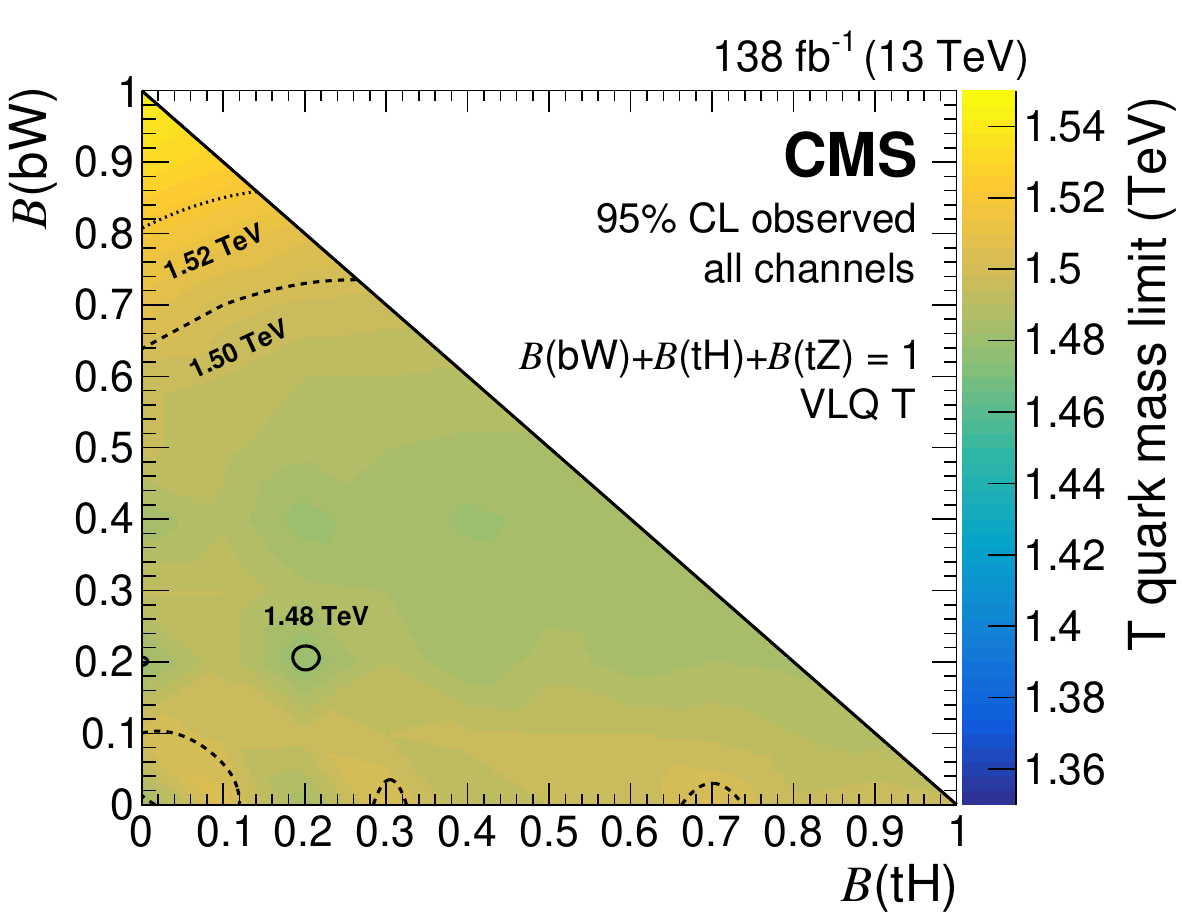}
  \includegraphics[width=0.48\textwidth]{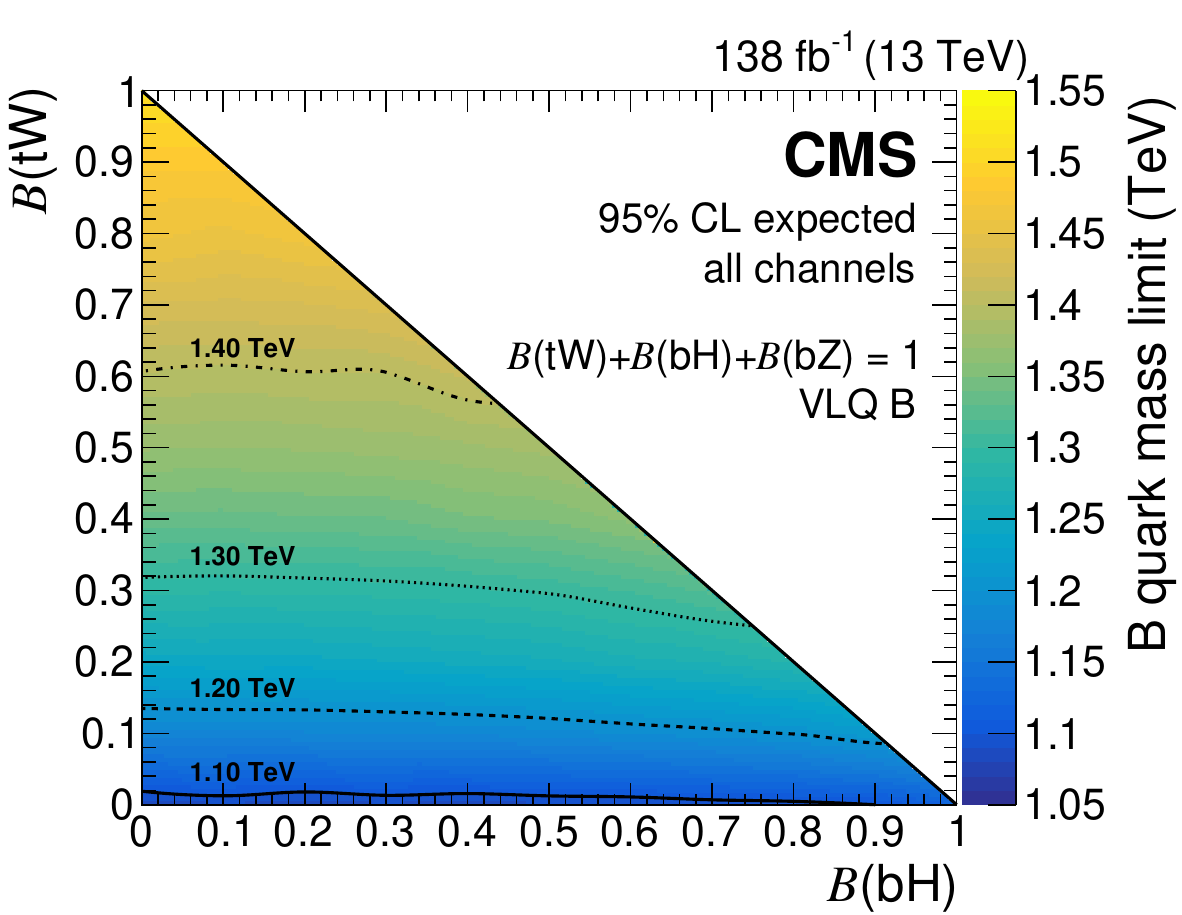}
  \includegraphics[width=0.48\textwidth]{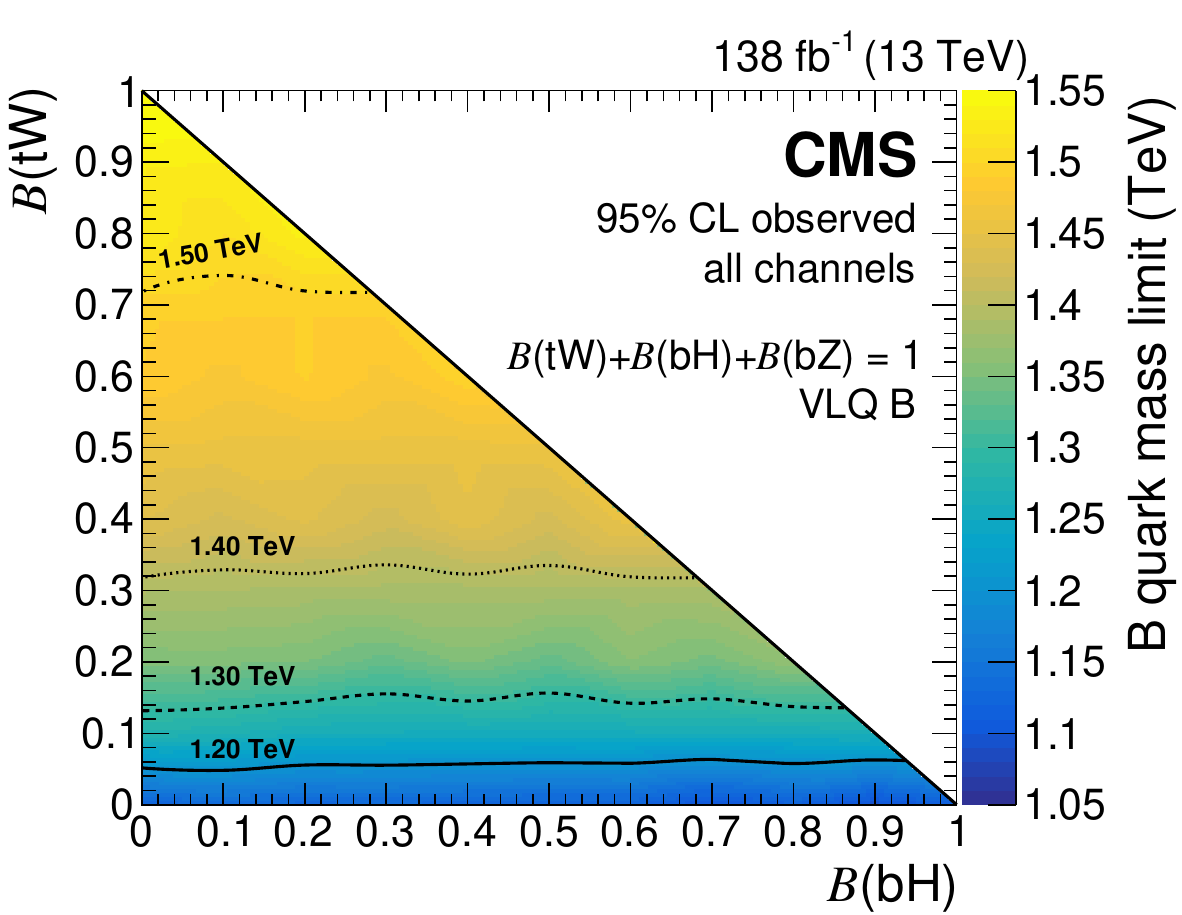}
  \caption{
    The 95\% \CL expected (left) and observed (right) lower mass limits on pair-produced \PQT (upper) and \PQB (lower) quark masses, from the combined fit to all channels, as functions of their branching ratios to $\PH$ and $\PW$ bosons.
    Mass contours are shown with lines of various styles.
  }
  \label{fig:boxLimits}
\end{figure}

Figure~\ref{fig:boxLimits} presents a scan over many possible branching fraction combinations, varying branching fractions in steps of 0.1 and requiring that $\mathcal{B}(\PQq\PW) + \mathcal{B}(\PQq\PH) + \mathcal{B}(\PQq\PZ) = 1$.
From the scan, we exclude \PQT quarks with masses below 1.48--1.54\TeV and \PQB quarks with masses below 1.12--1.56\TeV, depending on the branching fraction.
For both VLQs the strongest sensitivity is to decay modes with multiple top quarks: $\TTbar\to \PQt\PH\PQt\PH$ and $\BBbar \to \PQt\PW\PQt\PW$. The sensitivity of the search is dominated by the single-lepton channel, with important contributions derived particularly from the multilepton channel in branching fraction scenarios with significant $\PQt\PZ$ decays.

These limits are the strongest to date for \TTbar production and for \BBbar production with \PQB quark decays to \PW bosons. In this analysis, the use of the neural network jet identification and event classification methods in the single-lepton channel contributed significantly to limits that reach beyond those expected simply from the increased size of the present data set over that used previously.

\begin{table}[htbp]
  \centering
  \topcaption{Expected and observed 95\% \CL lower limits on the \PQT (upper section) and \PQB (lower section) quark masses.}
  \begin{tabular}{lccccc}
    Scenario  & $\mathcal{B}(\PQb\PW)$ & $\mathcal{B}(\PQt\PZ)$ & $\mathcal{B}(\PQt\PH)$ & Expected (\TeVns) & Observed (\TeVns) \\
    \hline
    $\PQb\PW$ & 1.0                    & \NA                    & \NA                    & 1.44              & 1.54              \\
    $\PQt\PZ$ & \NA                    & 1.0                    & \NA                    & 1.46              & 1.48              \\
    $\PQt\PH$ & \NA                    & \NA                    & 1.0                    & 1.55              & 1.50              \\
    Singlet   & 0.50                   & 0.25                   & 0.25                   & 1.41              & 1.48              \\
    Doublet   & \NA                    & 0.50                   & 0.50                   & 1.50              & 1.49              \\[0.8em]
    Scenario  & $\mathcal{B}(\PQt\PW)$ & $\mathcal{B}(\PQb\PZ)$ & $\mathcal{B}(\PQb\PH)$ & Expected (\TeVns) & Observed (\TeVns) \\
    \hline
    $\PQt\PW$ & 1.0                    & \NA                    & \NA                    & 1.51              & 1.56              \\
    $\PQb\PZ$ & \NA                    & 1.0                    & \NA                    & 1.08              & 1.12              \\
    $\PQb\PH$ & \NA                    & \NA                    & 1.0                    & 1.12              & 1.12              \\
    Singlet   & 0.50                   & 0.25                   & 0.25                   & 1.37              & 1.47              \\
    Doublet   & \NA                    & 0.50                   & 0.50                   & 1.09              & 1.12              \\
  \end{tabular}
  \label{tab:123lepLimits}
\end{table}

\section{Summary}
\label{sec:summary}

A search has been presented for vector-like \PQT and \PQB quark-antiquark pairs produced in proton-proton collisions at a center-of-mass energy of 13\TeV.
Data collected by the CMS experiment at the LHC in 2016--2018 are analyzed in the single-lepton final state, and data from 2017--2018 are analyzed in the same-sign charge dilepton and multilepton final states.
In the single-lepton channel, parent particles of large-radius jets are identified using the \textsc{DeepAK8} algorithm, and vector-like quark candidates are reconstructed.
A multilayer perceptron network is trained to separate signal events from standard model backgrounds.
In the same-sign charge dilepton and multilepton channels, low background rates and the large energy signature of the signal are exploited by studying jet and lepton momentum scalar sum distributions.
Pair production is excluded at 95\% confidence level for \PQT quarks with masses up to 1.54\TeV and for \PQB quarks with masses up to 1.56\TeV, depending on the branching fraction scenario, and \PQT quarks with masses below 1.48\TeV are excluded in any scenario.
The limits obtained in this search are the strongest limits to date for \TTbar production with all \PQT quark decay modes, and are the strongest limits to date for \BBbar production with \PQB quark decays to $\PQt\PW$.

\begin{acknowledgments}
We congratulate our colleagues in the CERN accelerator departments for the excellent performance of the LHC and thank the technical and administrative staffs at CERN and at other CMS institutes for their contributions to the success of the CMS effort. In addition, we gratefully acknowledge the computing centers and personnel of the Worldwide LHC Computing Grid and other centers for delivering so effectively the computing infrastructure essential to our analyses. Finally, we acknowledge the enduring support for the construction and operation of the LHC, the CMS detector, and the supporting computing infrastructure provided by the following funding agencies: BMBWF and FWF (Austria); FNRS and FWO (Belgium); CNPq, CAPES, FAPERJ, FAPERGS, and FAPESP (Brazil); MES and BNSF (Bulgaria); CERN; CAS, MoST, and NSFC (China); MINCIENCIAS (Colombia); MSES and CSF (Croatia); RIF (Cyprus); SENESCYT (Ecuador); MoER, ERC PUT and ERDF (Estonia); Academy of Finland, MEC, and HIP (Finland); CEA and CNRS/IN2P3 (France); BMBF, DFG, and HGF (Germany); GSRI (Greece); NKFIH (Hungary); DAE and DST (India); IPM (Iran); SFI (Ireland); INFN (Italy); MSIP and NRF (Republic of Korea); MES (Latvia); LAS (Lithuania); MOE and UM (Malaysia); BUAP, CINVESTAV, CONACYT, LNS, SEP, and UASLP-FAI (Mexico); MOS (Montenegro); MBIE (New Zealand); PAEC (Pakistan); MES and NSC (Poland); FCT (Portugal); MESTD (Serbia); MCIN/AEI and PCTI (Spain); MOSTR (Sri Lanka); Swiss Funding Agencies (Switzerland); MST (Taipei); MHESI and NSTDA (Thailand); TUBITAK and TENMAK (Turkey); NASU (Ukraine); STFC (United Kingdom); DOE and NSF (USA).
  
\hyphenation{Rachada-pisek} Individuals have received support from the Marie-Curie program and the European Research Council and Horizon 2020 Grant, contract Nos.\ 675440, 724704, 752730, 758316, 765710, 824093, 884104, and COST Action CA16108 (European Union); the Leventis Foundation; the Alfred P.\ Sloan Foundation; the Alexander von Humboldt Foundation; the Belgian Federal Science Policy Office; the Fonds pour la Formation \`a la Recherche dans l'Industrie et dans l'Agriculture (FRIA-Belgium); the Agentschap voor Innovatie door Wetenschap en Technologie (IWT-Belgium); the F.R.S.-FNRS and FWO (Belgium) under the ``Excellence of Science -- EOS" -- be.h project n.\ 30820817; the Beijing Municipal Science \& Technology Commission, No. Z191100007219010; the Ministry of Education, Youth and Sports (MEYS) of the Czech Republic; the Hellenic Foundation for Research and Innovation (HFRI), Project Number 2288 (Greece); the Deutsche Forschungsgemeinschaft (DFG), under Germany's Excellence Strategy -- EXC 2121 ``Quantum Universe" -- 390833306, and under project number 400140256 - GRK2497; the Hungarian Academy of Sciences, the New National Excellence Program - \'UNKP, the NKFIH research grants K 124845, K 124850, K 128713, K 128786, K 129058, K 131991, K 133046, K 138136, K 143460, K 143477, 2020-2.2.1-ED-2021-00181, and TKP2021-NKTA-64 (Hungary); the Council of Science and Industrial Research, India; the Latvian Council of Science; the Ministry of Education and Science, project no. 2022/WK/14, and the National Science Center, contracts Opus 2021/41/B/ST2/01369 and 2021/43/B/ST2/01552 (Poland); the Funda\c{c}\~ao para a Ci\^encia e a Tecnologia, grant CEECIND/01334/2018 (Portugal); the National Priorities Research Program by Qatar National Research Fund; MCIN/AEI/10.13039/501100011033, ERDF ``a way of making Europe", and the Programa Estatal de Fomento de la Investigaci{\'o}n Cient{\'i}fica y T{\'e}cnica de Excelencia Mar\'{\i}a de Maeztu, grant MDM-2017-0765 and Programa Severo Ochoa del Principado de Asturias (Spain); the Chulalongkorn Academic into Its 2nd Century Project Advancement Project, and the National Science, Research and Innovation Fund via the Program Management Unit for Human Resources \& Institutional Development, Research and Innovation, grant B05F650021 (Thailand); the Kavli Foundation; the Nvidia Corporation; the SuperMicro Corporation; the Welch Foundation, contract C-1845; and the Weston Havens Foundation (USA).  
\end{acknowledgments}

\bibliography{auto_generated}
\cleardoublepage \appendix\section{The CMS Collaboration \label{app:collab}}\begin{sloppypar}\hyphenpenalty=5000\widowpenalty=500\clubpenalty=5000
\cmsinstitute{Yerevan Physics Institute, Yerevan, Armenia}
{\tolerance=6000
A.~Tumasyan\cmsAuthorMark{1}\cmsorcid{0009-0000-0684-6742}
\par}
\cmsinstitute{Institut f\"{u}r Hochenergiephysik, Vienna, Austria}
{\tolerance=6000
W.~Adam\cmsorcid{0000-0001-9099-4341}, J.W.~Andrejkovic, T.~Bergauer\cmsorcid{0000-0002-5786-0293}, S.~Chatterjee\cmsorcid{0000-0003-2660-0349}, K.~Damanakis\cmsorcid{0000-0001-5389-2872}, M.~Dragicevic\cmsorcid{0000-0003-1967-6783}, A.~Escalante~Del~Valle\cmsorcid{0000-0002-9702-6359}, P.S.~Hussain\cmsorcid{0000-0002-4825-5278}, M.~Jeitler\cmsAuthorMark{2}\cmsorcid{0000-0002-5141-9560}, N.~Krammer\cmsorcid{0000-0002-0548-0985}, L.~Lechner\cmsorcid{0000-0002-3065-1141}, D.~Liko\cmsorcid{0000-0002-3380-473X}, I.~Mikulec\cmsorcid{0000-0003-0385-2746}, P.~Paulitsch, F.M.~Pitters, J.~Schieck\cmsAuthorMark{2}\cmsorcid{0000-0002-1058-8093}, R.~Sch\"{o}fbeck\cmsorcid{0000-0002-2332-8784}, D.~Schwarz\cmsorcid{0000-0002-3821-7331}, S.~Templ\cmsorcid{0000-0003-3137-5692}, W.~Waltenberger\cmsorcid{0000-0002-6215-7228}, C.-E.~Wulz\cmsAuthorMark{2}\cmsorcid{0000-0001-9226-5812}
% ...
\par}
\vskip\cmsinstskip
\dag:~Deceased\\
$^{1}$Also at Yerevan State University, Yerevan, Armenia\\
$^{2}$Also at TU Wien, Vienna, Austria\\
% ...
\end{sloppypar}
\end{document}